\documentclass[onecolumn,11pt]{article}
\usepackage[top=1in, bottom=1in, left=1in, right=1in]{geometry}
\usepackage{amsmath,amsfonts,amscd,amssymb}
\usepackage[utf8]{inputenc} 
\usepackage[T1]{fontenc}    
\usepackage{etoolbox}
\appto\UrlBreaks{\do\-}
\usepackage{url}            
\usepackage{booktabs}       
\usepackage{amsfonts}       
\usepackage{nicefrac}       
\usepackage{microtype}      
\usepackage{indentfirst}
\usepackage{hyphenat}
\usepackage{amsmath}
\usepackage{amssymb}
\usepackage{mathrsfs} 
\usepackage{mathtools}
 \usepackage{tocloft}

\usepackage{graphicx}
\usepackage{caption}
\usepackage{float}
\usepackage[abs]{overpic}
\usepackage{subfig}
\usepackage{epstopdf}
\usepackage{rotating}
\usepackage{multirow}
\usepackage{wrapfig}

\usepackage{enumerate}
\usepackage[numbers,sort&compress]{natbib}
\usepackage{diagbox} 
\usepackage{longtable} 
\usepackage{pdflscape}

\usepackage{algorithm}
\usepackage[noend]{algpseudocode}
\usepackage[utf8]{inputenc}
\usepackage{fontenc}
\usepackage{lipsum}

\usepackage[bottom,flushmargin,hang,multiple]{footmisc}
\usepackage{lipsum}
\usepackage{cancel}
\usepackage{url}
\usepackage{color}
\usepackage{mathtools}
\usepackage{subeqnarray}
\usepackage{setspace}
\usepackage{palatino} 
\usepackage{appendix}
\usepackage{tabu} 

\usepackage[breaklinks=true]{hyperref}
\hypersetup{colorlinks=true,linkcolor=blue,urlcolor=blue,citecolor=blue}

\definecolor{header1}{cmyk}{0,0,0,1}

\newcommand{\bY}{\mathbf{Y}}

\newcommand\blfootnote[1]{%
  \begingroup
  \renewcommand\thefootnote{}\footnote{#1}%
  \addtocounter{footnote}{-1}%
  \endgroup
}

\title{\vspace{-.4in}{\fontsize{17}{17}\selectfont \textbf{The Experimental Multi-Arm Pendulum on a Cart: \\
A Benchmark System for Chaos, Learning, and Control}}\vspace{-.15in}}

\begin{document}

\author{\normalsize{Kadierdan Kaheman$^{1,*}$,
Urban Fasel$^1$,
Jason J. Bramburger$^2$,
Benjamin Strom$^3$,}
\\
\normalsize{J. Nathan Kutz$^4$,
Steven L. Brunton$^1$}\\
\footnotesize{$^1$ Department of Mechanical Engineering, University of Washington, Seattle, WA 98195, United States of America}\\
\footnotesize{$^2$ Department of Mathematics and Statistics, Concordia University, Montr\'eal, QC H3G 1M8, Canada}\\
\footnotesize{$^3$ XFlow Energy Company, Seattle, WA, 98108, United States of America}\\
\footnotesize{$^4$ Department of Applied Mathematics, University of Washington, Seattle, WA 98195, United States of America\vspace{-.2in}}
}


\date{}

\maketitle

\blfootnote{$^*$ Corresponding author (kadierk@uw.edu); Code availalbe at \href{https://github.com/dynamicslab/MultiArm-Pendulum}{https://github.com/dynamicslab/MultiArm-Pendulum}}

\vspace{-.2in}
\begin{abstract}
The single, double, and triple pendulum has served as an illustrative experimental benchmark system for scientists to study dynamical behavior for more than four centuries. 
The pendulum system exhibits a wide range of interesting behaviors, from simple harmonic motion in the single pendulum to chaotic dynamics in multi-arm pendulums. 
Under forcing, even the single pendulum may exhibit chaos, providing a simple example of a damped-driven system. 
All multi-armed pendulums are characterized by the existence of index-one saddle points, which mediate the transport of trajectories in the system, providing a simple mechanical analog of various complex transport phenomena, from biolocomotion to transport within the solar system.  
Further, pendulum systems have long been used to design and test both linear and nonlinear control strategies, with the addition of more arms making the problem more challenging.  
In this work, we provide extensive designs for the construction and operation of a high-performance, multi-link pendulum on a cart system.  
Although many experimental setups have been built to study the behavior of pendulum systems, such an extensive documentation on the design, construction, and operation is missing from the literature.  
The resulting experimental system is highly flexible, enabling a wide range of benchmark problems in dynamical systems modeling, system identification and learning, and control.  
To promote reproducible research, we have made our entire system open-source, including 3D CAD drawings, basic tutorial code, and data. 
Moreover, we discuss the possibility of extending our system capability to be operated remotely to enable researchers all around the world to use it, thus increasing access.
\end{abstract}

\clearpage
 \begin{spacing}{.8}
\setlength{\cftbeforesecskip}{2.pt}
\tableofcontents
 \end{spacing}
\clearpage


\section{Introduction}

In its simplest form, the single gravity pendulum constitutes a body suspended by a cord or rod that swings back and forth under the influence of gravity~\cite{gitterman2010chaotic}. 
Investigations of this nonlinear system date at least to the seventeenth century and the work of Galileo Galelei~\cite{galilei1967dialogue,galilei1914dialogues,palmieri2007galileo}, with derivations and explanations of the dynamics now commonplace in introductory mechanics courses. 
Perhaps the most well known application of the single pendulum is in the measurement of time, with Christiaan Huygens' proposed pendulum clock concept of 1656 being the most accurate time keeping device until the 1930s. The predictable oscillatory motion has also been used to infer the constant of gravitational acceleration, $g$~\cite{baker2008pendulum}. 
As a mechanical benchmark system, the single pendulum has many notable variants: the single pendulum on a cart~\cite{aguilar2022self,oka2016nonlinear}, Foucault's pendulum~\cite{matthews2005pendulum,freidovich2007experimental}, Furuta's pendulum~\cite{acosta2010furuta}, the vertical take-off and landing (VTOL) single pendulum~\cite{abraham2017model}, the inertial wheel pendulum~\cite{block2007reaction,freidovich2009shaping}, the spherical pendulum on a puck~\cite{shiriaev2010transverse}, and the inverted wheel pendulum~\cite{meindl2021bridging}.   

In the eighteenth century, Daniel Bernoulli advanced the study of the pendulum by introducing a second mass suspended by a cord or rod from the first mass, resulting in the {\em double pendulum}\footnote{The double pendulum can also be referred to as a compound or linked pendulum.}~\cite{nolte_2020,euler1980rational,bernoulli1738theoremata}.  
Although the gravity pendulum leads to predictable periodic motion, the double pendulum is a prototypical example of a chaotic system~\cite{shinbrot1992chaos,chen2008chaos,rafat2009dynamics,stachowiak2006numerical}, requiring specialized numerical integration techniques such as symplectic~\cite{Yoshida1990pla} and variational~\cite{Marsden2001dmvi} integrators. 
The double pendulum has played a central role in the historical development of dynamical systems, attracting the attention of early pioneers, such as Johann Bernoulli and D'Alembert~\cite{nolte_2020}. 
Notable variants of the system include: the double pendulum on the cart~\cite{timmermann2011discrete}, the rotary double pendulum~\cite{wang2017periodic,wang2020almost,aastrom1999energy}, the ``acrobot"~\cite{spong1995swing}, which is a double pendulum with actuation torque on the second arm, 
and the ``pendubot"~\cite{fantoni2000energy,freidovich2008periodic}, which is a double pendulum with actuation torque on the first arm. 
The double pendulum remains relevant in the study of nonlinear dynamics and has emerged as an important benchmark problem in system identification~\cite{saad2016parameter,abraham2017model,bongard2007automated,schaeffer2017sparse,kaheman2020sindy} and machine learning~\cite{meindl2021bridging,langley1981data,langley1981bacon,abraham2017model,deisenroth2010efficient,kaiser2021data,schaeffer2017sparse,lillicrap2015continuous,burby2020fast}. 

\begin{figure}
    \centering
    \begin{overpic}[width=0.725\textwidth]{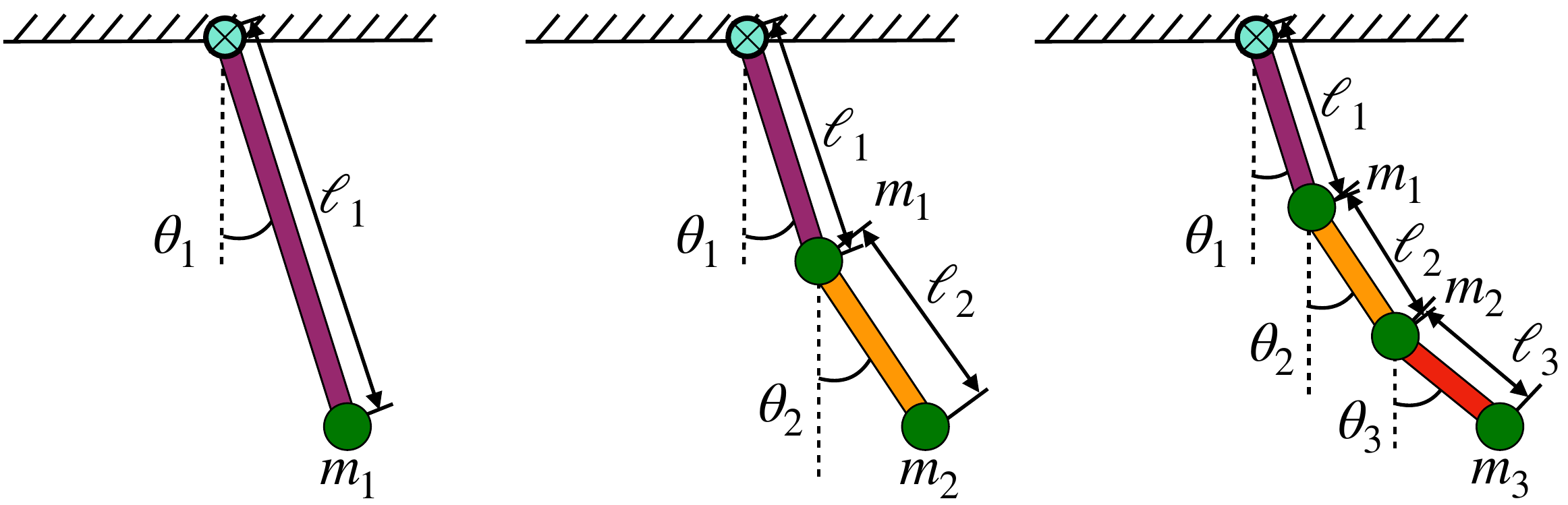}
    \put(10,120){\small Single Pendulum}
    \put(122,120){\small Double Pendulum}
    \put(240,120){\small Triple Pendulum}
    \end{overpic}
    \vspace{-.1in}
    \caption{A schematic illustration of the single, double, and triple pendulum.}
    \label{fig:PendulumIllustration}
    \vspace{-.05in}
\end{figure}

Beyond the double pendulum, one can continue adding ``arms'' (masses suspended by rods attached to the previous point masses) to form a chain resulting in the triple pendulum, and so on. The study of such pendulum systems have a long tradition in classical mechanics for a good reason: they are simple mechanical analogs that display much of the rich dynamical behavior observed in far more complex systems. Figure~\ref{fig:PendulumIllustration} illustrates the single, double, and triple pendulum.

The single, double, and triple pendulum are also widely used in the control community to develop and test new algorithms.  
The wide adoption of the multi-arm pendulum as a benchmark problem stems from the simple derivation of the equations of motion, the tunable complexity of behavior, and to the wide applicability in the physical sciences, including to robotics~\cite{ono2001control}, engineering~\cite{ramli2017control}, and biology~\cite{morasso2019quiet,bertram2001mechanical}. 
In the case of the single pendulum, controllers have been designed to swing the pendulum up and/or stabilize it in the inverted position~\cite{la2009new,aastrom2000swinging,kennedy2011inverted,takahashi2002swing,Furuta1991Swingup,Kapnisakis2016swingup,matsuda2009swinging,mills2009nonlinear,aastrom2000swinging,maeba2010swing,yoshida1999swing,graichen2008feedforward,abraham2017model,aranda2016control,driver2004design,chen2018adaptive}. 
Similar control methods have been developed to swing up the arms of the double and triple pendulums~\cite{driver2004design,zhong2001energy,tao2009adaptive,henmi2014unified,graichen2007swing,pamulaparthy2017evolutionary,xin2011analysis,aastrom1999energy,jaiwat2014real,rubi2002swing,xu2017swing,nakayama2010genetic,gluck2013swing,timmermann2011discrete,vsetka2017triple,liu2011stabilization}, stabilize their arms in various unstable vertical positions~\cite{driver2004design,zhong2001energy,eltohamy1997real,medrano1997robust,tsachouridis1999robust}, and perform time-periodic motion~\cite{wang2017periodic,luo2018period,wang2020almost,llibre2011periodic,freidovich2008periodic,jahn2021design}. 
Due to the chaotic nature of multi-armed pendulums, the sensitivity increases as more arms are added to the pendulum, thus making it an increasingly difficult control benchmark. 
The multi-arm pendulum is characterized by the existence of index-one saddle points (i.e., saddle points with exactly one unstable direction), that mediate the chaotic transport of trajectories in phase space; this provides an analog for several more complex systems, such as transport in the solar system~\cite{koon2000heteroclinic,gomez2004connecting,koon2008dynamical} and chemical reaction kinetics~\cite{Gabern:2005}.  

With such interest in understanding and controlling the dynamics of pendulums, a number of researchers have build physical models to visualize and test their theoretical calculations; these experimental demonstrations include the single~\cite{acosta2010furuta,block2007reaction,fantoni2000energy,driver2004design,mandal2022machine,mandal2022machine,aguilar2022self,oka2016nonlinear,sorensen2001friction,shiriaev2001stabilization,la2009new,freidovich2009shaping,kennedy2011inverted,Furuta1991Swingup,Kapnisakis2016swingup,mills2009nonlinear,saad2016parameter,aranda2016control,freidovich2007experimental}, double~\cite{spong1995swing,fantoni2000energy,driver2004design,knudson2012double,timmermann2011discrete,hesse2018reinforcement,shinbrot1992chaos,christini1996experimental,myers2020low,rubi2002swing,rafat2009dynamics,kaheman2019learning,freidovich2008periodic}, and triple pendulum~\cite{eltohamy1997real,medrano1997robust,tsachouridis1999robust,graichen2005fast,gluck2013swing,vsetka2017triple,vcevcil2016radio}.
Among these physically-realized pendulums, it is the pendulum on a moving cart that has received the most attention. 
The general idea is to use the motion of the cart to control and stabilize the motion of the pendulum arms, with benchmark control problems typically focusing on forcing the arms into the unstable upright vertical position. 
Such a control problem is a proxy for more complex upright stabilization, including human beings standing using their feet as the pivot and applying small muscular adjustments to remain in the upright position. 
This control procedure is becoming increasingly important for the development of robots that stand upright, and for personal transportation devices, such as self-balancing scooters and single-wheeled electric unicycles. 
Much like these physical balancing problems, the difficulty in controlling pendulums on a cart usually lies in the physical constraints of the system, such as the pendulum cart having a limited travel range and velocity. 
Other difficulties come from the specific hardware of the system, the control authority of the cart, and the placement of sensors for real-time measurements of the dynamics. 
The most common method to measure the rotational angle of each pendulum arm uses encoders and a motion capturing camera~\cite{myers2020low,schmidt2009distilling}, with variability in transmitting the collected data using hard wiring or wireless technology~\cite{vsetka2017triple,vcevcil2016radio}. In terms of the actuation of the cart, variants in the physical models use a rotational motor, a servo motor and belt drive, or a linear motor. 
All of these choices have a profound impact on the robustness of developed control methods.   

In this paper we introduce the process and materials needed for constructing a high-performance, fully instrumented, multi-link pendulum on a cart. With regards to 
the variations on the physical model discussed above, we detail reasons for our choice of sensors, signal transmission, pendulum arm design, and actuation method. In particular, our pendulum on a cart uses an optical encoder and slip-ring to record and transmit the rotational angle of each arm, a linear motor for cart actuation, and Speedgoat and Simulink for the real-time control interface. The goal of this work is to provide a reference guide and tutorial for the construction of the pendulum on a cart system with a flexible design that can further be mounted for nonlinear dynamical demonstrations or easily altered to various other pendulum models to initiate new studies. Our contributions are as follows: 
\vspace{-1.75em}
\begin{enumerate}\setlength\itemsep{-.35em}
    \item A detailed tutorial on how to build a multi-link pendulum on a cart system.
    \item Open-source design files, including 3D CAD files and the Simulink files for data collection and control of the system.
    \item Open access data sets of the pendulum system with its rotational angle and velocity recorded with and without control input.
\end{enumerate}
\vspace{-.5em}
These data sets could be of great benefit to the system identification, machine learning, and artificial intelligence communities to test their techniques and algorithms. Finally, in the conclusion we introduce the concept of cloud experiments so that researchers worldwide can access our model and run experiments without having to build their own system. 

We organize our work as follows: In Sec.~\ref{sec:Overview}, we present the major components of the multi-arm cart pendulum system. In Sec.~\ref{sec:PendulumArm}, we describe the design and manufacturing procedure of the pendulum arm. In Sec.~\ref{sec:PendulumCart}, we show the design process of the pendulum cart. 
In Sec.~\ref{sec:RealTimeSystem}, the selection of the real-time control interface is presented. In Sec.~\ref{sec:Electrical}, we discuss the electrical components of the system and illustrate the wiring specification of the entire system. Finally, Sec.~\ref{Sec:Conclusion} summarizes the work and explores the idea of cloud experiments. Sec.~\ref{Sec:Conclusion} also includes the discussion of software setup, operation and safety, and parameter estimation of the pendulum arms.


\section{Overview of the System}
\label{sec:Overview}
\begin{figure}
    \vspace{-.1in}
    \centering
    \begin{overpic}[width=0.89\textwidth]{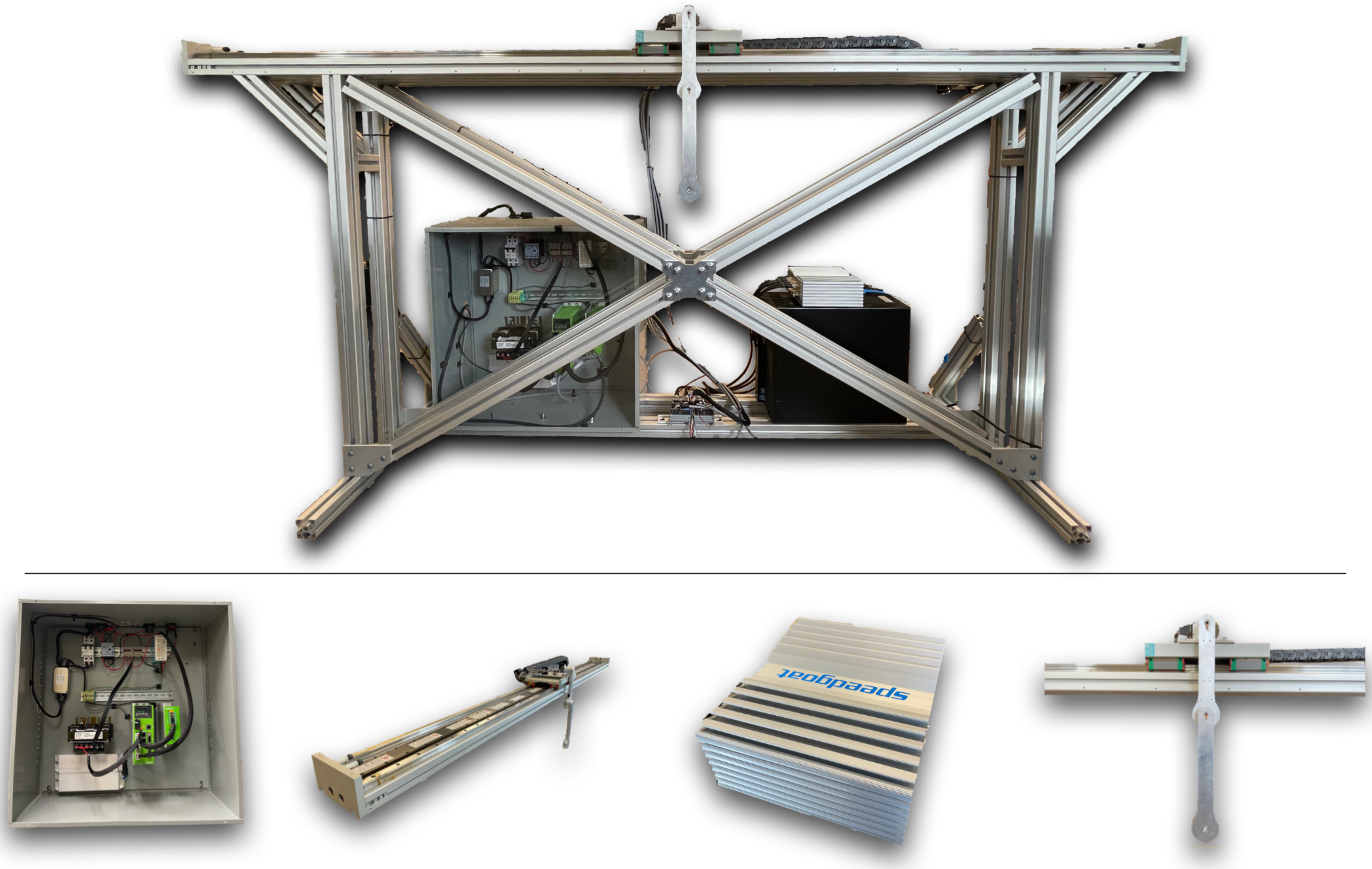}
        \put(110,105){\small Assembled Multi-Link Pendulum on the Cart}
        \put(12,-5){\small Servo Drive}
        \put(123,-5){\small Linear Motor}
        \put(222,-5){\small Real-Time System}
        \put(342,-5){\small Pendulum Arm}
    \end{overpic}
    \caption{Assembled multi-link pendulum on the cart system. The major components of the system are the servo drive, linear motor, real-time control system, and pendulum arm on the cart.}
    \label{fig:SystemOverview}
    \vspace{-.1in}
\end{figure}

In this section we introduce the main components, design, and closed-loop control of the multi-link pendulum on the cart system. The system can be used to gather experimental data of a single, double and triple pendulum's oscillatory motion, and can be used to validate different control laws and system identification algorithms. An overview of the system can be seen in Fig.~\ref{fig:SystemOverview}.

The proposed system consists of four major components: 1) the servo drive~(motor drive), which provides proper current to the linear motor to move it according to the desired speed; 2) the linear motor with a $1\mu m$ resolution magnetic encoder, which provides actuation forces to the pendulum;  3) the real-time system, which controls the motion of the pendulum cart and pendulum arm; and 4) the pendulum arm, which can rotate freely around its joint. 
The pendulum arm has an optical encoder with $10000$ counts per revolution~(CPR) to measure its angular position. 
The four main components are mounted on a frame, as shown in Fig.~\ref{fig:SystemOverview}. 

\begin{figure}
    \vspace{-.1in}
    \centering
    \begin{overpic}[width=0.9\textwidth]{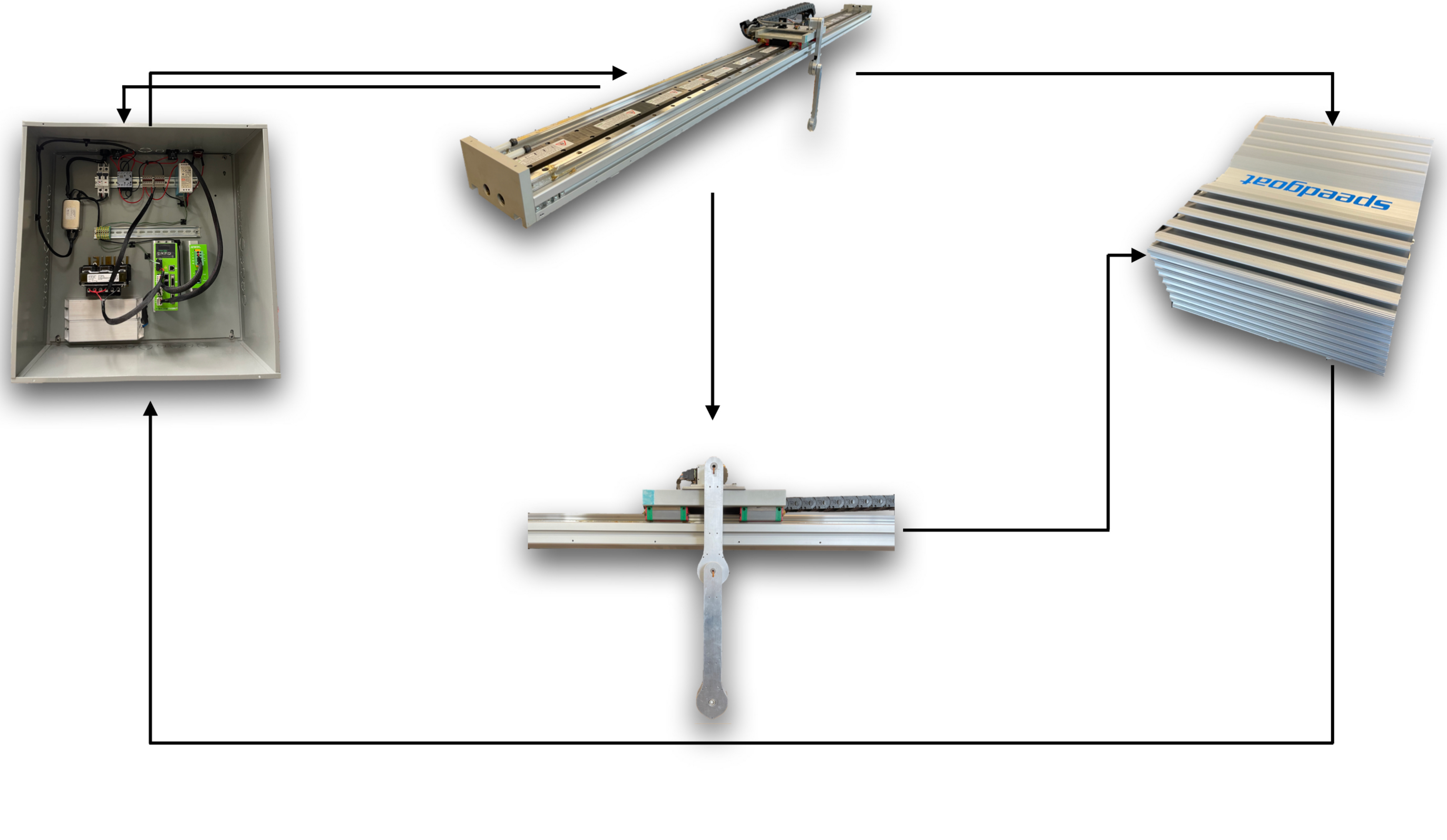}
        \put(120,20){\footnotesize {Controller Feedback (Analog Voltage Output)}}
        \put(120,70){\small{Pendulum Arm}}
        \put(95,160){\small{Servo Drive}}
        \put(86,150){\small{(Velocity Mode)}}
        \put(95,242){\footnotesize{Current}}
        \put(60,230){\footnotesize{Cart Position and Velocity}}
        \put(270,230){\footnotesize{Cart Position and Velocity}}
        \put(220,200){\small{Linear Motor}}
        \put(210,160){\footnotesize{Actuation}}
        \put(253,180){\small{Real-Time System}}
        \put(243,115){\footnotesize{Angular Position}}
        \put(253,105){\footnotesize{and Velocity}}
        
    \end{overpic}
    \vspace{-.25in}
    \caption{Closed loop control diagram of the proposed system. The linear motor and pendulum arm sends measurement data to the real-time system, which is used to calculate the corresponding control action needed to achieve the control object, such as stabilization, swing-up, stabilizing periodic orbit, etc. After the control command is received by the servo drive as an analog signal, the servo drive controls the linear motor to achieve the desired motion.}
    \label{fig:ControlLoop}
\end{figure}
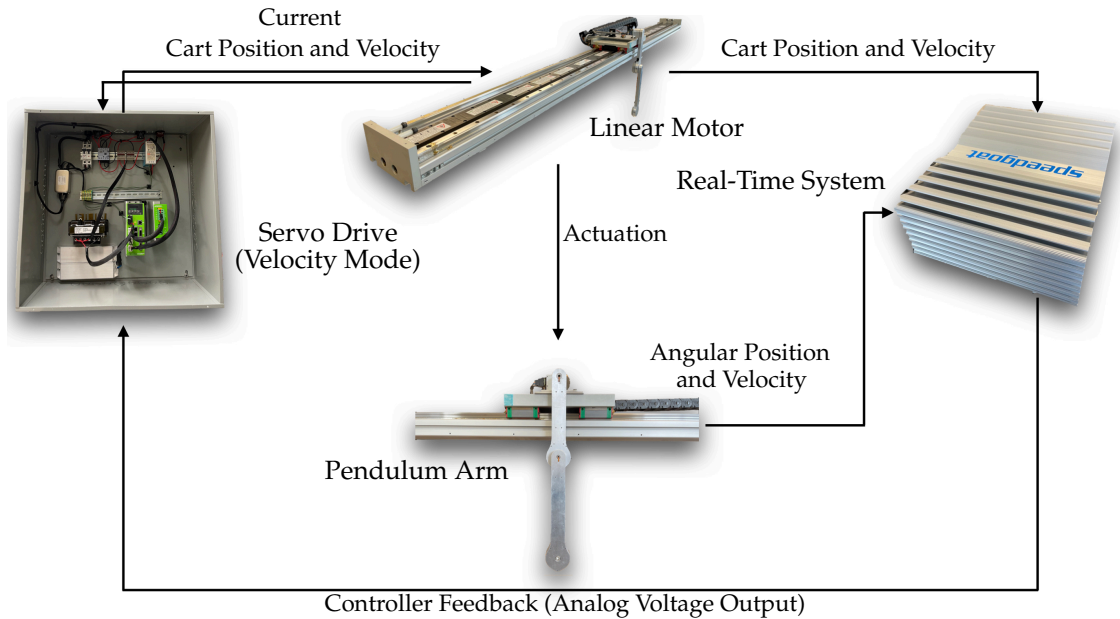

In order to form the closed-loop control, the measurements from the pendulum arm and linear motor are sent to the Speedgoat real-time system. 
The real-time system then runs the user defined controller and calculates the control action needed for the next time step given the control objective. 
The corresponding control value is then converted to an analog voltage output, and this voltage is sent to the servo drive. 
In velocity mode, the servo drive measures the voltage of the analog signal generated by the real-time system, and determines the desired velocity of the linear motor. 
To achieve the desired velocity required by the user, the servo drive uses encoder measurements from the linear motor to calculate its position and velocity. 
By comparing the actual and the user defined velocity of the linear motor, the servo drive {internally}\footnote{To the best of our knowledge, the servo drive uses a PID controller to achieve the target speed set by the user.} calculates the currents needed to achieve the target velocity. 
The closed loop control diagram is illustrated in Fig.~\ref{fig:ControlLoop}.

Our pendulum design enables simple manufacturing and reliable and accurate operation. 
It offers several advantages compared to alternative designs: 1) A linear motor overcomes the backlash that can occur in belt-driven cart systems, which poses a challenge for multi-link pendulum control. 2) Using a slip-ring to transfer the electrical signal in the rotating pendulum arm avoids latency that may occur in pendulum designs with wireless transmission. Moreover, using slip-rings avoids adding a battery to the pendulum arm and enables light weight designs~\cite{vsetka2017triple,vcevcil2016radio}. 
However, using slip-rings results in a more complicated pendulum arm design, which is more difficult to machine. 
Also, due to the limited number of channels the slip-ring provides (5 or 8 signal wires), it is difficult to add a gyro sensor to measure the acceleration of the pendulum arm. 3) The Speedgoat baseline machine used as our real-time control system fully supports the Simulink Real-Time software, which simplifies the controller validation and testing. In case the system is designed to only study a single pendulum, the Speedgoat machine can be replaced with a less expensive solution, discussed in Appendix.~\ref{Appendix:AlternativeRealTime}. 4) The same system can be used to study both controlled and uncontrolled behavior, and the user can add/detach pendulum arms to study single, double, or triple pendulums using one system. 5) The pendulum can be detached from the linear motor, and the system can be used to perform other experiments, such as oscillation studies. 6) Different pendulum arm designs can be installed and tested. Thus, the resulting system has great flexibility for future extensions. 
%

\section{Pendulum Arm}
\label{sec:PendulumArm}

The main component of the cart-pendulum system is the pendulum arm. The most simple design consists of an off-the-shelf rod with a mass attached to its end. This design is widely used for single and double pendulums due to its simplicity, but it limits the integration of sensors to measure angular position. For the triple pendulum, it is more common to machine custom aluminum pendulum arms with integrated sensors that can record the rotational angle of the pendulum. 
In this section, we detail our multi-link pendulum arm design based on custom machined parts with integrated sensors. We introduce the pendulum arm design and assembly, and refer to Appendix.~\ref{Appendix:PendulumArm} for a detailed description of the design and manufacturing of the arm.

\begin{figure}[t]
    \centering
    \includegraphics[width=0.95\textwidth]{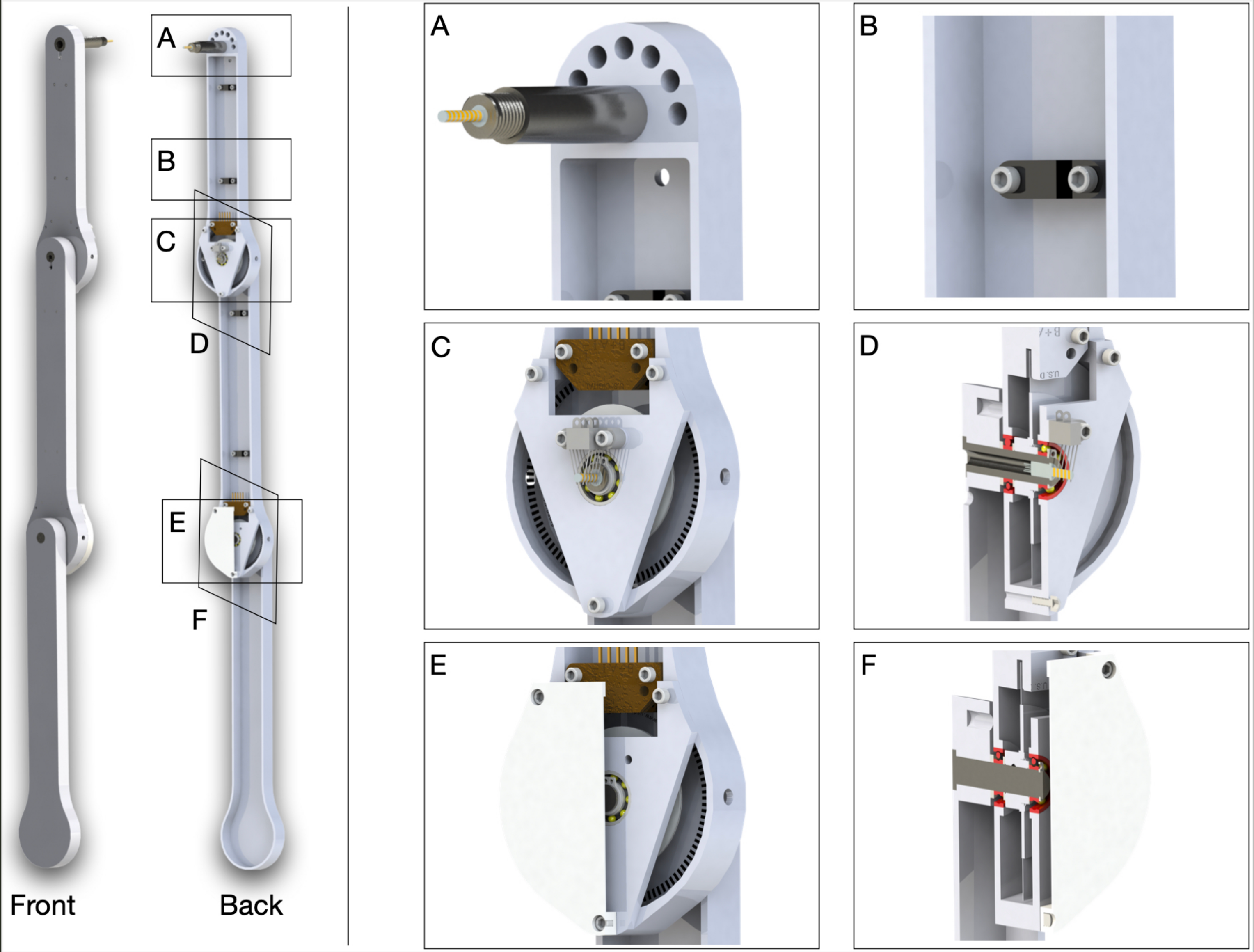}
    \caption{Design of the multi-link pendulum arm. The main components of the assembled pendulum arm are: 1) the pendulum arm body, which is used to install different parts of the pendulum arm, such as sensors, wires, slip-ring, etc.; 2) the pendulum shaft, which is mainly used to support the rotational movement of the pendulum arm; 3) the bearing plate, which is used to secure the pendulum shaft and connect two pendulum arms; and 4) the protection case, which is used to enclose the sensors and slip-ring and prevent them from being damaged.}
    \label{fig:PendulumArmOverview}
\end{figure}

\subsection{Design}


An overview of our pendulum arm design is illustrated in Fig.~\ref{fig:PendulumArmOverview}. The main components of the pendulum arm are: 1) pendulum body; 2) shaft; 3) bearing plate; and 4) 3D printed protection case. The overall structure of the pendulum arm is determined by how it transmits the rotational information measured by the encoders. 
In our design, a slip-ring sends the encoders' electrical signals to the real-time system. 
The advantage of the slip-ring design is the low latency in the signal transmission compared to vision-based and wireless communication systems.
Also, no additional computational resources are needed to determine the rotational angle of the pendulum, compared to vision-based tracking systems. 
This characteristic is particularly beneficial for achieving high sampling rates.
One drawback of the slip-ring design is the additional friction on the contact between the slip-ring shaft and brush block. 
This can be minimized by using a miniature slip-ring and slip-ring brush with gold contact surfaces, which also reduce the electrical noise during the rotational movement of the pendulum arm. 
An additional challenge of using slip-rings is the requirement for precision machining of the pendulum shaft and also the fixed number of channels to transmit signals. This may reduce the flexibility of the setup if new sensors (e.g. an inertial measurement unit~(IMU) sensor) are required for future experiments. 

\begin{figure}[t]
    \centering
    \includegraphics[width=0.95\textwidth]{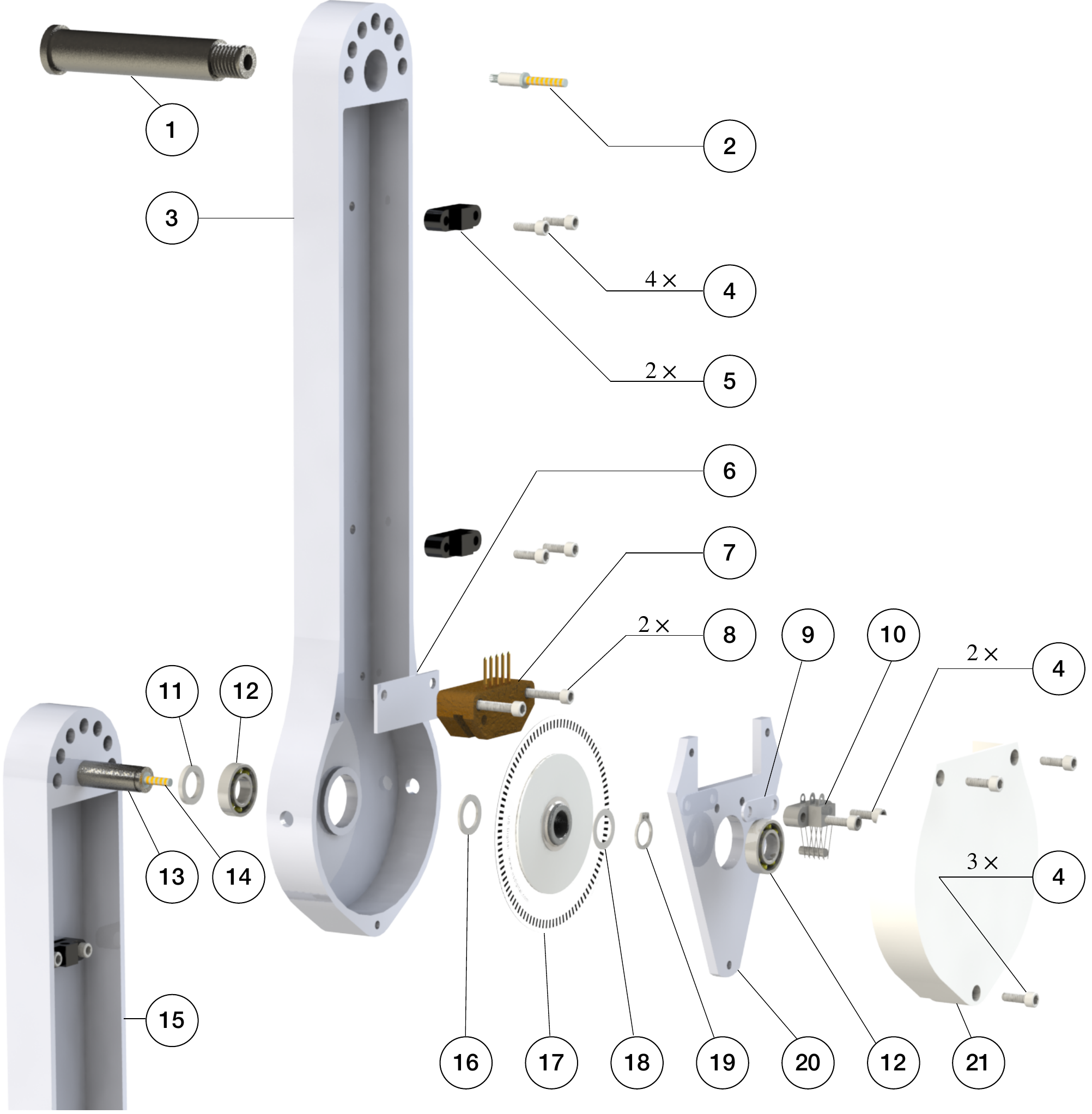}
    \caption{Components and assembly of the first and second pendulum arm.}
    \label{fig:Arm1Arm2Assemble}
\end{figure}

In our design, the shafts of the first and second pendulum arms are hollow to accommodate the slip-ring and the connections to the sensors. 
The slip-ring wires are clipped to the pendulum arm to prevent twining of the cables. 
Moreover, two holes on the side of the pendulum arm facilitate the installation of the encoder disk on the pendulum shaft. 
A stair case shoulder properly secures the bearing that is installed on the first and second pendulum arm. 
Ceramic bearings are used to minimize the friction during the rotational movement. 
The advantage of ceramic bearings is that they operate without lubrication. 
Two bearings are used to fully support the rotational movement of the pendulum arm. 
To avoid the pendulum shaft sliding out of the pendulum arm during operation, external retaining rings are used to secure the pendulum shaft. 
The main components of the pendulum arm are manufactured using CNC milling and turning. 
Further details are provided in Appendix.~\ref{Appendix:PendulumArm}.

\begin{figure}[t]
    \centering
    \includegraphics[width=0.95\textwidth]{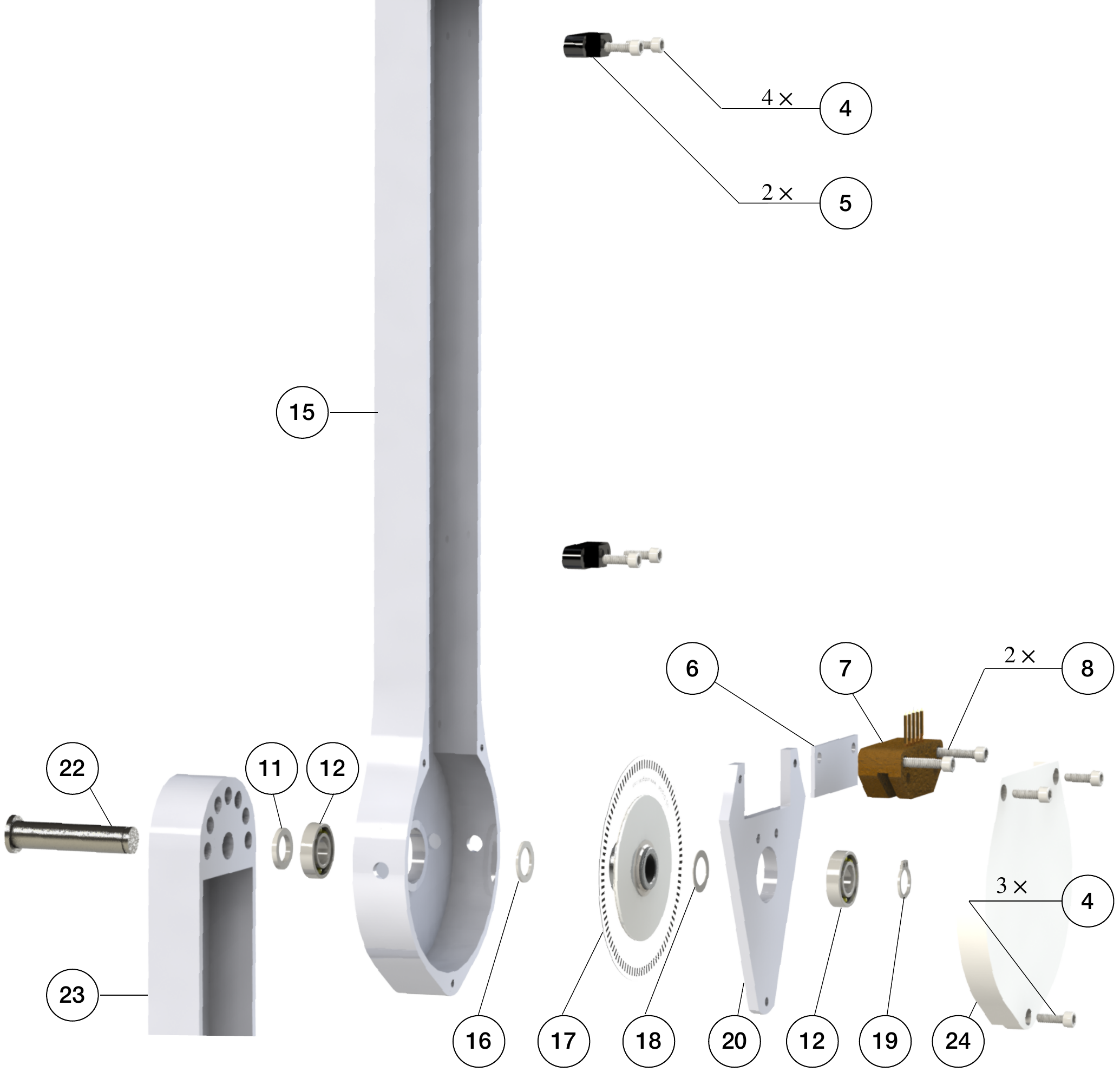}
    \caption{Components and assembly of the second and third pendulum arm.}
    \label{fig:Arm2Arm3Assemble}
\end{figure}

\subsection{Assembly}
The assembly of the pendulum arms is shown in Fig.~\ref{fig:Arm1Arm2Assemble} and Fig.~\ref{fig:Arm2Arm3Assemble}. The pendulum body and bearing plate are assembled first. In order to assemble the pendulum arm body, the following steps are performed: 1) the bearing~($12$) is installed to the pendulum arm body~($3$, $15$) (the transitional fit between the bearing and bearing hole should facilitate the assembly). To make sure the bearing stays in place, a thin layer of Loctite is applied to the outer ream of the bearing before it is push into the bearing hole. Once the bearing is installed, a paint tape is used to cover the bearing to prevent dust entering the assembly. 2) The pendulum shaft~($1$, $13$, $22$) is press fitted into the pendulum arm body~($3$, $15$, $23$). To make sure the installation is smooth, it is recommended to lubricate the pendulum shaft and shaft hole before the press fit. 3) The slip-ring~($2$, $14$) is then installed onto the pendulum shaft~($1$, $13$). Those steps complete the assembly of the individual pendulum arm. The assembly of the bearing plate follows a similar process. The outer ream of the bearing~($12$) is applied with a thin layer of Loctite, then pressed into the bearing plate~($20$). This step should also be straightforward, since the fit tolerance between the bearing and bearing plate is a transitional fit.

To assemble the double pendulum~(first and second arm), the following steps are performed: 1) the shim~($11$) is slid onto the shaft of the second arm~($13$). Next, the pendulum shaft is slid into the bearing of the first arm~($12$) until the shim~($11$) contacts the inner ring of the bearing. 2) The shim~($16$), encoder disk~($17$), and another shim~($18$) are slid onto the second arm's shaft~($13$). Then, the assembled bearing plate is slid onto the shaft~($13$) as well. Finally, the external retaining ring is clipped onto the shaft~($13$). This should hold everything together while still allowing adjustments, since the bearing plate is not screwed yet. 3) The 3D printed shim~($6$) and the encoder reader~($7$) are pushed onto the desired place by aligning the holes. Then, the screws~($8$) are positioned. 4) The 3D printed shim~($9$) and slip-ring brush block~($10$) are installed by aligning the holes on the bearing plate, and the screws~($8$) are tightened. After this step, the contact between the slip-ring and slip-ring brush block should be carefully observed. The slip-ring brush should be centered to its corresponding channel. 5) Next, the protection case is installed by aligning the holes on the pendulum arm body, bearing plate, and protection case. The screws~($4$) are mounted to fasten the assembly. 6) Finally, the wire clipper~($5$) is installed onto the pendulum arm~($3$) by aligning the holes, and the screws~($4$) are tightened. The above steps finish the assembly of the first and second pendulum arm.
Fig.~\ref{fig:Arm2Arm3Assemble} shows the assembly of the triple pendulum arm, following similar steps as for the assembly of the double pendulum arm. The difference though is that the third arm's shaft does not have a slip-ring installed. In Appendix.~\ref{Appendix:PendulumArm} the detailed steps to assemble the triple pendulum arm are introduced.

\section{Pendulum Cart}
\label{sec:PendulumCart}
In this section we introduce the design and assembly process of the pendulum cart. We first discuss the motor type selection and sizing, and then introduce the design and assembly process of the pendulum cart and the bearing house that is used to connect the pendulum arm to the motor, as shown in Fig.~\ref{fig:PendulumArmOverview}. The detailed design and manufacturing of the cart and linear motor support frame is introduced Appendix.~\ref{Appendix:PendulumCart} and~\ref{Appendix:SystemBase}.

\subsection{Design}
\begin{figure}[t]
    \centering
    \includegraphics[width=1\textwidth]{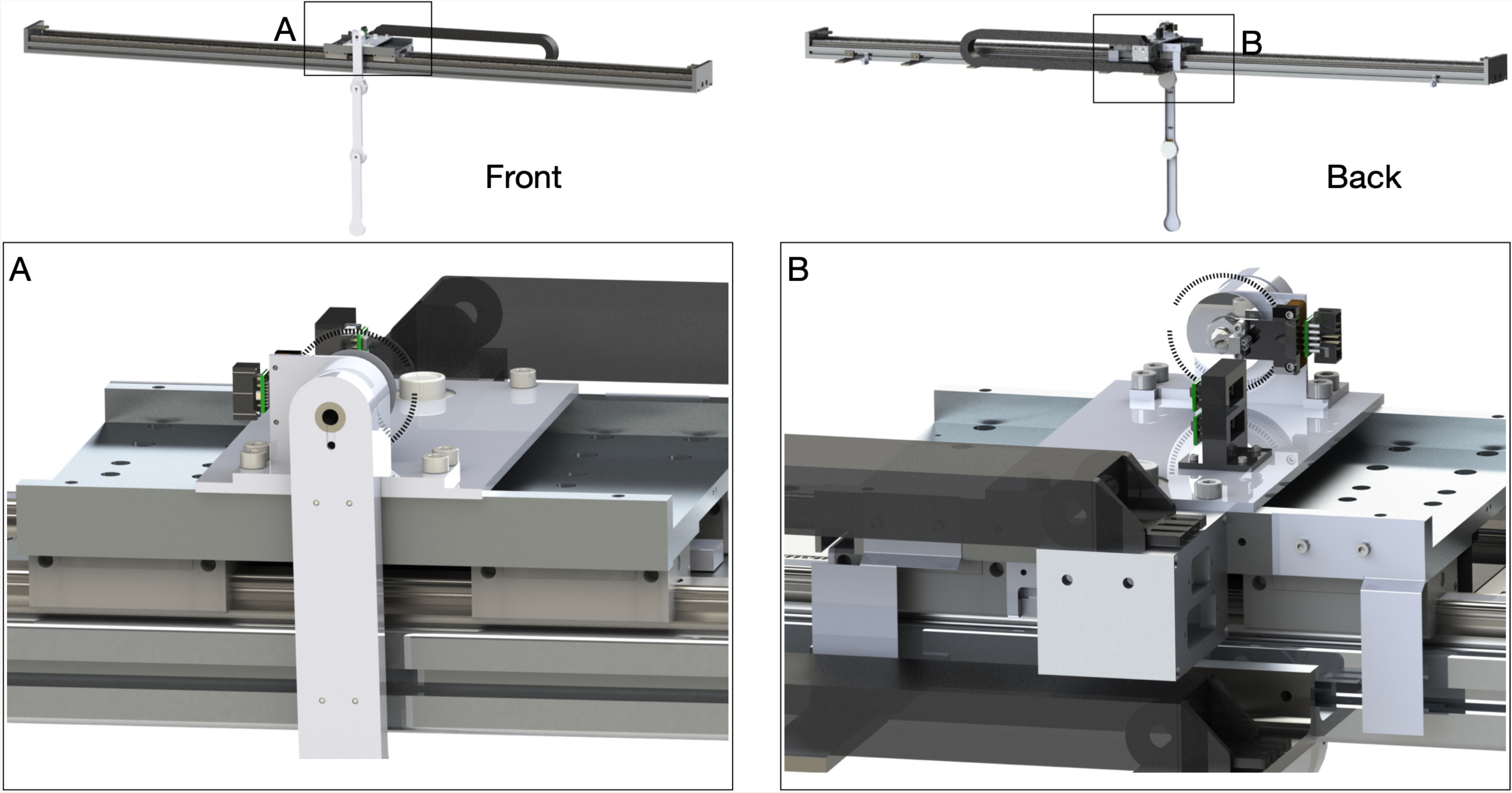}
    \caption{Overall assembly of the pendulum arm and pendulum cart. The main components of the pendulum cart are an aluminum plate and a bearing house.}
    \label{fig:PendulumCartOverview}
\end{figure}

The pendulum cart has two main functionalities: 1) mounting and supporting the pendulum arm; and 2) providing the actuation to the first pendulum arm to control the system dynamics. The first step in the design of the pendulum cart is the selection and sizing of the motor. Once the motor type and size are selected, a mechanism has to be designed to connect the pendulum arm and motor. Fig.~\ref{fig:PendulumCartOverview} illustrates the designed pendulum cart with the assembled pendulum arm.

The actuation of the pendulum arm is provided by a linear motor. The advantage of using a linear motor is that it does not have backslash issues, which frequently happen in belt-drive type servo motors. This allows accurate control of sensitive maneuvers, such as the swing-up of the double and triple pendulum. 
Once the type of linear motor is determined, the next step is to size the linear motor. 
First, the desired maximum speed and acceleration of the cart is defined. 
To determine the maximum speed and acceleration, we look at the pendulum cart's desired motion profile. We take the feed-forward trajectory~\cite{ControlPend} that is needed to swing up the double or triple pendulum. Using this desired motion profile, we determine the peak force and velocity required and use this information to size the linear motor. 
%
Here, we determine the pendulum cart's top speed and acceleration to be $5 m/s$ and $20 m/s^2$. 
The pendulum arm we designed in Fig.~\ref{fig:PendulumArmOverview} weights less than $0.5 kg$ and the mass of the linear motor stage is $5 kg$. Thus, the total continuous force provided by the linear motor should be around $110 N$. 
We choose a HIWIN linear motor system~\footnote{See Appendix A of the HIWIN linear motor system manual document "Linear Motor System(EN).pdf"~\cite{HIWIN_Linear}.} LMX1K-SA12-1-2000-PGS1-V103+HS. This linear motor can provide a peak force of $579 N$ with a peak current of $12.7 A_{rms}$ and a continuous force of $205 N$ with a continuous current of $4.2 A_{rms}$. This motor is powerful enough to provide the desired acceleration and speed. Moreover, the effective stroke of the linear motor is $2 m$ with a magnetic incremental encoder of $1 \mu m$ resolution. 
%

Once the linear motor has been selected, the next step is to design the connection between the pendulum arm and the linear motor stage. As illustrated in Fig.~\ref{fig:PendulumCartOverview}~(A) and (B), a bearing housing is needed to provide support for the first pendulum arm shaft and to secure it so that the pendulum arm can perform free swing. Moreover, an aluminum plate is machined so that the bearing housing can be connected with the linear motor stage. 
The detailed design of the bearing house is introduced in Appendix.~\ref{Appendix:PendulumCart}.
The final components we design for the pendulum cart are the limit switch plates, as shown in Fig.~\ref{fig:PendulumCartOverview}~(B). Fig.~\ref{fig:PendulumCartOverview}~(B) shows two limit switch plates installed on the back side of the linear motor stage. They are responsible to block the laser limit switch when the pendulum cart moves to the edge of the linear rail. 
%
%
Same as the pendulum arm, the main components of the pendulum cart are manufactured using CNC milling. 
The details of the design and manufacturing are provided in Appendix.~\ref{Appendix:PendulumCart}.

\subsection{Assembly}
\begin{figure}[t]
    \centering
    \includegraphics[width=1\textwidth]{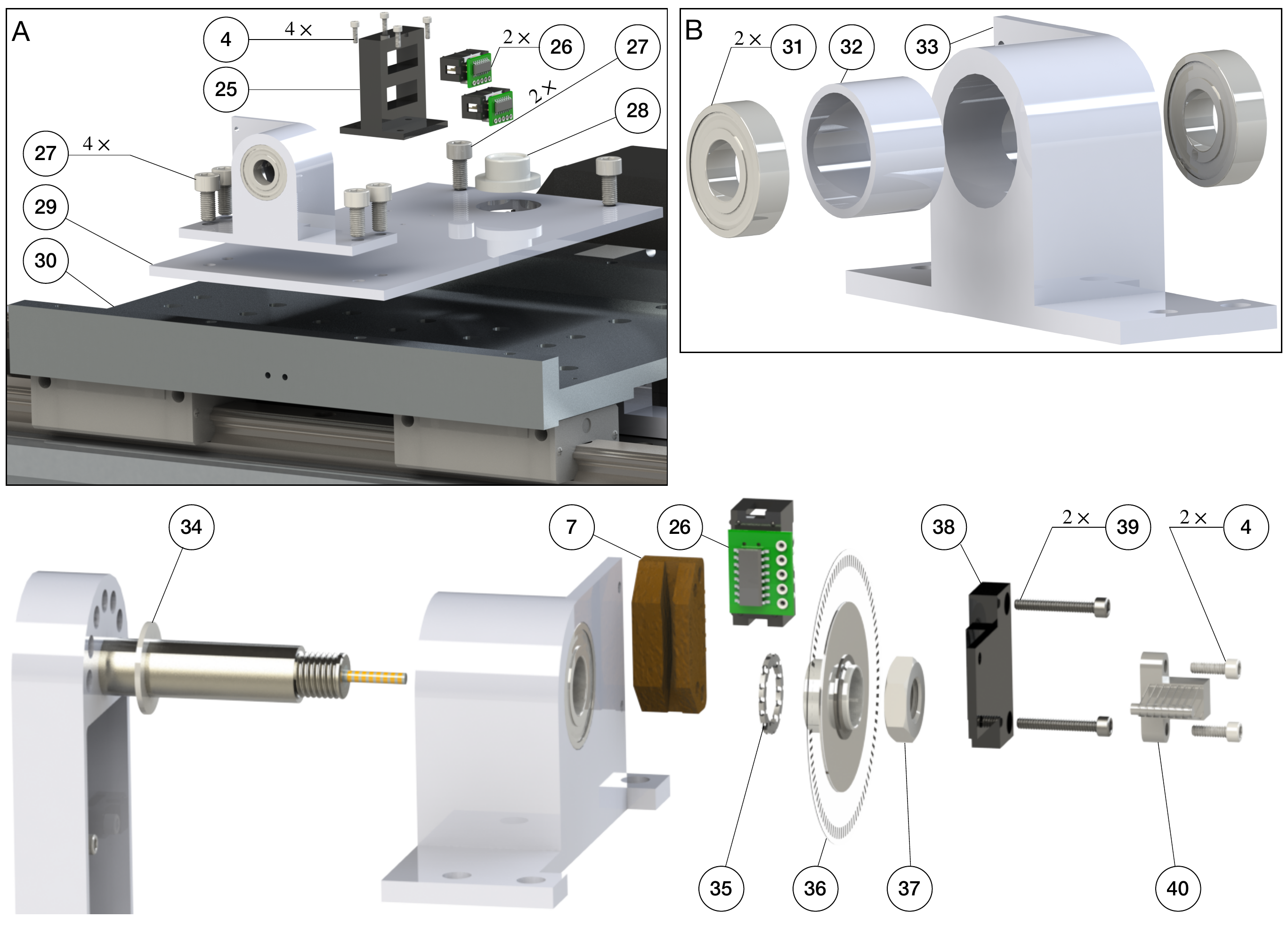}
    \caption{Assembly process of the pendulum cart, pendulum arm and bearing housing. (A) Installation of the bearing housing, cart plate and linear motor stage. (B) Assembly of the ceramic bearing and the bearing housing. (Bottom) Installation of the pendulum arm and bearing housing.}
    \label{fig:CartAssembleOverview}
\end{figure}

The assembly of the cart plate, bearing housing, pendulum arm, and the linear motor stage is illustrated in Fig.~\ref{fig:CartAssembleOverview}. First, the bearing housing is assembled, which includes two steps: 1) a 3D printed spacer~($32$) is press-fitted into the bearing hole~($33$). 2) A thin layer of Loctite is applied to the outer ream of the bearing~($31$). Then, the bearing is installed on the backside of the bearing house. Next, the same process is repeated to install the bearing on the front side. It is important to avoid excess Loctite to enter the bearing, which may damage it. Once finished, the bearing housing is left for 24 hours to ensure the bearing is securely installed. This process can be seen in Fig.~\ref{fig:CartAssembleOverview}~(B). Next, the cart plate is assembled. First, the circular level indicator~($28$) is installed to the cart plate~($29$). To secure it, a thin layer of Loctite is applied to the outer ream of the circular level. Next, the 3D printed cable management tool~($25$) is screwed~($4$) to the cart plate~($29$), and the differential driver~($26$) is installed on the cable management tool~($25$). Then, the cart plate~($29$ ) is mounted to the linear motor stage~($30$) using 4 screws~($27$). This completes the installation of the cart plate to the linear motor, as shown in Fig.~\ref{fig:CartAssembleOverview}~(A).

The installation of the pendulum arm to the bearing housing is shown in Fig.~\ref{fig:CartAssembleOverview}. 1) The spring steel ring shim~($34$) is slid onto the first pendulum arm shaft~($1$). Then the shaft~($1$) is slid into the bearing~($31$) until the shim contacts the inner ring of the bearing. 2) A spacing shim~($35$) is placed onto the shaft until it contacts the inner ring of the bearing~($31$). The encoder disk~($36$) is installed onto the shaft, which allows the measurement of the first arm's rotational angle. 3) A nut~($37$) is tightened using the thread on the first pendulum arm. This allows the application of axial force to the bearing housing, which helps to reduce the oscillation of the pendulum arm on the axial direction. The nut should not be tightened too much, to prevent damage of the bearing and to reduce the friction between the bearing ball and bearing case. 4) The encoder reader~($7$) is placed so that its holes are aligned with the holes on the bearing housing. Next, the holes are aligned on the 3D printed slip-ring brush base~($38$) with the holes on the bearing housing. Once the holes are aligned, the bearing housing, encoder reader, and 3D printed slip-ring brush block base are mounted with screw~($39$). 5) The differential driver~($26$) is installed onto the encoder reader~($7$). 6) The slip-ring brush block~($40$) is installed onto the 3D printed base~($38$) with screw~($4$). It is important to make sure that each brush is centered with the corresponding channel on the slip-ring~($2$). The above steps complete the assembly of the pendulum arm and the linear motor.

\begin{figure}[t]
    \centering
    \includegraphics[width=1\textwidth]{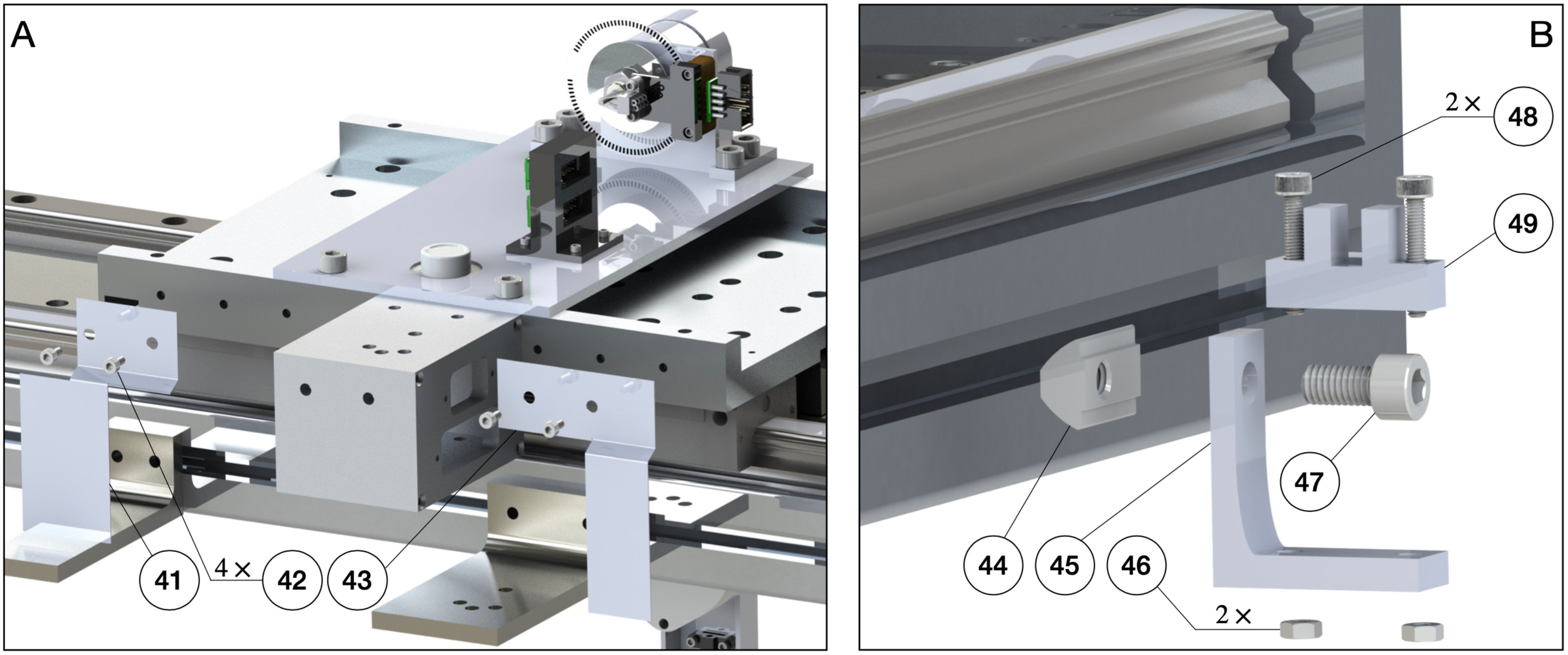}
    \caption{Assembly of the limit switch plate and limit switches.}
    \label{fig:LimitSwitchAssemble}
\end{figure}

The assembly of the limit switch plate~($41,43$) and the linear motor stage~($30$) is shown in Fig.~\ref{fig:LimitSwitchAssemble}~(A). The left and right limit switches~($49$) are connected to the linear motor frame with the 3D printed limit switch base~($45$). The 3D printed base~($45$) is first connected with the linear motor frame using drop in T-slotted framing fasteners~($44$) and screws~($47$). Next, the limit switch~($49$) is connected with the 3D printed base using screws~($48$) and nuts~($46$). The detailed assembly process is illustrated in Fig.~\ref{fig:LimitSwitchAssemble}~(B).

\section{Real-Time System}
\label{sec:RealTimeSystem}
\begin{figure}[t]
\vspace{-.1in}
    \centering
    \includegraphics[width=.825\textwidth]{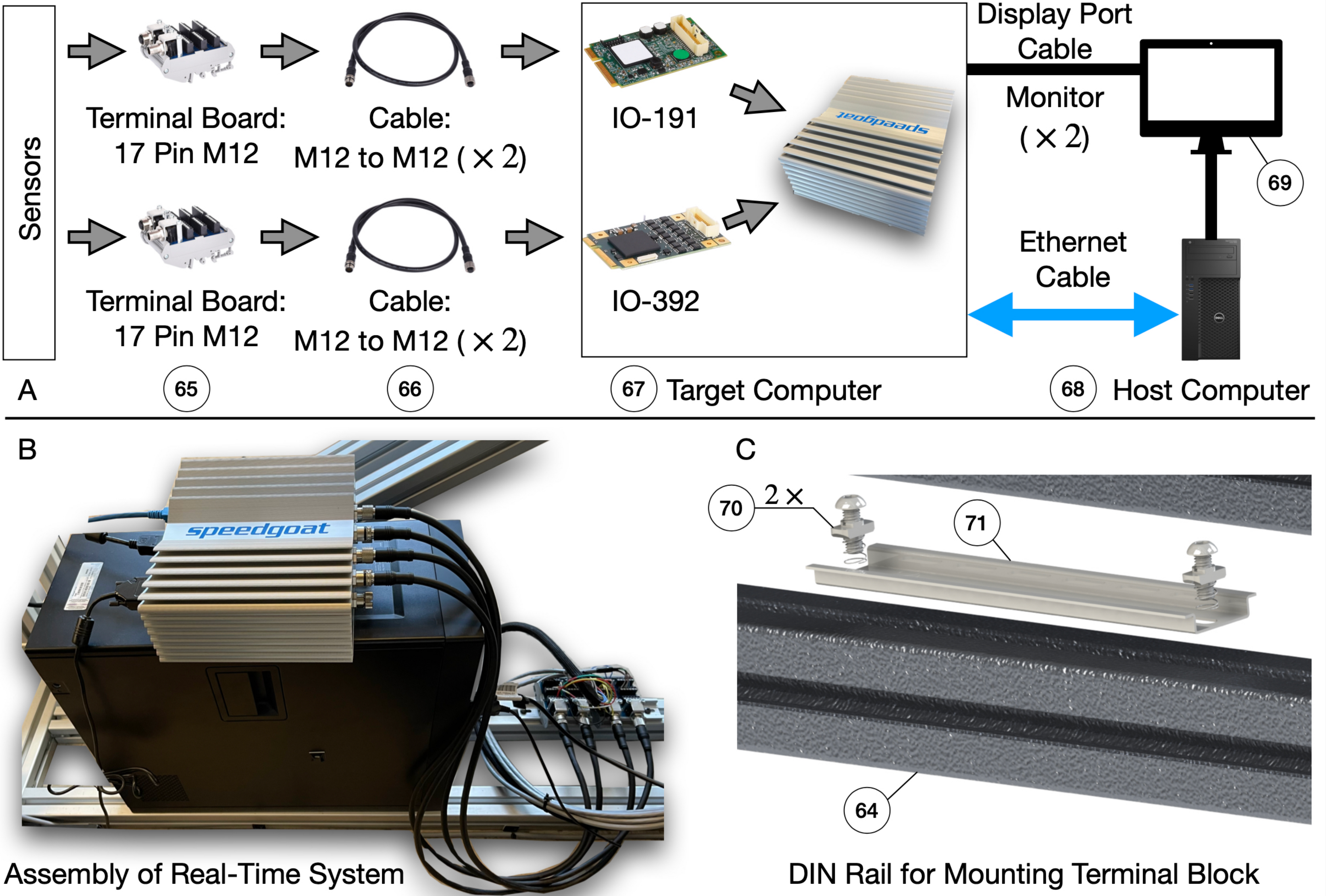}
    \vspace{-.05in}
    \caption{Assembly of the real-time system. (A) rough connection illustration of the target system. The terminal block is used to connect with the sensors. A cable is used to send the signals to the target system. The running status of the target machine is shown on the monitor and the target system and host computer is connected using an Ethernet cable which allows the update of the real-time program to be executed. (B) Assembly of the real-time system. The target machine is place on top of the host machine while the host computer is place on top the system frame. (C) Assembly of the DIN rail that is used to mount the terminal block of the real-time system.}
    \label{fig:RealTimeSysOverview}
    \vspace{-.075in}
\end{figure}

Sec.~\ref{sec:PendulumArm} and~\ref{sec:PendulumCart} introduce the major hardware components of the pendulum on the cart system. In this section, we introduce the real-time system that is responsible for: 1) processing the signal sent by the pendulum arm, linear motor, and other sensors; and 2) using the received signal to determine the control action in real-time based on the user program. The major components of the real-time system include: 1) baseline real-time target machine by Speedgoat. 2) User-selected I/O modules, including modules installed inside the real-time system, terminal boards connected to the sensors, and cables connecting the terminal boards and the real-time system. 3) Computer used to program the controller using Simulink Real-Time. 4) Other components such as the power cord, power adapter, Ethernet cable, software drivers, etc. Fig.~\ref{fig:RealTimeSysOverview}~(A) illustrates the overall Real-Time system. We use the Speedgoat machine with Simulink Real-Time for several reasons: 1) Matlab and Simulink have a wide application in both industry and academia. 2) Speedgoat and Simulink Real-Time simplify rapid prototyping and hardware in the loop control. 3) The Speedgoat machine has a responsive customer service that helps users to solve their software and hardware problems while using the Real-Time machine. 
One drawback of the system is that it is comparably expensive. 
In Appendix.~\ref{Appendix:AlternativeRealTime}, we introduce an alternative custom-made Real-Time system using the National Instrument Data Acquisition~(DAQ) board with Simulink Desktop Real-Time. Other alternatives of the Real-Time system include dSPACE, Typhoon HIL, National Instrument, and others~\cite{kumar2016alternative}.
\vspace{-.025in}

\subsection{System Choice and Specifications}
The number and types of I/O channels and the desired sampling frequency determines which specific Speedgoat Real-Time system can be used. In our design, the Real-Time system needs to read at least four quadrature differential encoder signals~(three from the pendulum arm and one from the linear motor). We need to have at least two digital input channels to read the limit switch signal, one digital output channel to enable/disable the linear motor drive, and one analog output to control the linear motor in velocity mode. The IO-191-EDU-Baseline as our Digital and Analog I/O module meets these requirements. The IO-191-EDU-Baseline is a 16-bit analog I/O module with eight single-ended or four differential sequential sampling analog inputs. The supported voltage ranges of the analog inputs are $\pm0.64V$, $\pm1.28V$, $\pm2.56V$, $\pm5.12V$, $\pm10.24V$, $\pm12.288V$, $\pm20.48V$, and $\pm24.576V$. Moreover, it has four single-ended analog outputs with supported voltage ranges of $\pm10V$, $\pm5V$, $\pm2.5V$, $0-10V$, or $0-5V$. Both the input and output range of the analog signal is software configurable. Finally, it also has 16 x general-purpose digital TTL I/O lines. Thus, the module meets our requirements for digital and analog I/O capability. More details can be found in the IO-191-EDU-Baseline manual. As for the encoder reading, we select the IO-392-Baseline configurable FPGA-based I/O module with 50k Artix 7 FPGA and 13x RS422 digital I/O lines. This module can read four quadrature differential encoder signals while providing a $5V$ power supply to the encoder sensor. The Pin-Out map for this module while using the driver IO-392-QAD4RS422 can be seen in Table.~\ref{table:IO-392-PinOut}. More details can be found in the user manual\footnote{All the user manuals can be found on \href{https://github.com/dynamicslab/MultiArm-Pendulum}{https://github.com/dynamicslab/MultiArm-Pendulum}}. 

After selecting the I/O modules, the next step is to determine which specific Speedgoat system is used. This is mainly determined by the sampling frequency, which depends on the user-specific program/algorithm. The desired sampling of the rotational information of the pendulum arm is $5kHz$ when no control is applied. When stabilizing the single, double and triple pendulum, the sampling rate should be at least $1kHz$~(using LQR with Kalman filter). 
According to the user manual of the Baseline Speedgoat system, the I/O latency for reading and sending signals is around $31.5\mu s$ when using IO-191-EDU-Baseline and IO-392-Baseline. For the baseline machine, the algorithmic calculation takes about $56\mu s$ when running a Simulink model with $1550$ blocks and $250$ continuous states~(equals to $25$  Simulink benchmark model F14). Thus, the total latency time is around $87.5\mu s$, which theoretically enables a sampling rate of $11.4kHz$, which meets our requirements. 
Therefore, we choose a baseline real-time target machine by Speedgoat as our real-time controller. 
After testing, we found a maximum sampling rate of around $12.5kHz$ for pure data collection, and $5kHz$ during double/triple pendulum stabilization~(using a time-varying LQR controller and a Kalman filter). 
The sampling rate is problem-specific and using a highly optimized code can further increase the maximum sampling rate. 
Finally, the host machine used for developing the control law has an Intel(R) Core(TM) i7-3770 CPU with 3.40GHz frequency and 32GB of RAM. The host machine runs on Windows 10 Pro with Matlab 2021b.

\subsection{Assembly}
The assembly of the real-time system is straightforward, as shown in Fig.\ref{fig:RealTimeSysOverview}~(B). The host computer~($68$) is directly placed on the aluminum extrusion~($64$), while the target computer~($67$) is placed on top of the host computer. The Speedgoat target machine can be mounted on the system frame following the steps specified in the user manual of the real-time system. By directly placing the host and target machine on the system frame, the center of mass of the whole system is lowered, making the system frame more stable. A DIN rail~($71$) is mounted on the system frame~($64$) using a drop-in fastener. This DIN-rail allows the mounting of the terminal block~($65$). The bill of materials of the real-time system can be found in Table.~\ref{table:RealTimeSys} in Appendix.~\ref{Appendix:BillofMaterials}.

\section{Electrical System}
\label{sec:Electrical}
In this section we introduce the major electrical components of the experimental setup. We introduce the functionality of the components and illustrate the wiring specification of the entire system. The electrical system of the whole setup can be divided into two parts: 1) Linear motor power supply. 2) Pendulum arm sensor system. In the following, we introduce each part separately and illustrate the entire wiring diagram of the electrical system. {\color{red}{WARNING: None of the authors of this paper is a certified electrician. The wiring diagram of the system is provided as a reference and it has not been examined by an electrician. To the best of the author's knowledge, the provided wiring diagram is safe to be used but NONE of the authors GUARANTEE its safety.}}

\begin{figure}[t]
    \centering
    \includegraphics[width=.95\textwidth]{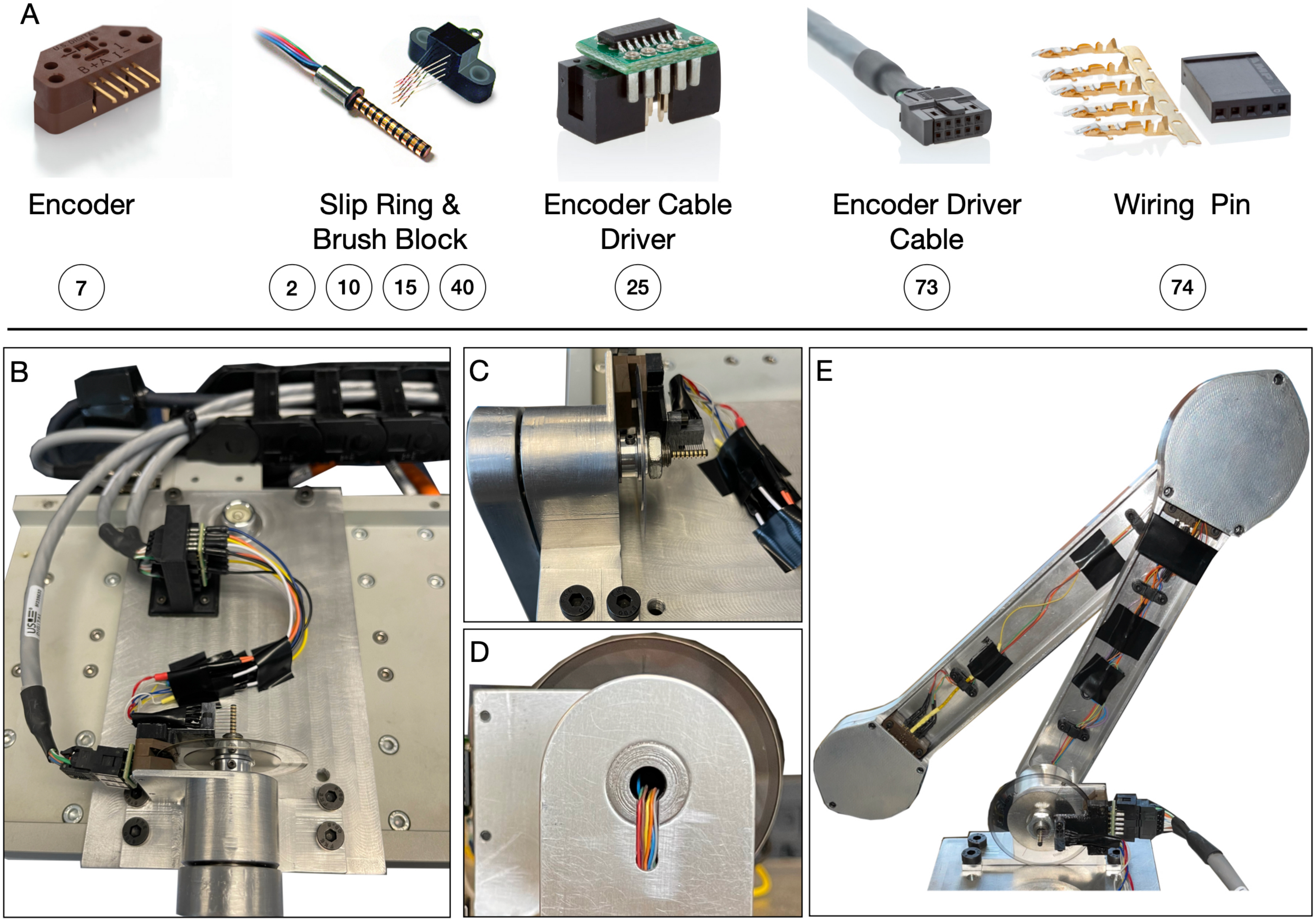}
    \caption{Electrical wiring and assembly of the pendulum arm. (A) Major electrical components of the pendulum arm. (B) connection method of differential driver and encoder reader. (C) to (E) wiring of the slip-ring.}
    \label{fig:PenArmElectric}
\end{figure}

\subsection{Electrical Component of Pendulum Arm}
The electrical component of the pendulum arm mainly consists of an encoder reader~($7$), slip-ring~($2,15$), slip-ring brush block~($10,40$), differential driver~($25$), and differential driver cable~($73$). Fig.\ref{fig:PenArmElectric}~(A) illustrates all those components. The encoder reader's single-ended A, B, and C/Index channel is transferred into the differential signal using differential drivers to improve the noise robustness of the encoder measurements. Using a differential signal helps to avoid the effect of electrical noise generated by the linear motor. It is generally recommended to use a differential signal whenever possible. Three US Digital CA-C10-SH-NC 10 feet cables are used~(differential driver cable) to transfer the differential signals to the target computer. One end of the differential driver cable is connected to a 10-pin female standard (non-latching) connector. This connector is then inserted into the differential driver. The other end of the differential driver cable is unterminated. Thus, it is stripped and inserted into the corresponding channels on the real-time system terminal block~($65$). This process completes the connection of the differential driver and target system. Last, the differential driver and encoder reader are connected, as shown in Fig.\ref{fig:PenArmElectric}~(B). 

The first pendulum arm's encoder reader (measuring the arm's rotational angle) is directly connected to the differential driver, since the encoder reader is mounted on the bearing housing. This mounting position simplifies the connection of the differential driver to the encoder reader.
However, the encoder reader and differential driver connection on the second and third arm are located inside the pendulum arm. This mounting position prohibits the direct connection of the driver and encoder reader. Therefore, two slip-rings are used that connect the encoder reader inside the first and second pendulum arm, as shown in Fig.~\ref{fig:PenArmElectric}~(E). This allows an indirect connection of the differential driver and encoder reader through the slip-ring brush block, which conducts the encoder reader signal out of the pendulum arm. Next, the breadboard jumper cable connects the slip-ring brush block and differential driver. One end of the breadboard jumper cable is stripped away using a wire stripper and soldered onto the slip-ring brush block. The other end is directly inserted into the differential driver, as shown in Fig.\ref{fig:PenArmElectric}~(B). 
As for the slip-ring, it is first installed onto the pendulum shaft. The unterminated slip-ring wire is striped and then pushed into the pendulum arm using the groove and hole in the front of the pendulum arm, as shown in Fig.\ref{fig:PenArmElectric}~(C, D). 
Inside the pendulum arm, the slip-ring wire is clipped using a 3D printed cable management tool~($5$). 
Next, the other end is soldered with the locking clip contact pin, and the locking clip contact pin is insulated with a heat shrink tube to avoid the shortage of the wire connection. Finally, it is connected with the encoder to allow the transmission of the encoder signal through the slip-ring. Fig.\ref{fig:PenArmElectric}~(C, D, E) illustrates the slip-ring wiring. Details of the wiring diagram are in Appendix.~\ref{Appendix:ElectricalSystem_Pendulum}.

\begin{figure}[t]
    \centering
    \includegraphics[width=.925\textwidth]{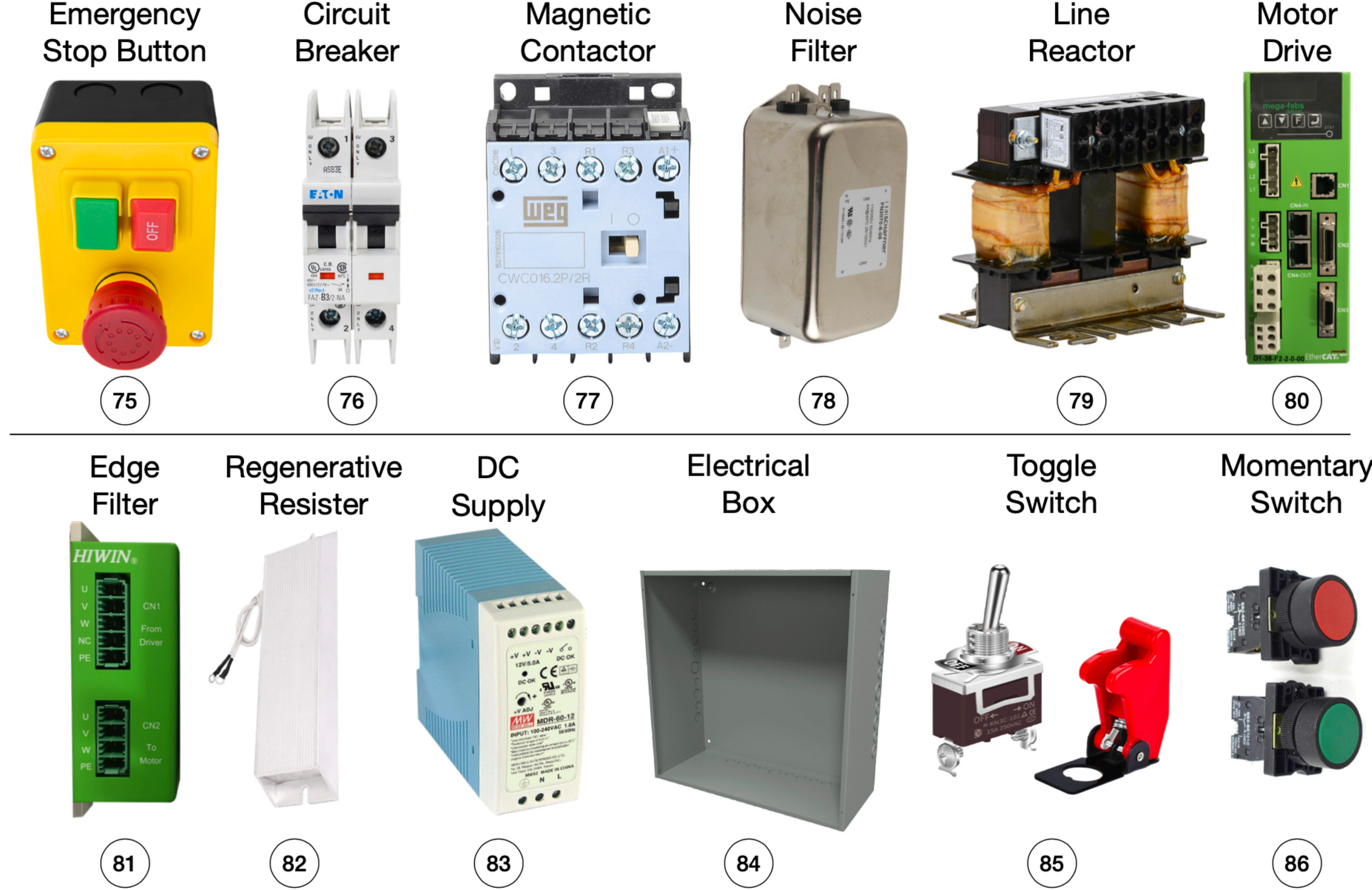}
    \caption{Electrical components needed to provide power supply to the linear motor. An electrical box is used to organize all the components.}
    \label{fig:MotorEletricalPartOverview}
\end{figure}

\subsection{Electrical Part of Linear Motor Connection}
The electrical part of the linear motor mainly concerns the following: 1) properly connecting the motor drive and motor; and 2) connecting the motor drive to the target computer.
The connection guidance of the linear motor and linear motor drive is detailed in the user manual of Mega-Fabs D1 Drive\footnote{Available on \href{https://github.com/dynamicslab/MultiArm-Pendulum}{https://github.com/dynamicslab/MultiArm-Pendulum}}. The user manual also shows the pin out of the control signal cable, which enables the communication between the motor drive and target computer. 
%
The main challenge of the electrical part of the linear motor is to reduce the effect of electromagnetic interference (EMI) generated by the linear motor and motor drive. 
Several techniques can be used for this: 1) use a ground filter to filter out the noise generated by the linear motor and the drive in the ground line; 2) use shielded cables to transmit the signal; 3) use twisted pairs of wires to transmit differential signals; and 4) proper grounding of the all electronic components. With all these techniques combined, EMI is reduced and a proper and safe connection of the linear motor electrical parts is achieved.


Several electrical components are required to connect the linear motor. We use the components that are suggested in the user manual of the D1 motor drive: 1) emergency stop switch; 2) circuit breaker; 3) noise filter; 4) magnetic contactor; 5) line reactor; 6) motor drive; 7) edge filter; 8) regenerative resistor; 9) DC voltage supply; 10) linear motor; 11) electrical box and mounting plate; and 12) other miscellaneous parts that help connecting the electrical components, such as wires, connectors, cables, etc. A summary of the major electrical components can be seen in Fig.~\ref{fig:MotorEletricalPartOverview}. 
The details of these parts are introduced in Appendix.~\ref{Appendix:ElectricalSystem_LinearMotor}, along with a detailed wiring diagram.

\section{Conclusions and Discussion}
\label{Sec:Conclusion}
In this paper, we have introduced an experimental multi-link pendulum on a cart system. 
This system can be used to collect experimental data from the single, double, and triple pendulum. Moreover, the user can control the pendulum motion by actuating the cart via a linear motor. 
This makes the experimental system a powerful tool for studying the control of chaotic systems. 
Our experimental setup is open source, with all the detailed design choices and CAD files freely available, which allows reproduction of the system. 
We have also collected experimental data sets of the single, double, and triple pendulum and made them freely available. 
We believe this data will be valuable for the machine learning and modeling communities to test various algorithms. All code and design files can be downloaded on our Github page\footnote{Code available at \href{https://github.com/dynamicslab/MultiArm-Pendulum}{https://github.com/dynamicslab/MultiArm-Pendulum}}. 

To make it possible to  reproduce this pendulum on a cart system, we have provided a detailed description of the assembly and manufacturing process. 
Furthermore, the wiring diagrams of the electrical components and the software set up are also documented. 
Although the detailed manufacturing process of our setup is documented, we realize that many labs may not have the time and funding to replicate this system. 
Beyond this, the time needed to manufacture the experiment and the effort needed to maintain it may not be worth the investment for groups that only need occasional use of the experimental setup. 
This high cost can be mitigated by using more standard parts and 3D printing technology, but it does not solve the maintenance effort and the time consuming manufacturing and assembly process. 
This drawback motivates a future modification to our experimental set up to allow cloud access for remote users, discussed below, so that they can remotely access the system for data collection and control experiments. To conclude this manuscript we also briefly discuss the system operation procedures, safety instructions, and the estimated parameter values associated to the pendulum arms, which are explored in more depth in the appendices.

 \begin{figure}[t]
    \centering
    \includegraphics[width=\textwidth]{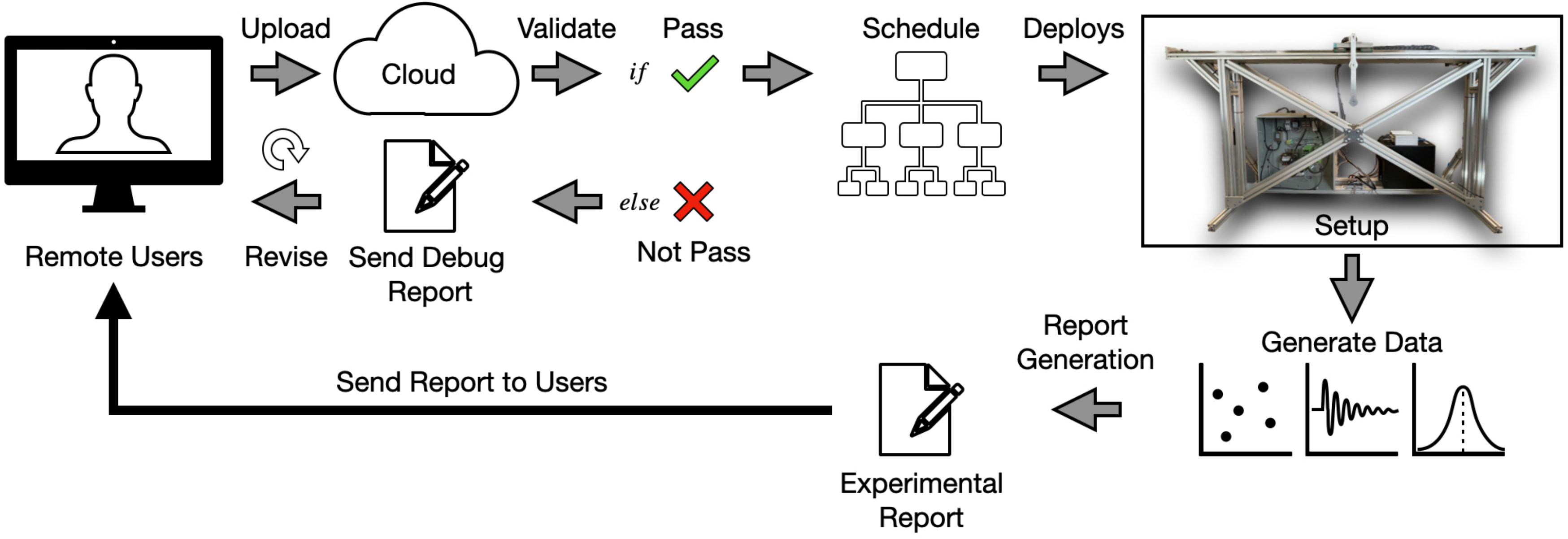}
    \caption{Overall schematic of a cloud experiment of the pendulum on the cart system.}
    \label{fig:CouldExp}
\end{figure}

\subsection{Cloud Access to Experiments}
\label{sec:Cloud}
Figure~\ref{fig:CouldExp} shows the remote cloud experiment concept. This is similar to cloud services where infrastructure, hardware, or software can be accessed by remote users through the internet. In our case, the user accesses the pendulum hardware. 
The user must first develop their own program to achieve some objective using the experimental system. 
The program could be a controller to move the pendulum cart for data collection or for a control objective. 
This program should be coded according to a given template. 
After the user uploads their program to the cloud, it is necessary to test for any compiling errors and controller errors. 
Testing whether the proposed controller is safe is a difficult task. 
For example, some controllers might require the pendulum cart to move at unreasonable speeds or accelerations. 
If the controller is unstable, then it might damage the experimental setup.
To avoid this, it may be necessary to test the controller in a digital twin of the pendulum. 
If position, velocity, or acceleration limits are violated, then a debug report will be sent to the user to allow revision and resubmission.
Once the user program passes all  tests on the digital twin, it may be deployed on the real experimental system. 
In the physical experiment, failsafe hardware limits may also be imposed, operating outside the user specified code.  

During the experiment, the system should record all available data, including pendulum arm rotational angle and angular speed, linear motor position and velocity. The system should also record user defined variables. 
The collected data will be sent to the user to allow further analysis. 
A video camera  will also record the experiments. 
The user will then be responsible for determining whether the desired objective is achieved by reviewing the recorded data and videos. The user should also be allowed to send a bug report to allow the maintenance of the hardware system.

There are several challenges that must be addressed for the cloud experiments. 
First is to deploy the user program to the real-time system automatically. 
In the current design, we use the Speedgoat machine as our real-time controller. 
Every time a controller is deployed, the user manually starts Simulink. 
A pipeline to automatically read, load, compile, and start the code must be developed. 
The second challenge is to automatically self-check the system. 
The real-time system should be able to perform the self-check and determine if any components must be replaced.

\subsection{Software Setup}
\label{sec:Software}
Two major software packages are used for successful and safe experiments: 1) The HIWIN Lightning software and 2) Simulink Real-Time model. The former is used to set up the parameters of the linear drive,  while the latter is used to develop the real-time control algorithm for the system. The motor drive's Lightning software allows the selection of motor type and motor parameters, for which the motor drive uses this information to determine the control parameters needed to move the linear motor in the desired motion profile. The Lightnight software also configures the linear motor's encoder reading, allows the setup of the linear motor's Hall sensor, allows the user to select which in mode the linear motor should operate~(position, velocity, force/torque, or stand-alone mode), configures the programmable I/Os of the D1 drive's CN2 channel, and finally, can set up software safety mechanisms. Appendix~\ref{Appendix:Lightning} enumerates the details to setup these functionalities. 
The Simulink Real-Time software is used to develop the controller for the real-time pendulum experiments. To read the sensor signal and output control signal, the Simulink Real-Time and Speedgoat needs to be configured properly. The main aspects to be considered when setting up the Simulink model are: 1) setting up the Simulink blocks to allow the desired digital I/O functionalities, 2) setting up the FPGA module to enable encoder reading, 3) utilizing the digital I/Os to start the linear motor and stop it when the limit switches are triggered, 4) switching of the operating mode of the linear motor drive and activating homing function, 5) sending the analog signal to control the velocity of the linear motorm and 6) reading encoder sensors. These aspects are detailed in greater depth in Appendix.~\ref{Appendix:Simulink}.

\subsection{Operation and Safety}
\label{sec:OperationAndSafety}
The operation procedures of the system include pre-experiment preparations, operations required during the experiments, and post-experiment operations. During the pre-experiment preparations, the operator should check all the wiring connections of the system and make sure there is no electrical shortage and hardware damage before starting. Then the controller file is prepared, and the system is turned on while ensuring safety. The main operation that must be performed during the experiment is to check whether the experiments are going as planned. 
If not, the operator should stop the experiments immediately and cut off all the power supply to the system. After the experiment, inspect any damage to the setup and then cut off all power to avoid any electrical hazards. For a detailed step-by-step operation guidance see Appendix.~\ref{Appendix:OperationDetail}. While operating the system, safety should always be the number one priority. To ensure the safe use of the experimental setup, mechanical, electrical, software, and personal safety measures must be implemented. A critical mechanical safety measure is to install a shock absorber on the side of the linear motor rail. This will help absorb the extra kinetic energy of the linear motor in case of controller failure. We further recommend purchasing a protective panel to surround the experimental setup. This can reduce the risk of personal injury caused by pendulum parts detaching while the linear motor is in motion. Appendix~\ref{Appendix:SafetyNotes} describes other safety measures that must be followed. Besides the safety notes in this paper, the reader must also follow all the safety instructions written in the individual components' user manuals. We close this section by emphasizing that the proposed design of the multi-link pendulum on a cart has been tested and used safely in the lab environment by the authors. One should always exercise caution when building and operating the system and none of the authors can be held responsible for any damage or injury caused by reproducing the design shown in this paper.

\subsection{Parameter Estimation of Pendulum Arms}
\label{sec:ParEs}
Estimating the parameters of the pendulum system requires data collected from the designed system. An optimization problem is solved to find the pendulum system's parameters that best predict the data. This also requires an analytical model of the pendulum, which is shown in Appendix~\ref{Appendix:ParameterEstimation}.Parameter estimation is a standard task in  pendulum control experiments where the model of the system is needed along with its parameters, while the model derivation of the single, double, and triple pendulum all follows through a similar process. We omit the model derivation for single and triple pendulums since they are similar. The former is standard and the latter can be derived through similar methods to the double pendulum~\cite{graichen2007swing,myers2020low,gluck2013swing}.

In the case of the experimental pendulum, there are several parameters that must be identified, including the mass of the pendulum arm $m$ and the position of center of mass, defined as $a$. We also require the length of the pendulum arm $l$ and the inertial of the pendulum arm $J$, as depicted in Fig.~\ref{fig:PendulumIllustration}. Although frequently ignored in parameter estimation, we also seek to determine the local constant of gravitational acceleration, $g$, which plays an important role on the chaotic dynamics of the double and triple pendulum. 
Some parameters do not show up in the derived equation of motion of the cart-pendulum, such as the mass of the cart $M$ when the control is taken to be the acceleration of the cart. We summarize these definitions using the double pendulum as an example in Figure~\ref{Fig:DP_Illustrate} and Table.~\ref{table:EstimatedParameters} provides details of the estimated pendulum arm parameters. Finally, Table.~\ref{table:EstimatedParameters} also provides the values of the the friction coefficients $\varepsilon_i$ which lead energy dissipation in the physical system. For more details on the method of parameter estimation, please see Appendix.~\ref{Appendix:ParameterEstimation} for an exposition with the double pendulum.

\section*{Acknowledgments}
The authors acknowledge funding support from the Army Research Office (ARO W911NF-19-1-0045) and National Science Foundation AI Institute in Dynamic Systems (grant number 2112085). The authors also would like to thank Eamon McQuaide and Veasna Thon for providing useful tutorials and instructions during the machining process and Zachary G. Nicolaou for helping set up the slow-motion camera.

\clearpage
 \begin{spacing}{.77}
 \small{
 \setlength{\bibsep}{.8pt}
\addcontentsline{toc}{section}{References}
\bibliographystyle{IEEEtran}
\bibliography{PaperReference}
 }
 \end{spacing}

\clearpage
\appendix

\section{Real-Time System Using Simulink Desktop Real-Time}
\label{Appendix:AlternativeRealTime}
Besides using the Speedgoat Real-Time machine and Simulink Real-Time, it is possible to use a low-cost alternative as the Real-Time system. This solution uses the National Instrument Data Aqcusition~(DAQ) device PCIe-6341 as the data collection device~(I/O module). PCIe-6341 can read up to four quadrature encoders at the same time. Moreover, it has multiple digital I/Os and analog I/Os. For more details on the device specification, please check the user manual. In this setup, the sensors are connected to the National Instrument terminal connector SCB-68A. Then, the terminal connector and DAQ are connected using cable SHC68-68-EPM. This connection completes the sensor data acquisition task. To read sensor signals in real-time and send control command~(analog signal) to the motor drive~(velocity mode), the Simulink Desktop Real-Time is used as the real-time operating system. In this case, only one computer is needed to develop the user program and perform real-time control. This real-time system setup has been tested in our previous paper~\cite{kaheman2019learning}. When using Simulink Desktop Real-Time, the real-time program only runs on a single core instead of multi-cores as Simulink Real-Time does. This drawback decreases the maximum sampling frequency of the Simulink Desktop Real-Time setup. Our previous experiments show the maximum sampling rate in pure data-collection mode is around $5 kHz$ when using Simulink Desktop Real-Time with the host machine we mentioned in Sec.~\ref{sec:RealTimeSystem}. The maximum sampling rate of the double pendulum stabilization task~(using time-varying LQR and Kalman filter) is around $2 kHz$ in this case.

\section{Pendulum Arm Details}
\label{Appendix:PendulumArm}

\subsection{Design Details}

\begin{figure}[t]
    \centering
    \includegraphics[width=0.8\textwidth]{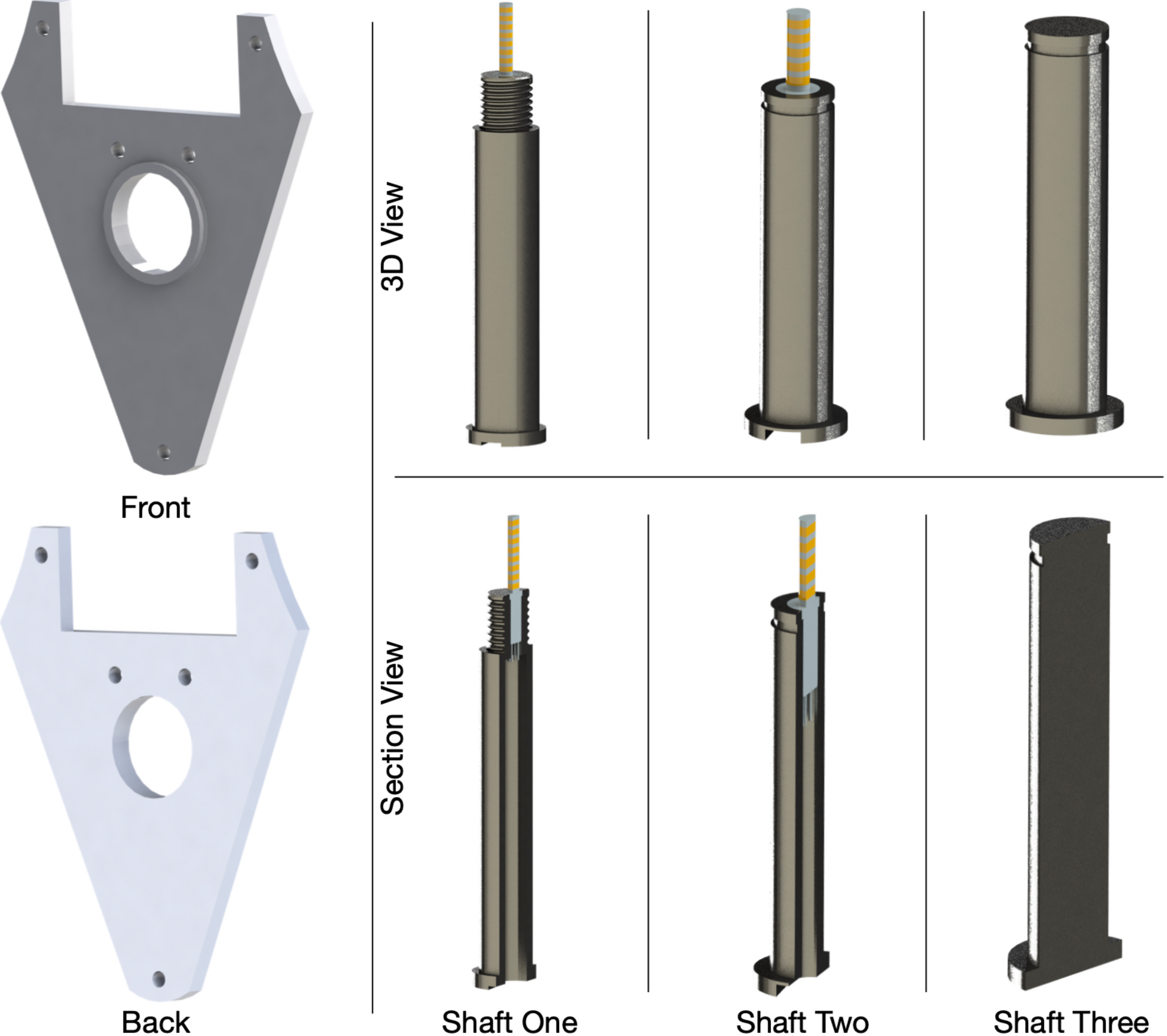}
    \caption{Design of the bearing plate and pendulum arm's shaft. The bearing can be installed on the bearing plate and together support the free rotational movement of the pendulum shaft and pendulum arm. The pendulum arm's shaft is designed hollow to accommodate the slip-ring wires. The groove on the bottom of the pendulum shaft is used to place guide the slip-ring wires into the pendulum arm.}
    \label{fig:PlatesAndShaft}
\end{figure}

An overview of our pendulum arm's design is described in section~\ref{sec:PendulumArm}. Here, we describe the design of the pendulum arm in more detail. The main components of the pendulum arm are: 1) pendulum body, 2) shaft, 3) bearing plate, and 4) 3D printed protection case. The overall structure of the pendulum arm is determined by how it transmits the rotational information measured by the encoders. 
In our design, a slip-ring sends the encoders' electrical signals (mainly the A-, B-, and I-channel signals) to the real-time system, and conducts currents to the encoder sensors on the rotational component. 
The slip-ring design introduces friction on the contact between the slip-ring shaft and brush block. 
To minimize this additional effect of friction, a miniature slip-ring and slip-ring brush are used with gold contact surfaces that reduce the electrical noise during the rotational movement of the pendulum arm. 
Other designs based on vision tracking or wireless technology to measure the pendulum arm rotational angle could avoid these additional friction effects. However, the slip-ring design is simple and does not have latency in the signal transmission that occur in vision-based and wireless communication type pendulum arms. 
Also, no additional computational resources are needed to determine the rotational angle of the pendulum compared to vision-based tracking systems. 
This characteristic is particularly beneficial for achieving higher sampling rates. 
The challenge of using a slip-ring is that it requires precision machining of the pendulum shaft. 
Moreover, the slip-ring has a fixed number of channels to transmit signals. This may reduce the flexibility of the setup if new sensors (e.g. an inertial measurement unit~(IMU) sensor) are required for future experiments. 
To accommodate the slip-ring and to connect it with the sensors, the first and second pendulum arm's shaft is designed to be hollow, as shown in Fig.~\ref{fig:PlatesAndShaft}. 
As the first pendulum arm shaft takes most of the static and dynamic load, it has a larger diameter than the second and third pendulum arm shafts. 
Moreover, to minimize the axial oscillation of the pendulum arm, a thread is machined on the first pendulum shaft such that the axial load of the pendulum arm is adjustable. When more axial force is applied, we can reduce the axial oscillation of the pendulum arm. 

\begin{figure}[t]
    \centering
    \includegraphics[width=0.9\textwidth]{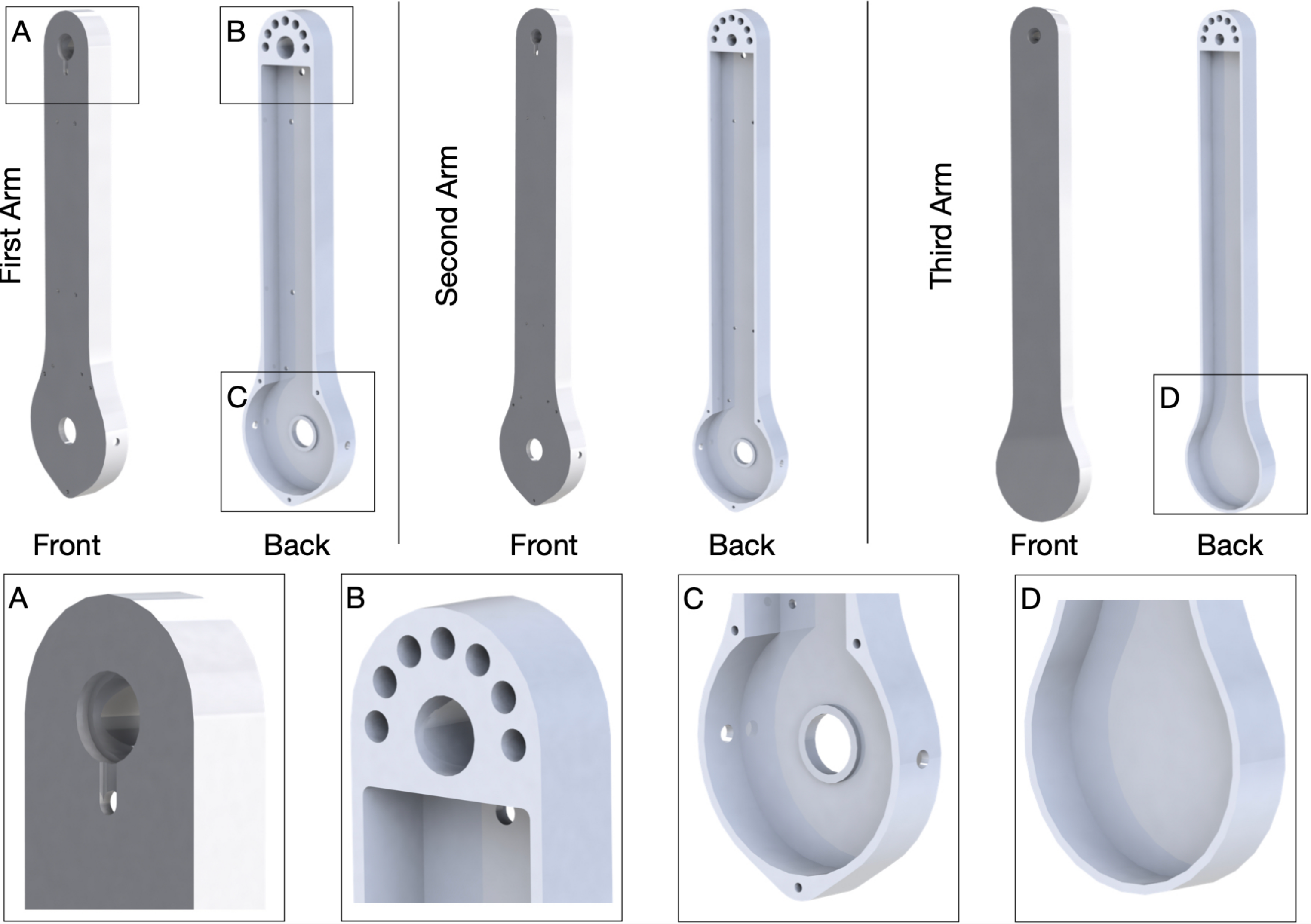}
    \caption{Design of the pendulum arm. (A) design features necessary to allow proper wiring of the slip-ring, and the hole where the pendulum shaft is installed. (B) back side of the pendulum where the shaft is installed, and a sequence of drilled holes that reduce the weight of pendulum arm. (C) place where the bearing and encoder are installed. (D) third pendulum arm without bearing hole.}
    \label{fig:ArmsDetails}
\end{figure}

The slip-ring wires enter the pendulum arm through a small hole on the front side of the pendulum arm, near the position where the pendulum shaft is mounted, as shown in Fig.~\ref{fig:ArmsDetails}~(A). The slip-ring wire enters the pendulum arm through this hole and connects with the encoder located on the backside of the pendulum arm. The groove in Fig.~\ref{fig:ArmsDetails}~(A) is aligned with the groove on the pendulum arm's shaft, as shown in Fig.~\ref{fig:PlatesAndShaft}. Together, they guide the slip-ring wire and provide space to store it and avoid the cable to stick out of the front face of the pendulum arm. To properly manage the slip-ring wires inside the pendulum arm, a 3D printed cable clip is designed, shown in Fig.~\ref{fig:PendulumArmOverview}~(B). This cable clip holds the slip-ring wire to its place and prevents twining of the slip-ring wire. To reduce the weight of the pendulum arm, several holes are drilled near the shaft hole, indicated in Fig.~\ref{fig:ArmsDetails}~(B). Moreover, two holes on the side of the pendulum arm facilitate the installing of the encoder disk on the pendulum shaft, shown in Fig.~\ref{fig:ArmsDetails}~(C). In Fig.~\ref{fig:ArmsDetails}~(C), the stair case shoulder is shown that properly secures the bearing that is installed on the pendulum arm. This stair case shoulder is not needed in the third pendulum arm since no other pendulum arm is attach to it, shown in Fig.~\ref{fig:ArmsDetails}~(D).

Ceramic bearings are used to minimize the friction during the rotational movement. They are installed on the bearing plate shown in Fig.~\ref{fig:PlatesAndShaft} and the pendulum arm shown in Fig.~\ref{fig:ArmsDetails}. The great advantage of the ceramic bearing is that it operates without lubrication, effectively decreasing the drag force caused by lubricants. Moreover, a bearing without sealing is used to further reduce the friction force, since the pendulum setup is used in a clean lab space. Thus, special care must be taken when assembling and dissembling the pendulum arm to ensure that no dust or chips enter the bearing during the process. Two bearings are used to fully support the rotational movement of the pendulum arm. Moreover, the bearings are aligned by connecting the bearing plate and pendulum arm body together. Together, they form a bearing housing where the shaft is installed. To make sure the pendulum shaft does not slide out during the rotational movement of the pendulum arm, external retaining rings are used to secure the pendulum shaft. To install the external retaining rings, grooves are machined onto the top of the second and third pendulum arm shafts, shown in Fig.~\ref{fig:PlatesAndShaft}.

\subsection{Manufacturing Details}
Several components of the pendulum arm are manufactured: 1) pendulum arm (Fig.~\ref{fig:ArmsDetails}), 2) pendulum shaft (Fig.~\ref{fig:PlatesAndShaft}), 3) bearing plate (Fig.~\ref{fig:PlatesAndShaft}), 4) protection case (Fig.~\ref{fig:ArmsDetails} (E) and (F)), and 5) wire clipper (Fig.~\ref{fig:ArmsDetails} (B)). In this section, we talk about the manufacturing details of those components. 
\begin{figure}[t]
    \centering
    \includegraphics[width=0.9\textwidth]{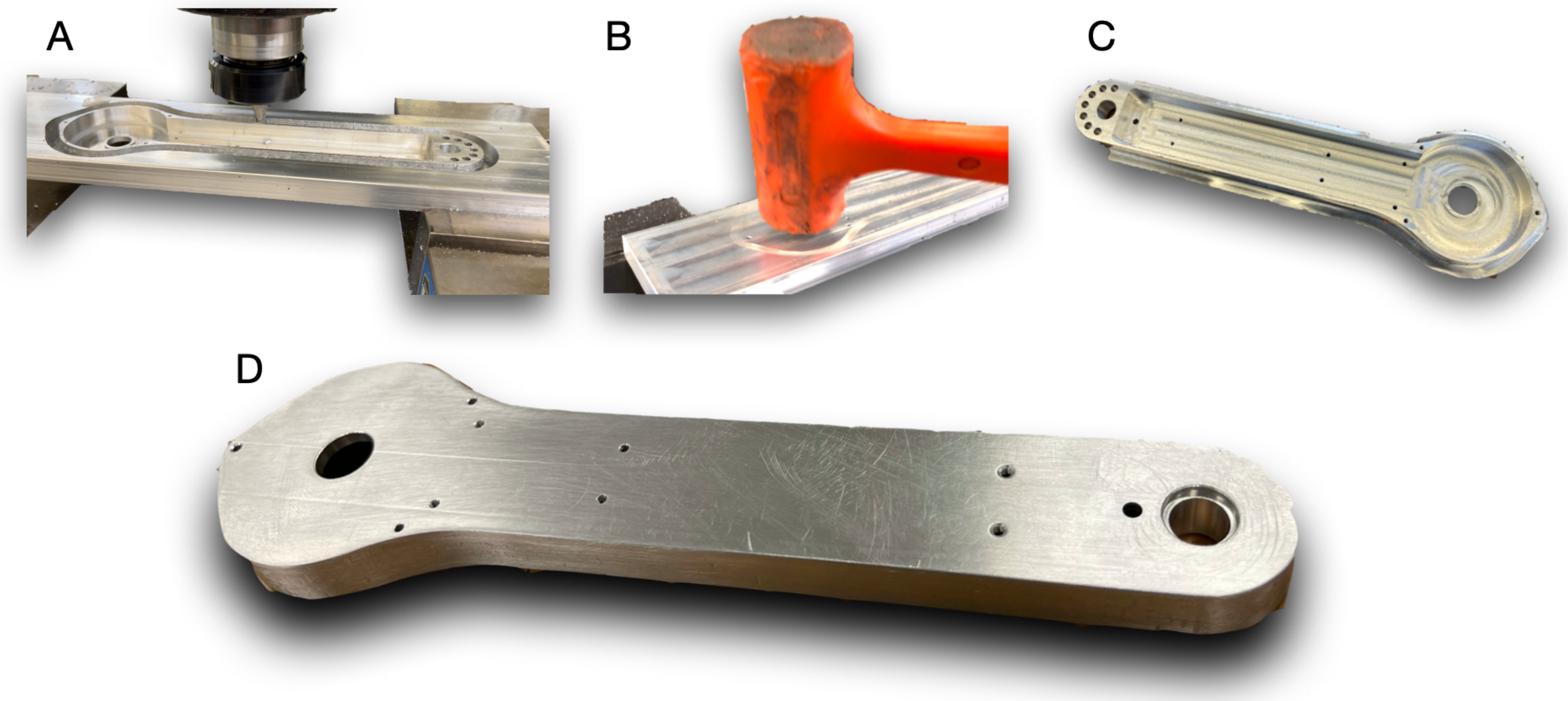}
    \caption{Overall steps needed to manufacture the pendulum arm: (A) the inside features of the pendulum arm is machined, and the outline of the pendulum arm is processed until only a thin layer of aluminium connects the pendulum arm and aluminium block. (B) and (C) the machined pendulum arm is knocked off using a robber hammer. (D) the machined pendulum arm is polished to get rid of any burr on the edge of the pendulum arm.}
    \label{fig:ArmMachining}
\end{figure}

The pendulum body is CNC milled from a multipurpose 6061 aluminum block with a dimension of 5/8" thick, 3" wide, 1 foot long. The aluminum block is first face milled on the front and back faces. Then the edge of the aluminum block is machined. This step allows to make the faces of the aluminum block parallel to each other~(front and back, edge to edge). Next, the aluminum block is secured on the CNC machine, and the center of the shaft hole is used as the origin of the x, y, and z coordinates. The back side of the pendulum arm is first machined. While more material remains in the aluminum bar, the hole for the pendulum shaft is first drilled to a diameter close to the actual diameter needed. Next, the CNC mill is used to refine the shape and dimension of the hole until it is close to the desired dimensions. Finally, the reamer is used to refine the size of the hole for a smooth installation of the shaft. We choose to first machine the hole for pendulum shaft when there is more material left, since this will increase the rigidity of the aluminum bar during the drilling and milling process, resulting in a more precise hole. After manufacturing the hole for the shaft, the hole for the bearing installation is machined using similar steps, where the hole is first drilled, then milled, and finally refined using a reamer. Next, the threaded hole for connecting the bearing plate and the pendulum arm is drilled, as shown in Fig.~\ref{fig:ArmsDetails}~(C). After this step, the weight reducing hole shown in Fig.~\ref{fig:ArmsDetails}~(B) is machined. Next, the aluminum block is flipped and recenter using edge finder. Then, the hole shown in Fig.~\ref{fig:ArmsDetails}~(A) is milled to correct shape and depth. Next, a threaded hole is drilled to install the wire clipper and encoder, shown in Fig.~\ref{fig:PendulumArmOverview}~(B) and Fig.~\ref{fig:ArmsDetails}~(C). Then the pendulum arm is flipped again to the back side to machine the inner part. The inner side of the pendulum arm is milled by programming the CNC machine to achieve the desired shape, shown in Fig.~\ref{fig:ArmsDetails}~\footnote{The inner side corner of the manufactured pendulum arm will not be a straight corner. Instead, its shape is an arc whose radius is determined by the radius of the end mill used to machine the inner side.}. Once this step is finished, the outer shape of the pendulum arm is milled, shown in Fig.~\ref{fig:ArmMachining}~(A). The TiAlN Coated, 2 flute, 1/4" mill diameter, 2-1/2" overall length end mill is used to mill the outer shape of the pendulum arm. At the final round of machining, a thin layer~(0.5 mm) of aluminium is left to connect the pendulum arm body and the remaining of aluminum block, then a rubber hammer is used to knock out the pendulum arm, as shown in Fig.~\ref{fig:ArmMachining}~(A) and~(B). The resulting pendulum arm will have rough edges, shown in Fig.~\ref{fig:ArmMachining}~(C). Thus, the pendulum arm is polished after the holes on the side of the pendulum arm shown in Fig.~\ref{fig:ArmsDetails}~(C) are drilled. The final pendulum arm is shown in Fig.~\ref{fig:ArmMachining}~(D).

The similar process is used to machine the bearing plate by using a 0.16" thick, 6" x 6" multipurpose 6061 aluminum sheet. After the aluminum sheet is secured on the CNC machine, the face mill is used to get the correct height of the bearing plate.  Next, the bearing hole is manufactured. Then the through holes shown in Fig.~\ref{fig:PlatesAndShaft} are drilled. Finally, the outer shape of the bearing plate is milled until there is only a thin layer of aluminum connecting the bearing plate and aluminum sheet, and the bearing plate is knocked out using the same approach shown in Fig.~\ref{fig:ArmMachining}~\footnote{Same as footnote 3, the straight corner in Fig.~\ref{fig:PlatesAndShaft} will be an arc whose radius is determined by the radius of the end mill.}.

The material we used for the pendulum shaft is 1566 carbon steel, which balance high strength and good machinability. For the first pendulum arm, 12" long material with 1/2" diameter raw material is used, while 12" long 3/8" diameter raw material is used for the second and third shaft. To manufacture the shaft, a lathe is used. When machining the first and second arm shaft, a through hole is first drilled. Then, a shallow hole is drilled to create a step which allows proper installation of the slip-ring. This can be seen in Fig.~\ref{fig:PlatesAndShaft}. Next, the main diameter of the shaft is lathed. After the major shape of the shaft is obtained, a groove is lathed for the second and third arm to allow installing the external retaining ring. For the first arm, a thread is lathed to allow application of the axial force by screwing a nut. The groove on the bottom of the first and second shaft is milled after it is press fitted into the pendulum arm body. After the machining, all the parts are polished and cleaned to remove dust or metal chips. Finally, the protection case and wire clip is 3D printed using Polylactic Acid~(PLA) material. The main functionality of the case is to protect the slip-ring inside the pendulum arm from accidental collision.

\subsection{Assembly Details of Second and Third Arm}

The assembly of the double pendulum is described in section~\ref{sec:PendulumArm}. For the assembly of the triple pendulum illustrated in Fig.~\ref{fig:Arm2Arm3Assemble}, the following steps are needed: 1) Follow all the assembly steps of the double pendulum. 2) Slide the shim~($11$) into the shaft of the third arm~($22$). Next, slide the third pendulum shaft~($22$) into the bearing of the second arm~($12$) and stop until the shim~($11$) contacts the inner ring of the bearing. 2) Slide the shim~($16$), encoder disk~($17$), and another shim~($18$) onto the third arm's shaft~($22$). Then, slide the assembled bearing plate onto the shaft~($22$). Finally, clip the external retaining ring~($19$) onto the shaft~($22$). This step is the same as mentioned before. 3) Slide the 3D printed shim~($6$) and encoder reader~($7$) into the desired place by aligning the holes. Then, install the screws~($8$) to secure its position. 4) Next, install the protection case by aligning the holes on the pendulum arm body, bearing plate, and protection case. Mount the screws~($4$) to fasten the assembly. 6) Finally, install the wire clipper~($5$) into the pendulum arm~($15$) by aligning the holes and tighten the screws~($4$). The above steps finish the assembly of the second and third pendulum arm, and results in the final assembly shown in Fig.~\ref{fig:PendulumArmOverview}. The complete assembly process of the second and third arm are illustrated in Fig.~\ref{fig:Arm2Arm3Assemble}.

\section{Pendulum Cart Details}
\label{Appendix:PendulumCart}

\begin{figure}[t]
    \centering
    \includegraphics[width=1\textwidth]{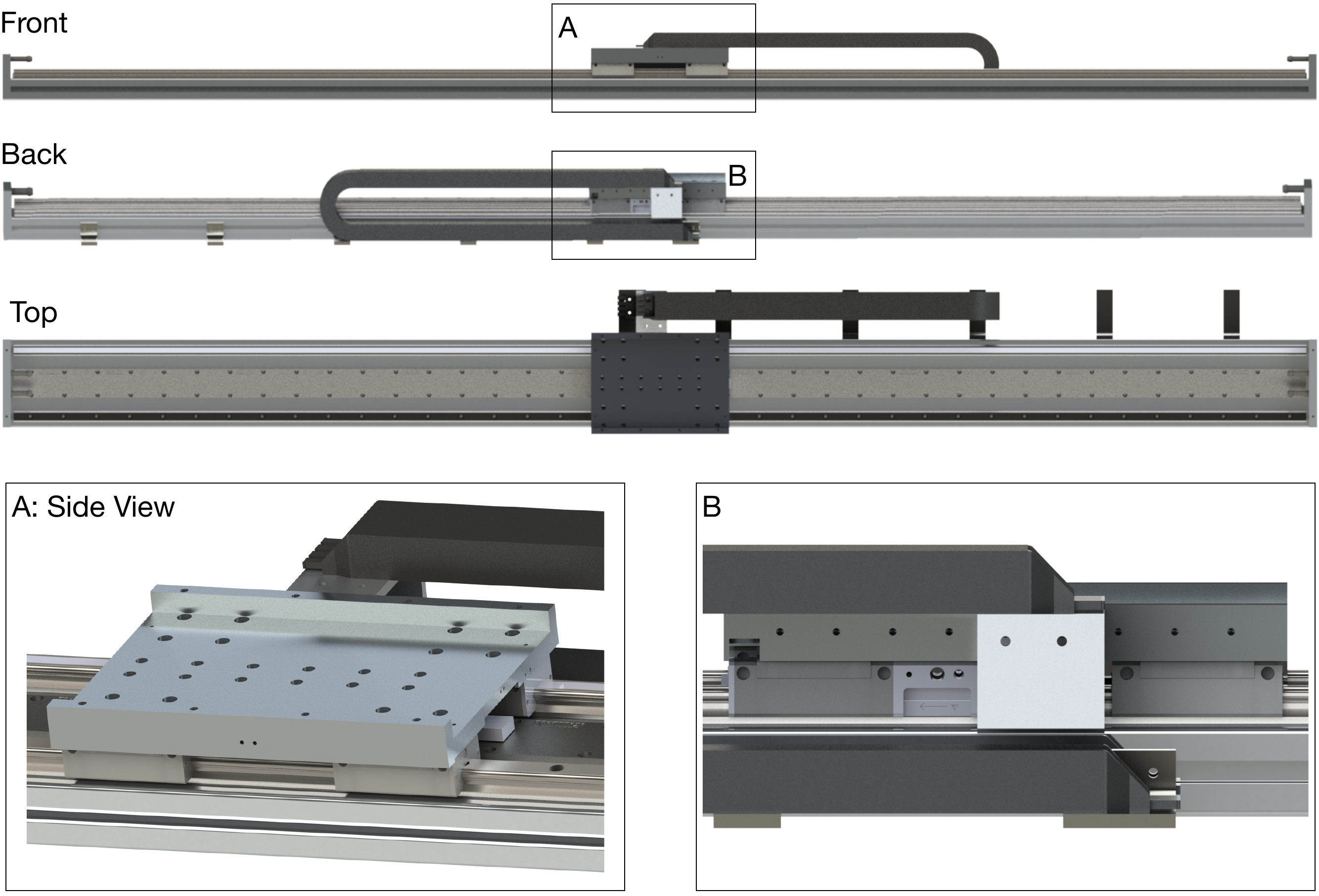}
    \caption{Overview of the linear motor that drives the pendulum cart, with the linear motor cable track and the stage with pre-drilled holes, which allow the installation of user designed parts. This linear motor stage is used to mount the aluminum plate that is used to install the bearing housing and the pendulum arm. (B) Limit switch plate installed on the back side of the linear motor stage.}
    \label{fig:LinearMotor}
\end{figure}

Our multi-link pendulum uses a linear motor to provide actuation to the pendulum arm. The advantage of using the linear motor is that it does not have backslash issues, which frequently happen in the belt-drive type servo motor. This benefit allows accurate control of sensitive maneuvers, such as the swing-up of double and triple pendulums. 
As discussed in section~\ref{sec:PendulumCart}, to size the linear motor, first, the desired maximum speed and acceleration of the cart are defined using desired motion profile of the pendulum. We choose a HIWIN linear motor system LMX1K-SA12-1-2000-PGS1-V103+HS. 
For more details on choosing the linear motor size, we refer the readers to see Appendix A of the HIWIN linear motor system manual~\cite{HIWIN_Linear}. 
The overall look of the selected linear motor can be seen in Fig.~\ref{fig:LinearMotor}. As Fig.~\ref{fig:LinearMotor} shows, the linear motor already comes with a cable management track and motor stage. Moreover, Fig.~\ref{fig:LinearMotor}~(A) suggests the motor stage is not flat, which suggests the proper design of connection between the pendulum arm and linear motor stage is needed. Fig.~\ref{fig:LinearMotor}~(B) indicates there are pre-drilled holes on the backside of the linear motor stage. Those pre-drilled holes can facilitate the installation of a switch plate. Finally, Fig.~\ref{fig:LinearMotor} also indicates that the both side of the linear motor rail has shock absorbers installed which prevents the direct collision of the motor stage and the end of linear motor rail.

\begin{figure}[t]
    \centering
    \includegraphics[width=1\textwidth]{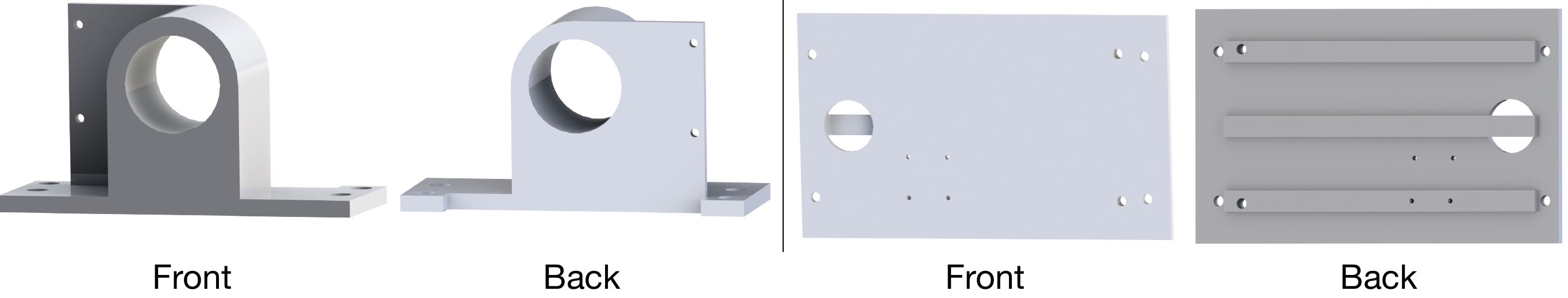}
    \caption{This figure illustrates the overall look of the bearing housing and cart plate. The main features on the bearing housing is the thorough hole that is used to install the bearing. To make the manufacturing process easier, a thorough hole is used and a spacer is later installed to maintain the distance of two bearings. As for the cart plate, it has simple shape and features. The four tiny threaded holes are used to install the cable management tool while the larger hole is milled to install the circular level indicator.}
    \label{fig:BearingHouseAndPlateDetails}
\end{figure}

The connection between the pendulum arm and the linear motor stage are shown in Fig.~\ref{fig:PendulumCartOverview}~(A) and (B). A bearing housing is needed to provide support for the first pendulum arm shaft and secure it so that the pendulum arm can perform free swing. Moreover, an aluminum plate is machined so that the bearing housing can be connected with the linear motor stage. 
The details of the bearing housing can be seen in Fig.~\ref{fig:BearingHouseAndPlateDetails}. As Fig.~\ref{fig:BearingHouseAndPlateDetails} shows, one thorough hole is drilled on the front and back side of the housing. This hole is used to install the bearings that mate with the first pendulum shaft. In order to maintain the relative distance of the bearings, a spacer needs to be installed. Same as the pendulum arm, the ceramic bearing is used to minimize the drag force. Four more thorough holes are been drilled on the bottom of the bearing house which is used to mount the bearing housing to the aluminum plate and linear motor stage. Two threaded holes are drilled on the side panel of the bearing house and are used to mount the encoder reader. To avoid collision of the encoder disk and bearing housing, the bottom of the bearing housing has been designed as a C shape. As for the aluminum plate, the design is straightforward as Fig.~\ref{fig:PendulumCartOverview} and Fig.~\ref{fig:BearingHouseAndPlateDetails} shows. It has four holes that is used to mate the aluminum plate with linear motor stage. The back side of the aluminum plate has multiple stripes that acts as a stiffener which is used to strengthen the plate. Four small threaded holes are drilled on the top of the aluminum plate to install 3D printed cable managing tool while one bigger hole is milled to install the circular level indicator. This circular level indicator is useful as it can help user to level the pendulum cart. The final piece we designed for the pendulum cart are the limit switch plate as it shown in Fig.~\ref{fig:PendulumCartOverview}~(B). Fig.~\ref{fig:PendulumCartOverview}~(B) shows two limit switch plates installed on the back side of the linear motor stage, they are responsible to block the laser limit switch when the pendulum cart moves to the edge of the linear rail. The laser limit switch is installed on the linear motor by designing a 3D printed base as Fig.~\ref{fig:LimitSwitchAssemble}~(B) shows.

\subsection{Manufacturing Details}
The major parts that need to be manufactured in the pendulum cart are: 1) Bearing housing. 2) Pendulum cart plate. 3) Limit switch plate. 4) 3D printed cable management tool. 5) 3D printed limit switch base. This section will go through the manufacturing process of those components.

\begin{figure}[t]
    \centering
    \includegraphics[width=1\textwidth]{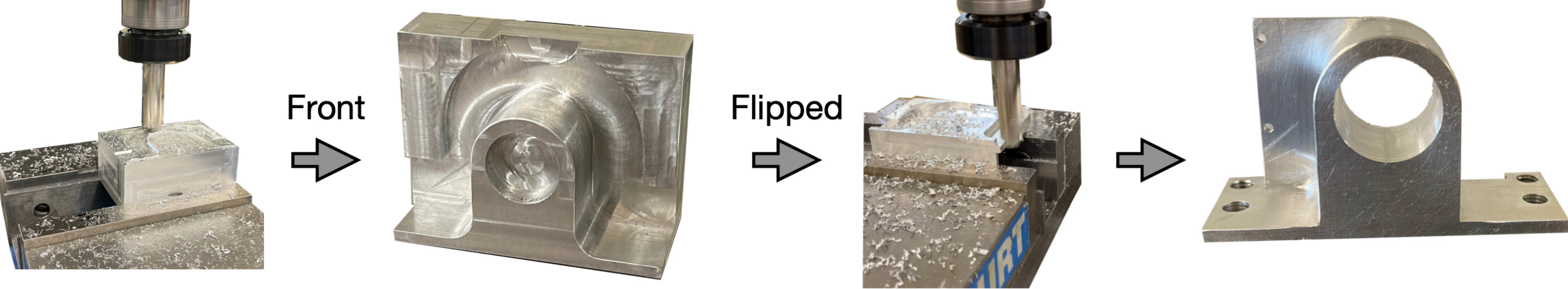}
    \caption{This figure illustrates the overall process of machining the bearing housing. To start with, the aluminum cube is face milled to create the datum face. Then the features on the front of the bearing housing is machined. Next, the aluminum cube is flipped over to machine the features on the backside of the aluminum cube and all the necessary holes needed are drill. Finally, the bearing housing is polished and sharp edges are smoothed.}
    \label{fig:BearingHouseMachineProcess}
\end{figure}

To manufacture the bearing housing, a 3" multi-purpose aluminum cube is machined.  Fig.~\ref{fig:BearingHouseMachineProcess} shows the overall machining process. 

1) To start with, the raw material is face milled on all fronts to create datum planes. Next, a shallow hole whose diameter is smaller than the diameter of the bearing is milled, and the center of the hole is set as the origin of the coordinate system. Then, the front of the bearing housing is milled to create the wanted shape. The milling continues until the end mill reaches the same height as the side panel used to mount the encoder reader. 2) Once the front shape of the bearing housing is machined, it is flipped over to manufacture the features on the backside. The aluminum cube is first milled until the end mill is at the same height as the total width of the bearing housing. Next, a through-hole is drilled whose diameter is smaller than the bearing to be installed. Then a reamer is used to refine the shape and diameter of the hole. The goal is to make the bearing hole and bearing have a transitional fit. Once this step is done, the milling is continued until the backside of the bearing housing is machined to desired shape. Then two threaded holes are drilled for installing the encoder reader. When drilling these two holes, the origin of the bearing hole is used as the center of the coordinated system since the relative distance between the bearing hole and the holes for installing the encoder reader is what matters. 3) Finally, the bearing housing is milled in the upright position to produce a straight corner. Once finished, it is polished using sandpaper, and the sharp edges are smoothed using a deburring wheel. It is noteworthy that this bearing housing can be 3D printed if it does not have too much load on it, for example, when only a single pendulum formulation is needed.

The manufacturing of the pendulum cart plate is easy to do. A  multipurpose 6061 Aluminum with 7/16" thickness and 8" x 8" length and width is used to manufacture the pendulum cart plate. The following steps are needed to manufacture it. 1) To start with, the raw material is face milled to create a datum plane. Next, it is milled to have the same width and length as the designed cart plate. 2) Next, the front face of the cart palate is face milled, and all the holes needed are drilled. This process includes the hole used to install the circular level indicator. 3) Once the features on the front of the cart plate are manufactured, it is flipped over to mill the stiffener. Finally, the post-processing is done to polish the cart plate, and the deburring wheel is used to make the smooth edges. If the pendulum cart does not have a considerable load and does not require high-speed movement, it is also possible to 3D print the pendulum cart plate.

The left and right limit switch plates are manufactured using a multipurpose 6061 aluminum sheet with 0.016" thickness. The overall outline of the limit switch plate is first drawn on the aluminum sheet using a marker pen and ruler. Then, the holes are drilled, which are needed to install the plates to the linear motor stage. Then, a metal brake and scissors cut the aluminum sheet to the desired shape. Finally, the aluminum sheet is bent, polished, and deburred. After using the above steps, the switch plates can be produced easily. As for cable management tools and limit switch base, they are 3D printed with PLA filament.

\section{System Frame Details}
\label{Appendix:SystemBase}

The linear motor selected in Sec.~\ref{sec:PendulumCart} is mounted on a support frame in order to minimize the unwanted vibration during the cart movement. This motivates us to design a system frame to mount the pendulum cart and provide support to it. This section introduces the design and manufacturing of the system frame we used. Moreover, we show how to install the linear motor to the system frame and provide a detailed bill of materials for reproduction.
 
\subsection{Design Details}
\begin{figure}[t]
    \centering
    \includegraphics[width=.95\textwidth]{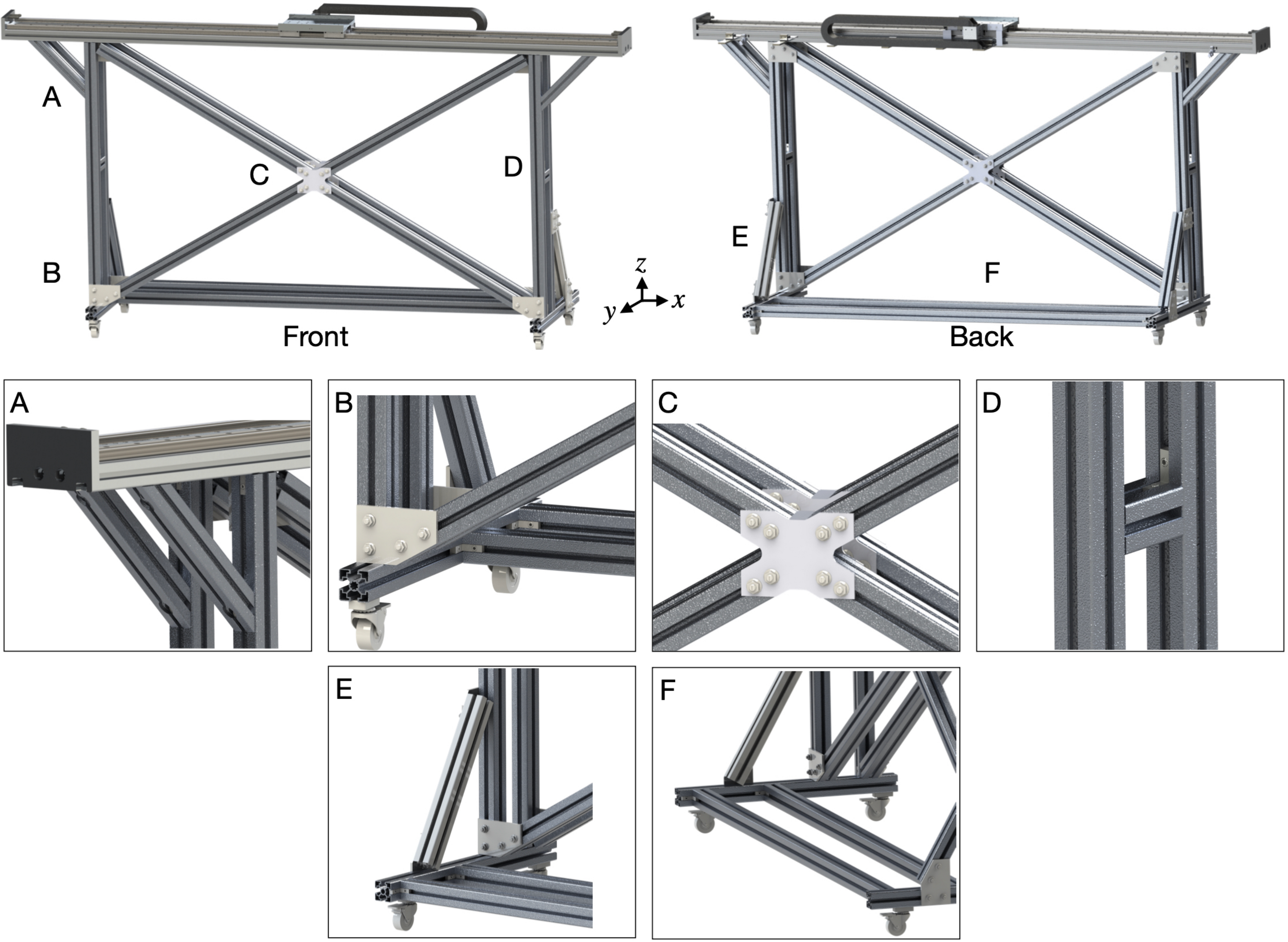}
    \caption{System frame. The linear motor is lifted up using four vertical aluminum extrusion. Different aluminum extrusions are installed to constrain the movement of system frame in the $x-z$ and $y-z$ plane. To allow easy moving of the experimental setup, four wheels can be installed to the system frame as shown in (E) and (F).}
    \label{fig:FrameOverview}
\end{figure}

Fig.~\ref{fig:FrameOverview} shows the overall view of the system frame we designed. Our design is similar to the one shown in~\cite{vcevcil2016radio,vsetka2017triple}. To make sure the pendulum arm won't collide with the ground, four vertical aluminum extrusions are connected with the linear motor using corner concealed brackets. This lifts up the linear motor from the ground and creates space for the pendulum arm. To make sure the vertical aluminum extrusion is stable in the $x-z$ plane, the diagonal brace is used to connect the linear motor and vertical aluminum extrusion as shown in Fig.~\ref{fig:FrameOverview}~(A). Moreover, four more aluminum extrusions are connected together using a machined connector plate to create a X shaped frame as shown in Fig.~\ref{fig:FrameOverview}~(C). By connecting the X shaped frame and the four vertical aluminum extrusions, the movement of the frame is further constrained in the $x-z$ plane. Next, the frame's movement is constrained in the $y-z$ plane. This is achieved by connecting the vertical frames using an aluminum extrusion as shown in Fig.~\ref{fig:FrameOverview}~(B) and (D). To provide more support, the vertical aluminum extrusion on the back side is further connected to another aluminum extrusion to form a triangle using T-slotted framing structural brackets, as shown in Fig.~\ref{fig:FrameOverview}~(E) and (F). To provide a space to place the electrical box, the real-time system and the development system, two horizontal aluminum extrusions are used to connect the left and right vertical aluminum extrusion as shown in Fig.~\ref{fig:FrameOverview}~(F). Finally, to allow easy movement of the experimental system from place to place, four wheels can be installed as an option which is shown in Fig.~\ref{fig:FrameOverview}~(F). However, it is recommended that the wheels are only installed when it is necessary to move the system frames. When performing actual experiments, the wheels should be removed to avoid unwanted oscillation of the system frame.

\subsection{Manufacturing Details}
\begin{figure}[t]
\vspace{-.1in}
    \centering
    \includegraphics[width=.95\textwidth]{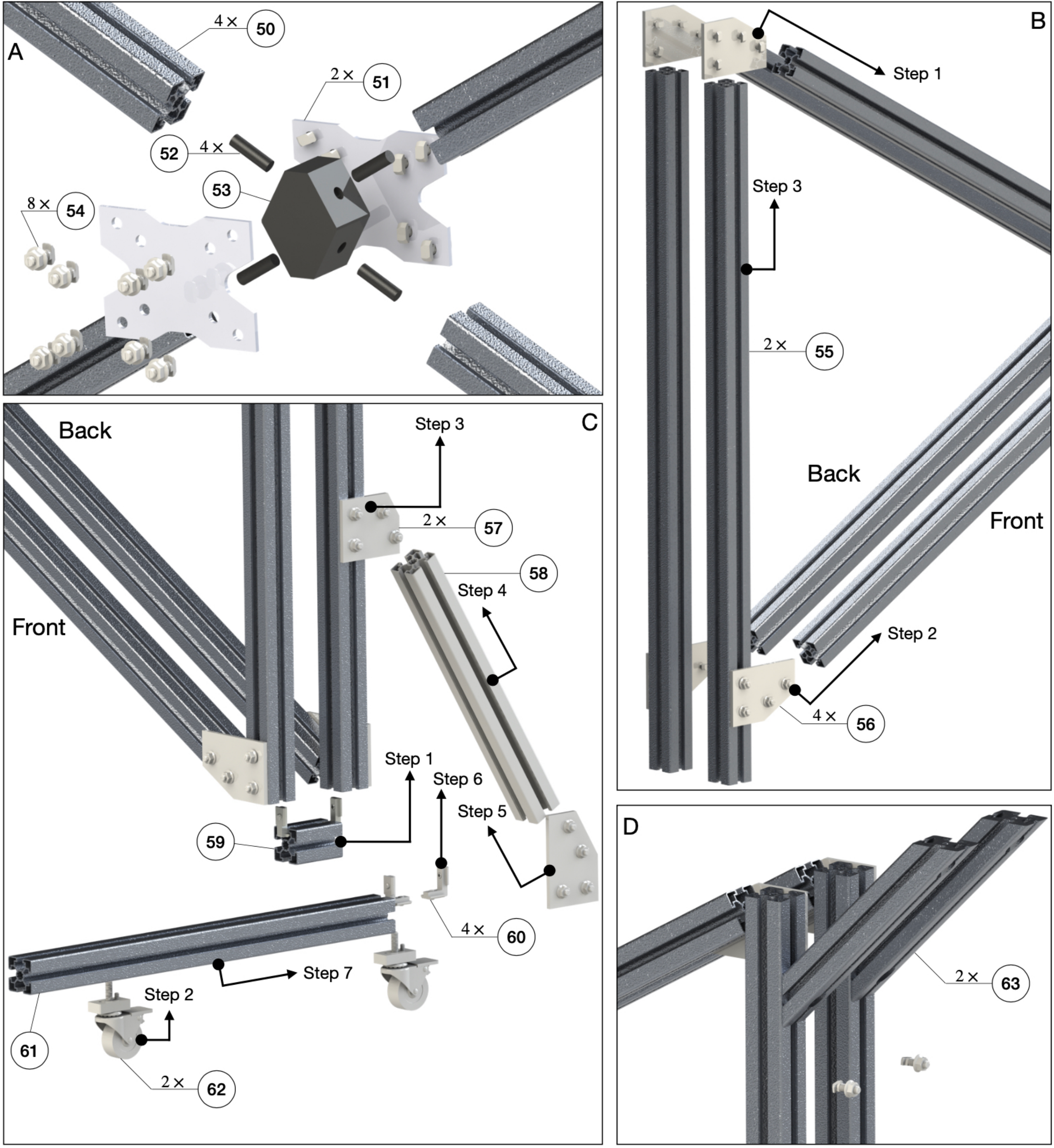}
    \vspace{-.1in}
    \caption{Assembly of the system frame. (A) Installation procedure of the X shaped frame. (B) Connection of the X shaped frame with the vertical aluminum extrusion~($55$). (C) Installation of the wheel to the system frame. (D) Installation of diagonal brace.}
    \label{fig:FrameAssemble_1}
    \vspace{-.1in}
\end{figure}

The manufacturing of the system frame is straightforward since most of the components can be directly purchased online. The main components that need to be manufactured are: 1) Aluminum extrusion, which has to be trimmed down to the correct length. 2) Aluminum extrusion connector plate, which needs to be machined. 3) 3D printed alignment tool, which is used to facilitate the assembly of the X shaped frame. In order to trim the aluminum extrusion to the correct length, the band saw is first used to trim the aluminum extrusion to a length that is slightly longer than the desired length. Next, the CNC mill is used to machine the cut surface of the aluminum extrusion to the correct length. Finally, the machined end is polished and unwanted burr is removed. To machine the connector plate shown in Fig.~\ref{fig:FrameOverview}~(C), a 0.16" thick, 6" x 6" aluminum plate~(89015K255) is machined. First, the through holes that is used to mate with the bolts and nuts are drilled. Then, a simple CNC program is written to mill the outer shape of the connector. As for the alignment tool that helps the assembly of the X frame, it is printed using PLA filament.

\subsection{Assembly Details}
To assemble the system frame, four major steps are needed: 1) The X shaped frames are assembled. 2) The X shaped frame is connected with the four vertical aluminum extrusions. 3) The assembled aluminum frame is connected with the aluminum  extrusion in the $x-y$ plane and diagonal braces for single rails. 4) Finally, the linear motor is installed with the system frame and all the screws are fastened accordingly. The above mentioned steps are illustrated in Fig.~\ref{fig:FrameAssemble_1}. 

As Fig.~\ref{fig:FrameAssemble_1}~(A) shows, to better assemble the X shaped frame, a 3D printed alignment tool~($52,53$) is used. The alignment tool is first assembled by inserting the 3D printed cylinder~($52$) into the 3D printed block~($53$). The 3D printed block~($53$) serves as a reference to make sure each aluminum frame~($50$) has the correct angle to each other. This is achieved by inserting the aluminum frame~($50$) into the 3D printed cylinder~($54$), making sure it is perpendicular to the surface of 3D printed block~($53$). Once the correct angle between the aluminum frame~($50$) is achieved, the connector plate~($51$) is used to connect each frames using drop-in bolts and nuts~($54$). This finishes the step 1, assembly of the X shaped frame. 

In step 2, 60 degree bracket~($56$) is first slid into the aluminum frame~($50$) as shown in Fig.~\ref{fig:FrameAssemble_1}~(B) step 1 and step 2. Then the vertical aluminum frame~($55$) is slid into the 60 degree bracket as shown in step 3 of Fig.~\ref{fig:FrameAssemble_1}~(B). This completes the assembly of the X shaped frame and vertical frame. To avoid the collision of the pendulum arm and 60 degree bracket~($56$), the top right and top left 60 degree bracket~($56$) is installed on the back side of front X shaped frame, as shown in Fig.~\ref{fig:FrameAssemble_1}~(B). 

\begin{figure}[t]
    \centering
    \includegraphics[width=\textwidth]{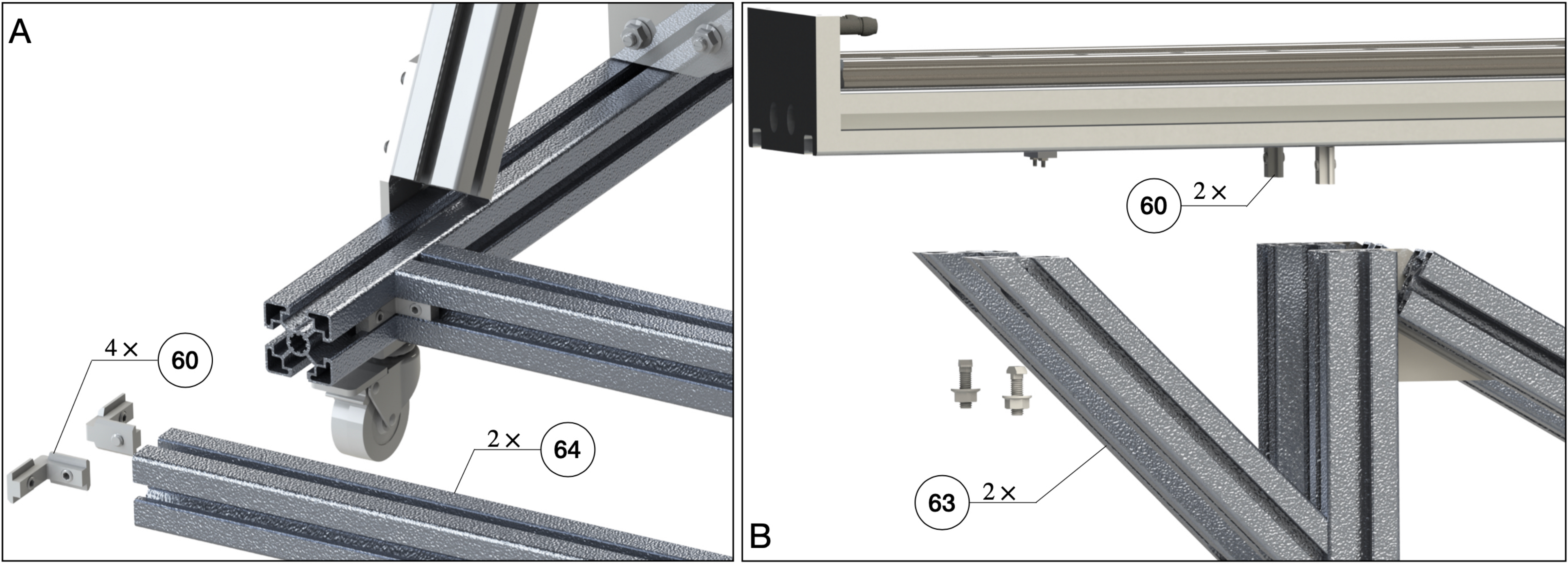}
    \caption{Assembly process of the system frame~(A) and linear motor~(B).}
    \label{fig:FrameMotorAssemble}
\end{figure}

Fig.~\ref{fig:FrameAssemble_1}~(C) and (D) illustrates the details of step 3. First, the corner concealed bracket~($60$) is slid into the aluminum spacer~($59$). Then, this aluminum spacer is slid into the middle of front and back X shaped frame as shown in Fig.~\ref{fig:FrameAssemble_1}~(C) step 1. In step two, the wheel~($62$) of the system frame is assembled with the aluminum frame~($61$). In step 3, the 30 degree bracket~($57$) is first slid into the vertical aluminum frame of the back X shaped frame~($55$) and the bolts and nuts are slightly tightened to keep them in place. Then, step 4 is conducted in which the aluminum frame~($58$) is slid into the 30 degree plate~($57$). Next, in step 5 a 60 degree bracket~($56$) is slid into the aluminum frame~($58$) and slightly tightened. In step 6, the corner concealed bracket~($60$) is installed with the vertical aluminum frame~($55$) and in step 7 the aluminum frame~($61$) is slid into the corner concealed bracket~($60$) and 60 degree plate~($56$). Next, the diagonal brace in installed into the vertical frame as shown in Fig.~\ref{fig:FrameAssemble_1}~(D), and the screws are slightly tightened to make sure the diagonal brace stays in place. The procedure to install the horizontal aluminum extrusion~($64$) to the aluminum extrusion with wheel~($61$) can be seen in Fig.~\ref{fig:FrameMotorAssemble}~(A). As Fig.~\ref{fig:FrameMotorAssemble}~(A) shows, the corner concealed bracket~($60$) is first installed into the aluminum extrusion~($64$) then extrusion~($64$) is slid into the extrusion~($61$). This further stabilizes the system frame and extrusion~($64$) can be used to place the development computer and electrical box, as shown in Fig.~\ref{fig:SystemOverview}. The above steps finish the assembly of the system frame. After the system frame is assembled, the linear motor~($30$) can be installed onto the system frame shown in Fig.~\ref{fig:FrameMotorAssemble}~(B). To do so, the corner concealed bracket~($60$) is installed onto the aluminum base of the linear motor. Then the corner concealed bracket~($60$) is slid into the vertical aluminum frame~($55$). Next, the diagonal brace is connected with the linear motor. Finally, all the screws and nuts are tightened. Since the linear motor and aluminum frame can be heavy, all the installation process mentioned above should be finished by at least two persons.

\section{Electrical System Details}
\label{Appendix:ElectricalSystem}
\subsection{Pendulum Arm Wiring Details}
\label{Appendix:ElectricalSystem_Pendulum}

\begin{figure}[t]
    \centering
    \includegraphics[width=\textwidth]{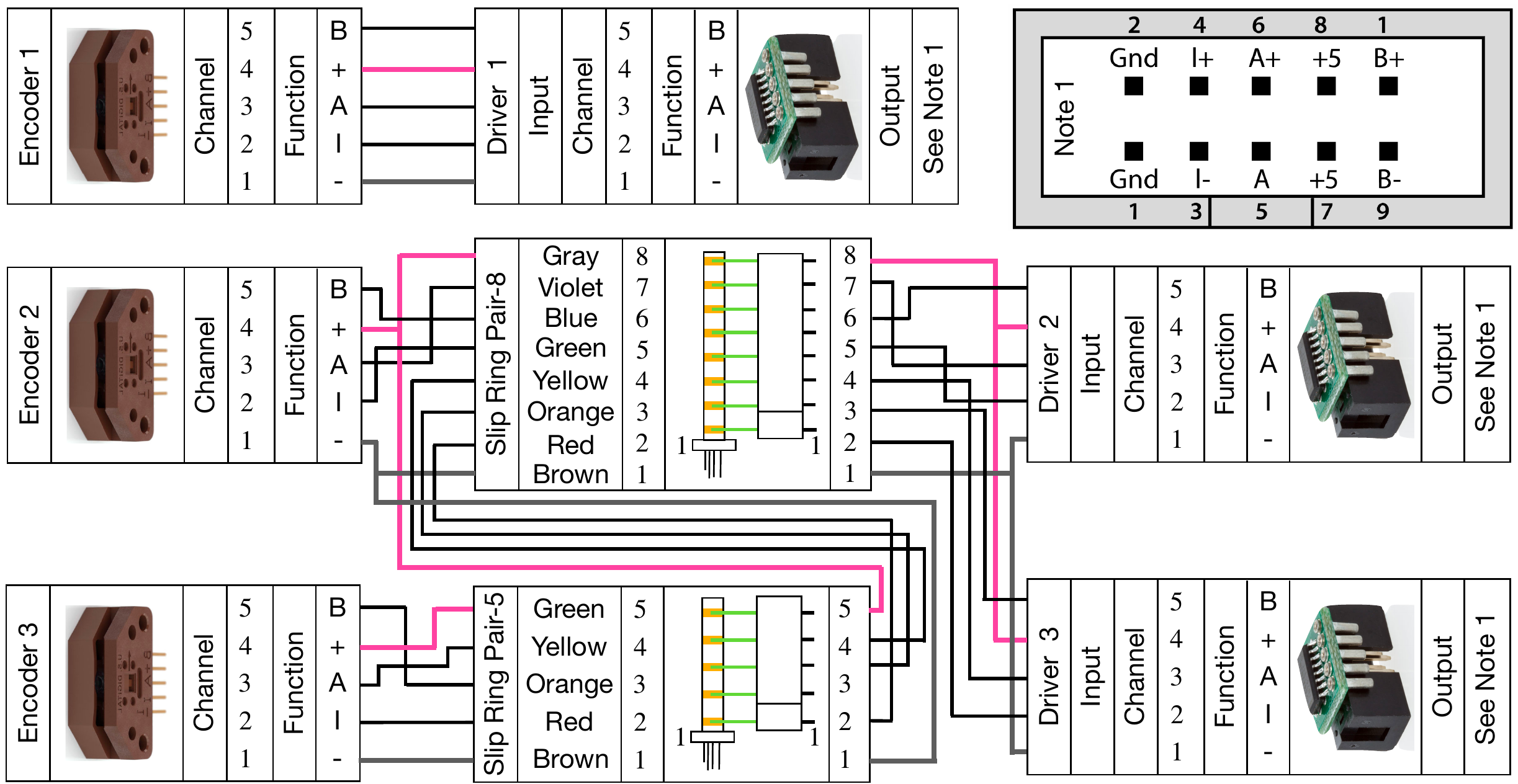}
    \caption{This figure illustrates the overall electrical wiring of the pendulum arm encoder. The first encoder is directly connected to the first differential driver. To connected the second and third encoder to the differential driver, two slip-rings~(slip-ring and brush block) are used. The second and third differential driver shares the 5V and ground channel of the first slip-ring. Moreover, the first slip-ring's 5V and ground channel is connected to the second slip-ring, which then conducts electricity to the third encoder. The third encoder uses the both the first and second slip-ring to connect with the differential driver. The output of the differential driver is connected to the US Digital CA-C10-SH-NC 10 feet cables, and those cables are then connected to the IO-392 module as Table.~\ref{table:IO-392-PinOut} shows.}
    \label{fig:PenArmWiring}
\end{figure}

Fig.\ref{fig:PenArmWiring} and Table.~\ref{table:IO-392-PinOut} show the detailed wiring specification of the pendulum arm. Fig.\ref{fig:PenArmWiring} shows the first encoder is directly connected to the first differential driver while the second and third encoders are connected with the differential driver using slip-rings. Moreover, the output of the differential driver connects to the US Digital CA-C10-SH-NC 10 feet cables, which then connect to the target computer terminal board as Table.~\ref{table:IO-392-PinOut} suggests. It is important to point out that the US Digital CA-C10-SH-NC 10 feet cables are shielded cables. Thus, the shield also connects to the ground of the IO-392 ground pin, which helps reduce the effect of EMI noise.

\begin{table}[ht]
\centering
\caption{This table shows the pin out of IO-392 module for encoder reading and its connections with the differential driver. D1 to D3 represents the differential driver 1 to 3 shown in Fig.~\ref{fig:PenArmWiring}. "-x" represents the corresponding "x"-th pin of the specific electrical component. The ground of the encoder sensor is connected to "$0$ V" instead of "ground" as suggested by Speedgoat.}
\label{table:IO-392-PinOut}
\resizebox{\textwidth}{!}{%
\begin{tabular}{|c|c|c|c|c|c|}
\hline
Pin &
  \begin{tabular}[c]{@{}c@{}}Code Module \\ Channel\end{tabular} &
  Functionality &
  Direction &
  Transceiver &
  Connects To \\ \hline
1a & \diagbox[width=\dimexpr \textwidth/8+4\tabcolsep\relax, height=0.47cm]{}{}
   &
  Ground & \diagbox[width=\dimexpr \textwidth/8+8\tabcolsep\relax, height=0.47cm]{}{}
   &
   \diagbox[width=\dimexpr \textwidth/8+8\tabcolsep\relax, height=0.47cm]{}{} & 
  \diagbox[width=\dimexpr \textwidth/8+12\tabcolsep\relax, height=0.47cm]{}{} \\ \hline
2a  &      \diagbox[width=\dimexpr \textwidth/8+4\tabcolsep\relax, height=0.47cm]{}{}              & Ground                           &          \diagbox[width=\dimexpr \textwidth/8+8\tabcolsep\relax, height=0.47cm]{}{}           &        \diagbox[width=\dimexpr \textwidth/8+8\tabcolsep\relax, height=0.47cm]{}{}      &     \diagbox[width=\dimexpr \textwidth/8+12\tabcolsep\relax, height=0.47cm]{}{}              \\ \hline
3a  & \multirow{2}{*}{1} & \multirow{2}{*}{QAD-A}           & \multirow{2}{*}{IN} & RS422/485(+) & D1-6             \\ \cline{1-1} \cline{5-6} 
4a  &                    &                                  &                     & RS422/485(-) & D1-5             \\ \hline
5a  & \multirow{2}{*}{1} & \multirow{2}{*}{QAD-B}           & \multirow{2}{*}{IN} & RS422/485(+) & D1-10            \\ \cline{1-1} \cline{5-6} 
6a  &                    &                                  &                     & RS422/485(-) & D1-9             \\ \hline
7a  & \multirow{2}{*}{1} & \multirow{2}{*}{QAD-C/Index}     & \multirow{2}{*}{IN} & RS422/485(+) & D1-4             \\ \cline{1-1} \cline{5-6} 
8a  &                    &                                  &                     & RS422/485(-) & D1-3             \\ \hline
9a  & \multirow{2}{*}{2} & \multirow{2}{*}{QAD-A}           & \multirow{2}{*}{IN} & RS422/485(+) & D2-6             \\ \cline{1-1} \cline{5-6} 
10a &                    &                                  &                     & RS422/485(-) & D2-5             \\ \hline
11a & \multirow{2}{*}{2} & \multirow{2}{*}{QAD-B}           & \multirow{2}{*}{IN} & RS422/485(+) & D2-10            \\ \cline{1-1} \cline{5-6} 
12a &                    &                                  &                     & RS422/485(-) & D2-9             \\ \hline
13a & \multirow{2}{*}{2} & \multirow{2}{*}{QAD-C/Index}     & \multirow{2}{*}{IN} & RS422/485(+) & D2-4             \\ \cline{1-1} \cline{5-6} 
14a &                    &                                  &                     & RS422/485(-) & D2-3             \\ \hline
15a & \diagbox[width=\dimexpr \textwidth/8+4\tabcolsep\relax, height=0.9cm]{}{}                   & 0 V                              &   \diagbox[width=\dimexpr \textwidth/8+8\tabcolsep\relax, height=0.9cm]{}{}                   &     \diagbox[width=\dimexpr \textwidth/8+8\tabcolsep\relax, height=0.9cm]{}{}          &     \begin{tabular}[c]{@{}c@{}}D1-2, D2-2, D3-2, \\ Differential Cable Shield\end{tabular}             \\ \hline
16a &           \diagbox[width=\dimexpr \textwidth/8+4\tabcolsep\relax, height=0.47cm]{}{}         & 5 V                              &      \diagbox[width=\dimexpr \textwidth/8+8\tabcolsep\relax, height=0.47cm]{}{}               &     \diagbox[width=\dimexpr \textwidth/8+8\tabcolsep\relax, height=0.47cm]{}{}         & D1-7, D2-7, D3-7 \\ \hline
17a &     \diagbox[width=\dimexpr \textwidth/8+4\tabcolsep\relax, height=0.47cm]{}{}                 & Ground                           &    \diagbox[width=\dimexpr \textwidth/8+8\tabcolsep\relax, height=0.47cm]{}{}                 &       \diagbox[width=\dimexpr \textwidth/8+8\tabcolsep\relax, height=0.47cm]{}{}       &      \diagbox[width=\dimexpr \textwidth/8+12\tabcolsep\relax, height=0.47cm]{}{}              \\ \hline
1b  &      \diagbox[width=\dimexpr \textwidth/8+4\tabcolsep\relax, height=0.47cm]{}{}                & 0 V                              &    \diagbox[width=\dimexpr \textwidth/8+8\tabcolsep\relax, height=0.47cm]{}{}                 &       \diagbox[width=\dimexpr \textwidth/8+8\tabcolsep\relax, height=0.47cm]{}{}       &       \diagbox[width=\dimexpr \textwidth/8+12\tabcolsep\relax, height=0.47cm]{}{}             \\ \hline
2b  &         \diagbox[width=\dimexpr \textwidth/8+4\tabcolsep\relax, height=0.47cm]{}{}             & 5 V                              &    \diagbox[width=\dimexpr \textwidth/8+8\tabcolsep\relax, height=0.47cm]{}{}                 &      \diagbox[width=\dimexpr \textwidth/8+8\tabcolsep\relax, height=0.47cm]{}{}        &          \diagbox[width=\dimexpr \textwidth/8+12\tabcolsep\relax, height=0.47cm]{}{}          \\ \hline
3b  & \multirow{2}{*}{3} & \multirow{2}{*}{QAD-A}           & \multirow{2}{*}{IN} & RS422/485(+) & D3-6             \\ \cline{1-1} \cline{5-6} 
4b  &                    &                                  &                     & RS422/485(-) & D3-5             \\ \hline
5b  & \multirow{2}{*}{3} & \multirow{2}{*}{QAD-B}           & \multirow{2}{*}{IN} & RS422/485(+) & D3-10            \\ \cline{1-1} \cline{5-6} 
6b  &                    &                                  &                     & RS422/485(-) & D3-9             \\ \hline
7b  & \multirow{2}{*}{3} & \multirow{2}{*}{QAD-C/Index}     & \multirow{2}{*}{IN} & RS422/485(+) & D3-4             \\ \cline{1-1} \cline{5-6} 
8b  &                    &                                  &                     & RS422/485(-) & D3-3             \\ \hline
9b &
  \multirow{2}{*}{4} &
  \multirow{2}{*}{QAD-A} &
  \multirow{2}{*}{IN} &
  RS422/485(+) &
  HIWIN D1-CN2-16 \\ \cline{1-1} \cline{5-6} 
10b &                    &                                  &                     & RS422/485(-) & HIWIN D1-CN2-17  \\ \hline
11b &
  \multirow{2}{*}{4} &
  \multirow{2}{*}{QAD-B} &
  \multirow{2}{*}{IN} &
  RS422/485(+) &
  HIWIN D1-CN2-18 \\ \cline{1-1} \cline{5-6} 
12b &                    &                                  &                     & RS422/485(-) & HIWIN D1-CN2-19  \\ \hline
13b &
  \multirow{2}{*}{4} &
  \multirow{2}{*}{QAD-C/Index} &
  \multirow{2}{*}{IN} &
  RS422/485(+) &
  HIWIN D1-CN2-20 \\ \cline{1-1} \cline{5-6} 
14b &                    &                                  &                     & RS422/485(-) & HIWIN D1-CN2-21  \\ \hline
15b & \multirow{2}{*}{1} & \multirow{2}{*}{Interrupt Input} & \multirow{2}{*}{IN} & RS422/485(+) &       \diagbox[width=\dimexpr \textwidth/8+12\tabcolsep\relax, height=0.47cm]{}{}             \\ \cline{1-1} \cline{5-6} 
16b &                    &                                  &                     & RS422/485(-) &        \diagbox[width=\dimexpr \textwidth/8+12\tabcolsep\relax, height=0.47cm]{}{}            \\ \hline
17b &     \diagbox[width=\dimexpr \textwidth/8+4\tabcolsep\relax, height=0.47cm]{}{}                 & Ground                           &     \diagbox[width=\dimexpr \textwidth/8+8\tabcolsep\relax, height=0.47cm]{}{}                &       \diagbox[width=\dimexpr \textwidth/8+8\tabcolsep\relax, height=0.47cm]{}{}       &        \diagbox[width=\dimexpr \textwidth/8+12\tabcolsep\relax, height=0.47cm]{}{}            \\ \hline
\end{tabular}
}
\end{table}

\subsection{Components Choice of Linear Motor's Electrical System and Wiring Details}
\label{Appendix:ElectricalSystem_LinearMotor}

This section will detail the functionalities and specifications of the electrical parts used to power the linear motor. A complete wiring diagram of the entire system is also provided.

The first component introduced here is the emergency stop button module. During the multi-link pendulum experiments, the linear motor may perform undesired motion due to coding error or controller failure. It is also possible that unwanted electrical shortage happens due to the improper use of the system. In those undesired situations, the user must press the emergency stop button to cut all the power supplied to the linear motor, avoiding damage to the setup and protecting the operator around the equipment. It is also necessary to have an On/Off button that controls the system's power supply. The POWERTEC 71354 magnetic switch~($75$) is selected as our emergency stop button to meet those requirements. This emergency stop button can be used in a single-phase 120V power line, matching how we supply the linear motor power. By checking the specification of the linear motor, we find out the peak current of the linear motor is around $12.7 A_{rms}$, which is lower than the rated current of the On/Off switch~($16 A_{rms}$) and emergency switch~($18 A_{rms}$) installed inside the emergency stop button module. Thus, this emergency stop module is selected to control the power supply of the linear motor. The emergency stop button module should be installed somewhere that is easy to reach for the operator. This location may change based on the operator's preferred standing position. In our application, we decided to install the emergency button at the side of the system frame. Please check its user manual for details on the position holes of the module.

The next vital component is the circuit breaker (MCCB, $76$). The circuit breaker has the following functionalities: 1) Circuit breaker can cut off the power supply to the linear motor when there's a shortage in the electrical system. 2) It can stop the power supply to the linear motor when it is overloaded, which avoids damage to the linear motor. 3) It can simply serve as an On/Off switch that controls the power supply to the linear motor. The first and third functions of the circuit breaker are easy to understand, and we will illustrate the second functionality more. The overloaded situation might happen when the control algorithm requires the linear motor to perform unrealistic acceleration or de-acceleration. For example, during the feedforward swing-up control of the double or triple pendulum, a feedback force is usually needed to compensate for noise and model uncertainty. It has been shown~\cite{graichen2007swing,Gluck2013} that sometimes the LQR gain calculated to stabilize the feedforward trajectory is near singular. This indicates a considerable control input needs to be applied to the system to compensate for a slight deviation. Suppose the feedback control force is not thresholded. In that case, the controller may require the linear motor to apply unrealistic force to the pendulum arm, causing a massive amount of currents supplied to the linear motor resulting in overload. Another case where overload might happen, although unlikely in our case, is that the linear motor gets stuck mechanically. In this situation, the motor drive will try to apply more current to the linear motor to make it move, eventually causing over current. Thus, having a circuit breaker is crucial in controlling the linear motor for safety reasons. The Eaton miniature circuit breaker FAZ-B3-2-NA~($76$) is used in our application. According to the user's manual, its rated current is $3 A$, with the maximum allowed AC voltage being $277 V$. The tripping current is 3 to 5 times the rated current, with stripping time around 0.005 to 2 seconds. The resulting tripping current ranges from $9$ to $15 A$ and covers the peak current of the linear motor, which is $12.7 A$. Thus, using this circuit breaker can help us avoid linear motor damage when the current supplied to the linear motor is larger than the allowed peak current. Moreover, the selected circuit breaker uses a trip-free design and can not be defeated by manually holding the handles at the "On" position. Another plus is that the user can easily install it on the DIN rail. For more specifications on the circuit breaker, please check its user manual.

\begin{figure}[t]
    \centering
    \includegraphics[width=\textwidth]{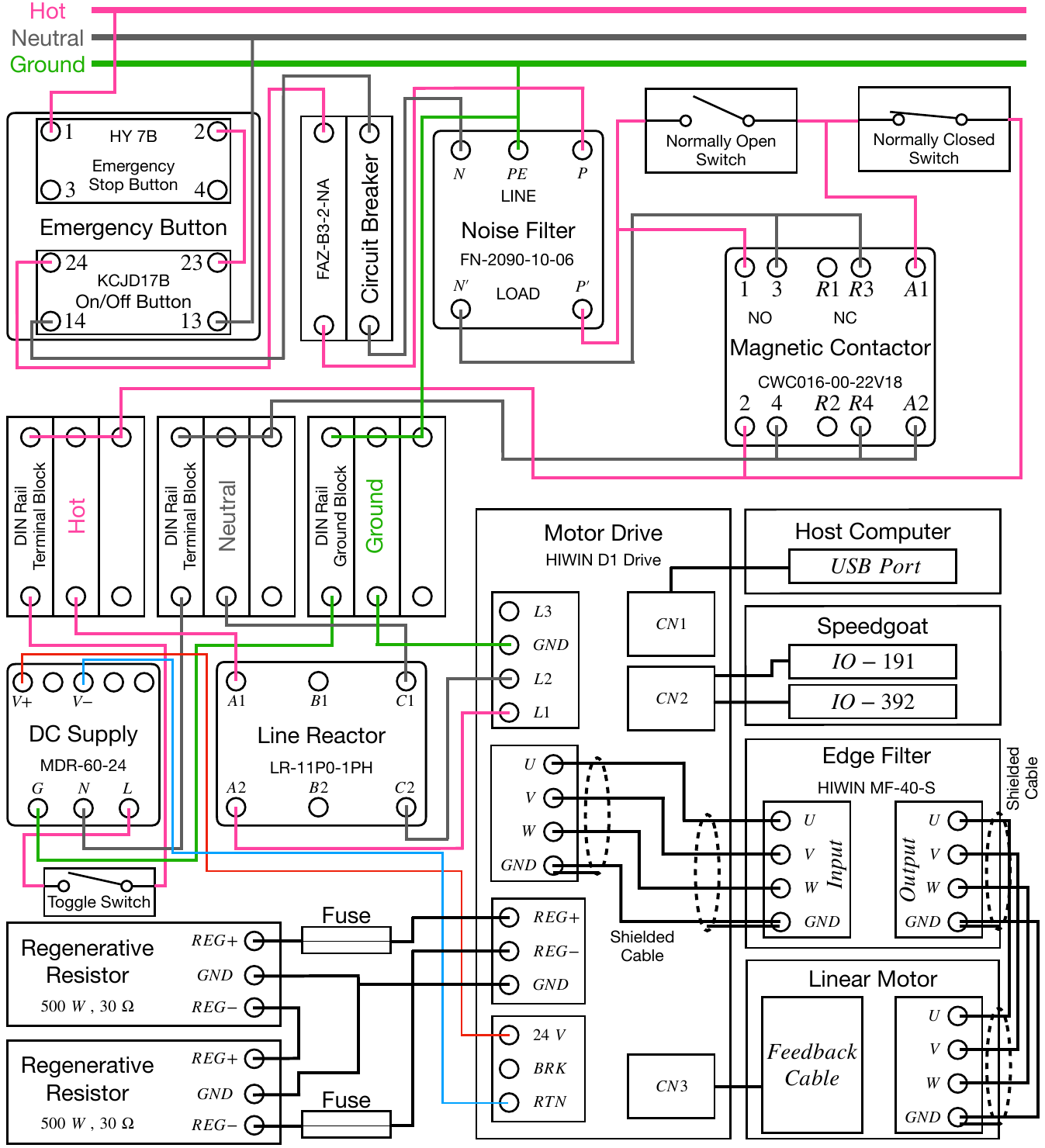}
    \caption{This figure illustrates the overall electrical components needed to provide power supply to the linear motor. An electrical box is used to put all the components together.}
    \label{fig:MotorWiring}
\end{figure}

The magnetic contactor~($77$) is needed to start and turn off the linear motor. The magnetic contactor is a middle path between the power source and the linear motor. It helps avoid the direct connection of the power source to the linear motor, which helps prevent the sudden change of state of the linear motor. Combined with a circuit breaker, they can be thought of as a motor starter. The WEG Electric CWC016-00-22V18~($77$) is used as a magnetic contactor in our design, allowing easy installation on the DIN rail. This contactor has two normally open and two normally closed contacts rated $16 A$ current input under $60 Hz$ $120 VAC$. This specification fits our needs since the power output in the US is $120 VAC$ at $60 Hz$, and the maximum current needed by the linear motor is below the rated current of the contactor. By checking the user manual of this contactor, a "locking" mechanism of the magnetic contactor is achieved by using a momentary switch. The LCLICTOP AC 660V Red and Green Momentary Switch~($86$) is used to accomplish this. It can operate up to 660 VAC, and the current rating is 10 A. Fig.~\ref{fig:MotorWiring} shows the overall wiring of the magnetic contactor and momentary switch. 

It is recommended to install a line reactor on the motor drive's input side to protect it from voltage spikes, surges, and transients. The LR-11P0-1PH~($79$) is used to achieve this. This line reactor has a rated current of $16 A$ at $120 VAC$. Thus, it satisfies our usage scenario in a single-phase case.

The linear motor drive~($80$) controls the movement of the linear motor by providing a high-frequency signal with sharp pulse edges. Those sharp edges of the signal can introduce unwanted ground current to the entire electrical system. The ground current can even cause interference between the different electrical components. This kind of interference is not wanted, especially when an accurate reading of the pendulum arm rotational angle is needed. The interference between the linear motor and encoder can cause additional unwanted pulse width signals to appear in the encoder's A, B, and Index channels, which might result in the wrong sensor reading. This is why the differential driver is used to transfer the single-ended encoder signal into the differential encoder signal to reduce the effect of electrical noise. Except for electrical interference, EMI can also cause electrical overstress (EOS) to sensitive components~\cite{kraz2016mitigating}. 

A multi-stage AC/DC EMI filter SCHAFFNER FN-2090-10-06~($78$) is used according to the D1 drive's user manual to reduce the ground current in the electrical system and protect all other appliances that share the same ground, such as the host and target computers and monitors. The HIWIN MF-40-S edge filter~($81$) is used to eliminate the noise effect further. The edge filter helps reduce the EMI noise on the output end of the motor drive. Two D1-EMC1 EMI cores~($87$) can also suppress the EMI noise. One EMI core is installed on the linear motor's power input cable near the motor drive end, while the other is installed on the motor cable near the linear motor's stage end. Besides using the approaches mentioned above to reduce the EMI interference, another easy way to reduce EMI interference is proper cable management. For example, make the power cable of the linear motor stay far away from the sensor cable.

\begin{table}[th]
\centering
\caption{This table shows the pin out of IO-191 module and its detailed connection with HIWIN D1 drive's control signal channel~(CN2).}
\label{table:IO-191-PinOut}
\resizebox{\textwidth}{!}{%
\begin{tabular}{|c|c|c|c|c|c|c|c|}
\hline
Pin & Functionality    & Direction & Connects To     & Pin & Functionality  & Direction & Connects To     \\ \hline
1a  & Analog Input 01  & In        & \diagbox[width=\dimexpr \textwidth/4+4\tabcolsep\relax, height=0.44cm]{}{}              & 1b  & Digital I/O 01 & Out       & HIWIN D1-CN2-3  \\ \hline
2a  & Analog Input 02  & In        & \diagbox[width=\dimexpr \textwidth/4+4\tabcolsep\relax, height=0.44cm]{}{}              & 2b  & Digital I/O 02 & Out       & HIWIN D1-CN2-4  \\ \hline
3a  & Analog Input 03  & In        & \diagbox[width=\dimexpr \textwidth/4+4\tabcolsep\relax, height=0.44cm]{}{}              & 3b  & Digital I/O 03 & Out       & HIWIN D1-CN2-5  \\ \hline
4a  & Analog Input 04  & In        & \diagbox[width=\dimexpr \textwidth/4+4\tabcolsep\relax, height=0.44cm]{}{}              & 4b  & Digital I/O 04 & Out       & HIWIN D1-CN2-6  \\ \hline
5a  & Analog Input 05  & In        & \diagbox[width=\dimexpr \textwidth/4+4\tabcolsep\relax, height=0.44cm]{}{}              & 5b  & Digital I/O 05 & Out       & HIWIN D1-CN2-7  \\ \hline
6a  & Analog Input 06  & In        & \diagbox[width=\dimexpr \textwidth/4+4\tabcolsep\relax, height=0.44cm]{}{}              & 6b  & Digital I/O 06 & Out       & HIWIN D1-CN2-12 \\ \hline
7a  & Analog Input 07  & In        & \diagbox[width=\dimexpr \textwidth/4+4\tabcolsep\relax, height=0.44cm]{}{}              & 7b  & Digital I/O 07 & Out       & \diagbox[width=\dimexpr \textwidth/4+4\tabcolsep\relax, height=0.44cm]{}{}              \\ \hline
8a  & Analog Input 08  & In        & \diagbox[width=\dimexpr \textwidth/4+4\tabcolsep\relax, height=0.44cm]{}{}              & 8b  & Digital I/O 08 & Out       & \diagbox[width=\dimexpr \textwidth/4+4\tabcolsep\relax, height=0.44cm]{}{}              \\ \hline
9a  & Analog Output 01 & Out       & HIWIN D1-CN2-24 & 9b  & Digital I/O 09 & In        & HIWIN D1-CN2-13 \\ \hline
10a & Analog Output 02 & Out       & \diagbox[width=\dimexpr \textwidth/4+4\tabcolsep\relax, height=0.44cm]{}{}              & 10b & Digital I/O 10 & In        & HIWIN D1-CN2-14 \\ \hline
11a & Analog Output 03 & Out       & \diagbox[width=\dimexpr \textwidth/4+4\tabcolsep\relax, height=0.44cm]{}{}              & 11b & Digital I/O 11 & In        & HIWIN D1-CN2-15 \\ \hline
12a & Analog Output 04 & Out       & \diagbox[width=\dimexpr \textwidth/4+4\tabcolsep\relax, height=0.44cm]{}{}              & 12b & Digital I/O 12 & In        & \diagbox[width=\dimexpr \textwidth/4+4\tabcolsep\relax, height=0.44cm]{}{}              \\ \hline
13a & Analog Ground    & In        & HIWIN D1-CN2-25 & 13b & Digital I/O 13 & In        & NC              \\ \hline
14a & Analog Ground    & In        & \diagbox[width=\dimexpr \textwidth/4+4\tabcolsep\relax, height=0.44cm]{}{}              & 14b & Digital I/O 14 & In        & \diagbox[width=\dimexpr \textwidth/4+4\tabcolsep\relax, height=0.44cm]{}{}              \\ \hline
15a & 0 V              & NONE       & \diagbox[width=\dimexpr \textwidth/4+4\tabcolsep\relax, height=0.44cm]{}{}              & 15b & Digital I/O 15 & In        & \diagbox[width=\dimexpr \textwidth/4+4\tabcolsep\relax, height=0.44cm]{}{}              \\ \hline
16a & 5 V              & Out       & \diagbox[width=\dimexpr \textwidth/4+4\tabcolsep\relax, height=0.44cm]{}{}              & 16b & Digital I/O 16 & In        & \diagbox[width=\dimexpr \textwidth/4+4\tabcolsep\relax, height=0.44cm]{}{}              \\ \hline
17a & Analog Ground & In & \diagbox[width=\dimexpr \textwidth/4+4\tabcolsep\relax, height=0.9cm]{}{} & 17b & Digital Ground & In & \begin{tabular}[c]{@{}c@{}}HIWIN D1-CN2-1\\ HIWIN D1-CN2-2\end{tabular} \\ \hline
\end{tabular}%
}
\end{table}

\begin{table}[th]
\centering
\caption{This table shows the pinout of the limit switch cable and its connection. The HIWIN D1 drive's CN2 channel is used to provide DC current to the limit switch. The limit switch's signal is then sent to the motor drive for the out of range protection.}
\label{table:LimitSwitchPinout}
\begin{tabular}{|c|c|c|c|}
\hline
D-Sub 9-Pin (Female) & Color & Function & Connects To \\ \hline
1 & Brown & V+ & HIWIN D1-CN2-22 \\ \hline
2 & Green & L-1 & NC \\ \hline
3 & Yellow & Out-1 & HIWIN D1-CN2-11 \\ \hline
4 & Violet & L2 & NC \\ \hline
5 & Gray & Out-2 & HIWIN D1-CN2-10 \\ \hline
9 & Blue & V- & HIWIN D1-CN2-23 \\ \hline
\end{tabular}
\end{table}

A regenerative resistor is usually needed to avoid the regenerative energy of the linear motor becoming too large and causing overvoltage to the motor drive. The motion profile of the linear motor and the weight of moving mass need to be known to pick a regenerative resistor. Then the designer can use the calculation shown in the HIWIN D1 drive's user manual~(see Section 8.2 of the user manual~\cite{HIWIN_D1}) to guide the selection of the regenerative resistor. The moving mass of the linear motor is around $5 kg$, and the weight of the pendulum arm and cart plate is around $0.5 kg$ in our design, resulting in a total moving mass of $M_t=5.5 kg$. Then, assume the linear motor is moving at $V_1=4m/s$ and needs to deaccelerate to $V_2=0m/s$ with $a=40m/s^2$. Using the above information and following the calculation shown in the user manual, we find there's no need to use a regenerative resistor since the energy return to the motor drive is smaller than the D1 drive amplifier's absorption capacity. However, we still decided to pick two 500W 30$\Omega$ regenerative resistors~($82$) and connect them in serial in case the returned energy is larger than the energy storage capacity.

The proper operation of the linear drive requires the supply of $24 V$ DC voltage. Thus, the MDR-60-24 AC to DC industrial DIN rail power supply~($83$) is used to output $24 V$ DC at $2.5 A$. A toggle switch~($85$) is used to control the On/Off state of the DC power supply to the linear motor drive. The Hammond CSKO242410 enclosure~($84$) is used to install electrical components. A Hammond CSFC2424 flush cover~($88$) is cut down to be used as an electrical panel that mounts some of the components mentioned above. Two DIN rails~($89$) are installed inside the electrical enclosure to facilitate the installation of different electrical components. This completes the selection process of all the electrical components related to the linear motor. Fig.~\ref{fig:MotorWiring} shows their wiring details.

As Fig.~\ref{fig:MotorWiring} shows, the emergency stop button is first connected to the single-phase power supply. Two 14 AWG power cord~($90$) is used as the emergency stop button's power input/output cable. Next, the output of the emergency stop button is connected with the circuit breaker, which serves as over-current protection. Then, the noise filter is connected to the circuit breaker. The connection wire used for the hot and neutral line is 12 AWG wire~($91$), while the connection wire for the ground is 10 AWG wire~($92$). The output of the noise filter connects to the magnetic contactor. The magnetic contactor is wired to self-lock when the normally open momentary switch is pressed. The output of the magnetic contactor is wired to the DIN rail terminal block~($93,94$) to facilitate cable management better. Then, the DC supply and line reactor input is connected to the DIN rail terminal block. The output of the DC supply is $24 V$ DC, and it is used to provide power to the motor drive. A toggle switch is used to make the supply of DC voltage controllable. Then, the output of the DC supply is connected to the Wago 721-103/026-000 connector~($95$). The output of the line reactor is connected to the AC power port of the motor drive using a single-phase connection formulation. The connector used is Wago 721-204/026-000~($96$). The U, V, W, and GND channel is connected to the corresponding channel on the edge filter using a shielded cable to reduce the EMI noise. The connector used on the motor drive is Wago 721-104/026-000~($97$). The connector used in the edge filter's input port is Phoenix Contact 1718517~($98$). The output of the edge filter then connects to the linear motor's power cable, which is a shielded cable as well. The output connector of the edge filter is Phoenix Contact 1718504~($99$). Two EMI cores~($87$) are installed to reduce the EMI noise further. One is near the motor connection end on the motor drive and the other near the motor stage end of the linear motor. Finally, two regenerative resistors are connected to the regenerative resistor port of the motor drive in serial. The connector used for the regenerative resistor port of the motor drive is Wago 723-603~($100$). The reader can choose to install two fuses as well for over-current protection. This completes the power supply to the linear motor. A D1-DNT07A USB232 to RJ11 adapter cable~($101$) is used to set up the motor drive parameters. The linear motor's feedback cable is connected to the CN3 of the motor drive to let the motor drive read the encoder and hall sensor signal from the linear motor. The encoder signal of the linear motor, along with other I/O signals of the motor drive, is connected to the Speedgoat's terminal block. This is achieved by making a custom cable using hook-up wires. This custom cable consists of 24 AWG hook up wires~($102$), cable shielding sleeve~($103$), cable sleeve~($104$), 3M 10126-3000 connector~($105$) and its cover 3M 10326-52F0-008~($106$). 
Table.~\ref{table:IO-392-PinOut} and  Table.~\ref{table:IO-191-PinOut} show the connections between the CN2 of the motor drive and the Speedgoat real-time system. Section 3 of the motor drive's user manual contains the connection details of the motor drive and its pinout of control signals~(CN2). The interested reader should reference the user manual if future system modifications are needed. The final piece to mention is the limit switches we used in the linear motor for out-of-range protection. A HIWIN 3-meter limit switch extension cable is used~($107$) to connect the limit switch and motor drive. Table.~\ref{table:LimitSwitchPinout} shows the pinout of the limit switch cable and its corresponding connection to the motor drive. Zip tie and wire clips are used to make proper cable management by securing the position of all the cables/wires we used.

\section{Software Setup Details}
\subsection{Software Setup of Motor Drive Details}
\label{Appendix:Lightning}

\begin{figure}[t]
    \centering
    \includegraphics[width=1\textwidth]{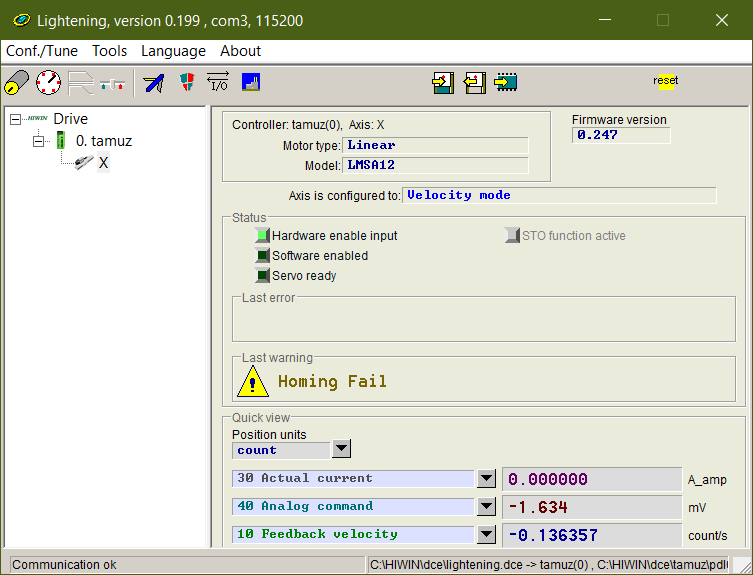}
    \caption{This figure illustrates the overall look of the Lighting software. }
    \label{fig:Lightning1}
\end{figure}

The HIWIN D1 drive's corresponding setup software is Lightning, and the reader can find a detailed introduction in sections 4 and 5 of the HIWIN D1 drive's user manual~\cite{HIWIN_D1}. The version of the software we are using is V-0.1999, while the latest version is V-0.20 at the time of writing this paper. First, we need to \href{https://www.mathworks.com/help/slrealtime/}{download} the Lightning software to the host computer to set it up. Next, the CN1 port of the D1 drive is connected to the USB port of the host computer using a communication cable~($101$). Once finished, the D1 drive is powered up, and the Lightning software is executed. Then, in the software's menu, select "Tools->Communication setup," then configure the "BPS" drop-down list to $115200$ and select the USB port that is used to connect the communication cable in the drop-down list. Finally, hit the "Apply" button, and the host computer and motor drive connection are complete. The above steps are needed to set up the necessary parameters of the motor drive. Fig.~\ref{fig:Lightning1} shows the overall look of the Lighting software after connecting to the host computer.

\begin{figure}[t]
    \centering
    \includegraphics[width=1\textwidth]{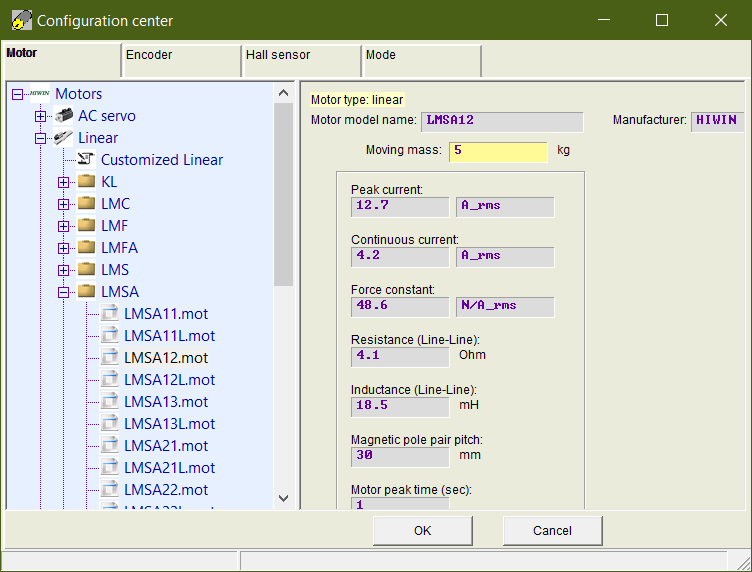}
    \caption{This figure illustrates the configuration of the motor parameters.}
    \label{fig:Lightning2}
\end{figure}

\begin{figure}[t]
    \centering
    \includegraphics[width=1\textwidth]{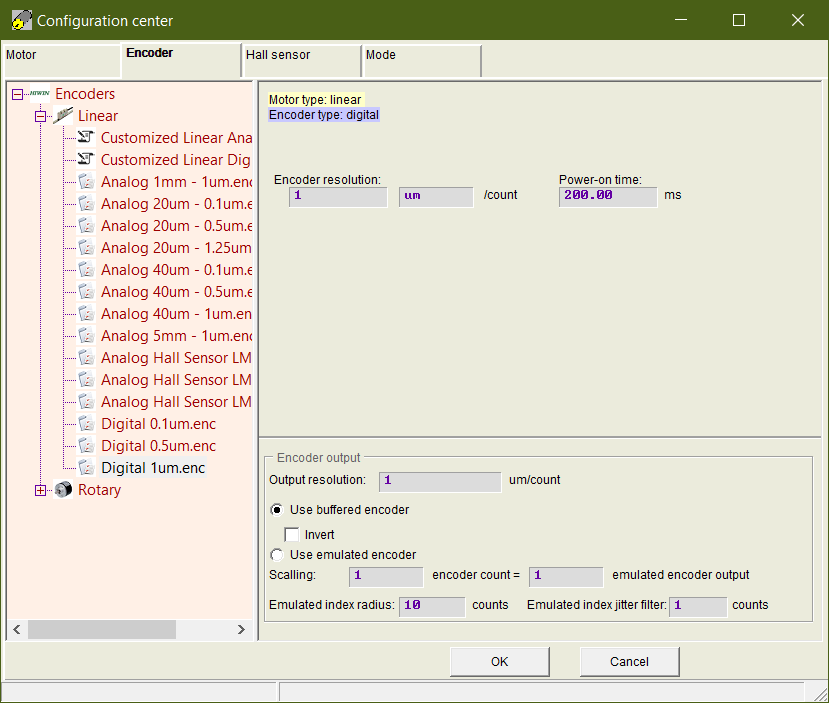}
    \caption{This figure illustrates the configuration of the linear motor's encoder.}
    \label{fig:Lightning3}
\end{figure}

To select the correct linear motor type, motor sensor, and operating mode, we need to select "Conf./Tune->Configuration Center," which will open a new window that allows the configuration of the motor parameters and sensors. We need to first click on the "Motor" tab to select the corresponding linear motor we are using. The LMSA12 linear motor is selected in the menu box on the left. This will bring out the pre-stored motor parameters. If the linear motor from a different brand is used, the user must enter all those parameters manually. In the "Moving mass" blank, one can enter the mass of the linear motor stage. In our case, it is $5kg$. Fig.~\ref{fig:Lightning2} illustrates this process. After selecting the correct linear motor, we need to click on the encoder tab to set up the encoder reading options, as Fig.~\ref{fig:Lightning3} shows.

\begin{figure}[t]
    \centering
    \includegraphics[width=1\textwidth]{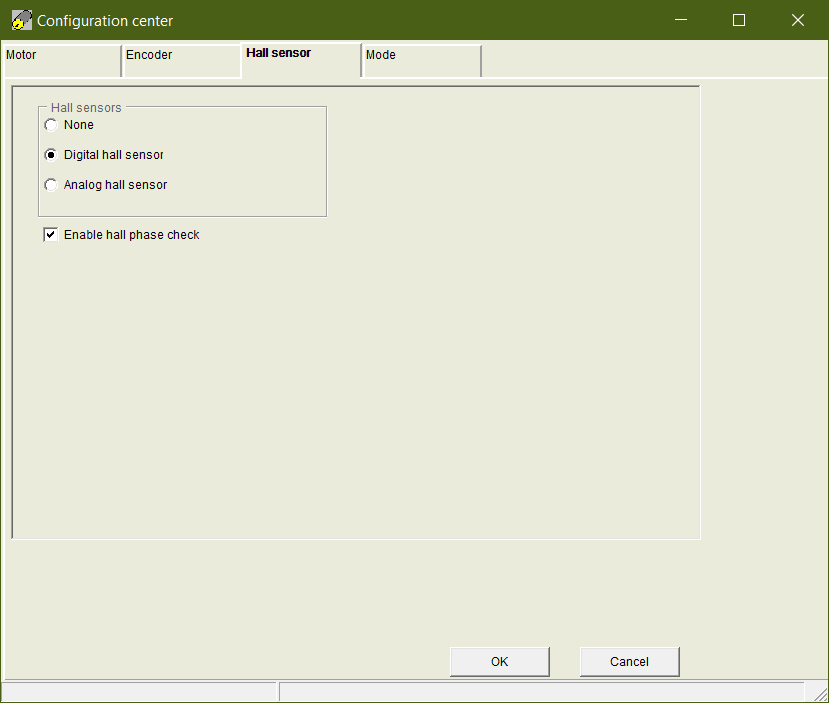}
    \caption{This figure illustrates the configuration of the linear motor's hall sensor.}
    \label{fig:Lightning4}
\end{figure}

In Fig.~\ref{fig:Lightning3}, the "Digital $1\mu m$.enc" is selected. The "Encoder output" panel in Fig.~\ref{fig:Lightning3} allows us to set up the encoder resolution output of the motor drive. Users can scale the encoder output of the motor drive to allow the successful reading of the encoder signal if another custom real-time system is used that can not read an encoder with fine resolution. This completes the setup of the linear motor encoder. Fig.~\ref{fig:Lightning4} shows the configuration of the hall sensor. This is achieved by clicking the "Hall Sensor" panel, and select the type of the Hall sensor as "Digital Hall sensor," and clicking on the "Enable hall phase check" option. The final parameter that needs to be configured in "Configuration Center" is the operating mode of the linear motor.

There are multiple operating modes of the linear motor. The first one is the position mode. The position mode takes the user desired position as a reference and controls the linear motor to stay at the reference position. The second mode is velocity mode. In velocity mode, the velocity of the linear motor is controlled. The user usually provides an analog signal to the motor drive, and scaling is used to transfer this analog signal to the desired velocity of the linear motor. Next, the motor drive will control the linear motor to achieve the desired speed. The third mode is force/torque mode. In the force/torque mode, the user again provides an analog voltage signal to the motor drive. Then, this voltage signal is scaled to the corresponding force/torque generated by the motor. The force/torque mode directly controls the acceleration of the motor. The final mode available is the stand-alone mode which means the linear motor can only be controlled using the Lightning software instead of the user-provided signal. In this paper, we follow the work of~\cite{graichen2007swing,Gluck2013} and use the velocity mode as our control method of the pendulum cart. As Appendix.~\ref{Appendix:ParameterEstimation} shows, this results in a nicer equation of motion for the pendulum on the cart system. 

\begin{figure}[t]
    \centering
    \includegraphics[width=1\textwidth]{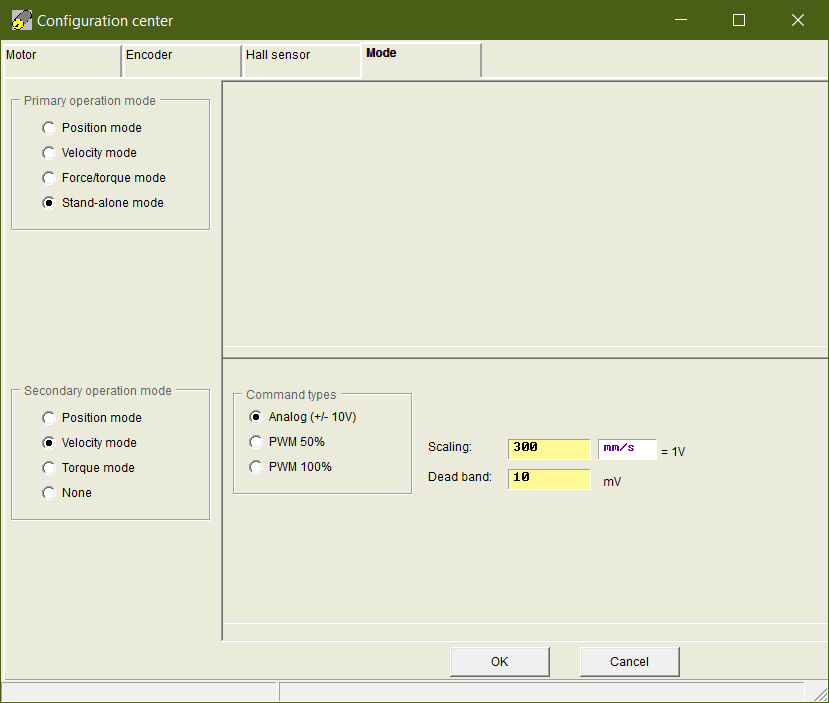}
    \caption{This figure illustrates the selection of the linear motor's operating mode.}
    \label{fig:Lightning5}
\end{figure}

The HIWIN D1 drive supports two operating modes to be selected. The first one is set to stand-alone mode as our default operating mode. This helps avoid the unwanted movement of the linear motor when the analog signal is accidentally drifting. The secondary mode is chosen to be velocity mode, which allows us to control the velocity of the linear motor given analog input. Fig.~\ref{fig:Lightning5} shows the configuration of the linear motor's operating mode. During the pendulum experiments, we usually desired a speed range of $\pm 3m/s$ for the linear motor. Given the output range of the analog signal is $\pm 10V$, the scaling we used is $0.3m/s=1V$. Moreover, we set a dead band of $10mV$. This is necessary since the analog signal is always going to be noisy. Thus, it is desired to put a dead band and filter out low magnitude noise to prevent the unwanted motion of the linear motor. The dead band we used filters out any velocity command lower than $3mm/s$. 

The dead band should not be too large, or it will affect the performance of the pendulum stabilization task, especially during the stabilization of the multi-link pendulum, where the fine movement of the pendulum cart is needed. The maximum value of the dead band that the user can select is determined by the parameters of the pendulum arm, sampling rate, scaling factor, etc. 
\begin{figure}[t]
    \vspace{-.1in}
    \centering
    \includegraphics[width=0.7\textwidth]{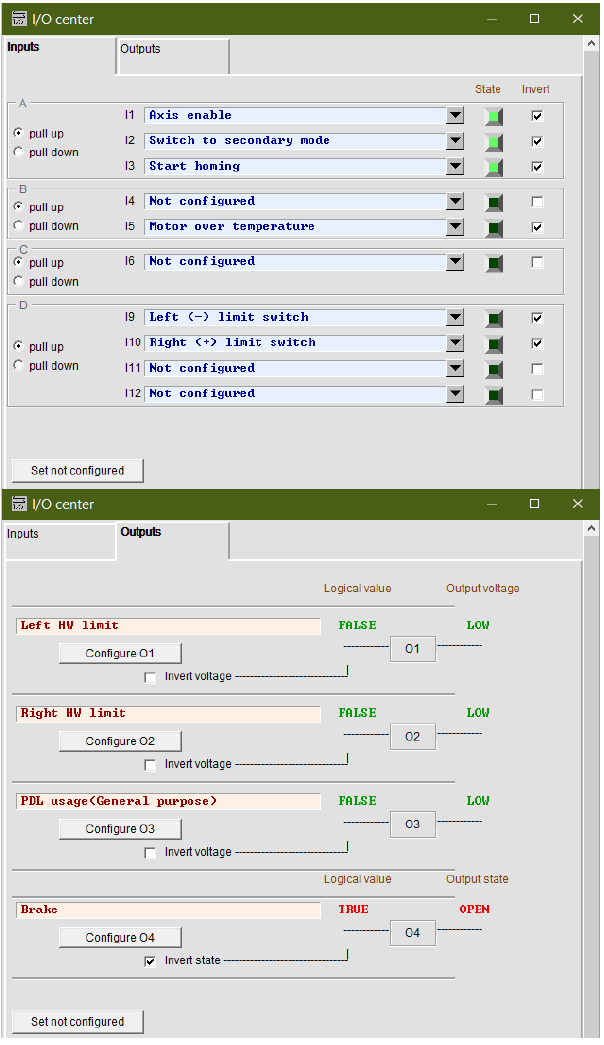}
    \vspace{-.1in}
    \caption{This figure illustrates the configuration of the HIWIN D1 drive's I/O functionalities.}
    \label{fig:Lightning6}
    \vspace{-.1in}
\end{figure}
No analytical analysis is given here for adequately selecting the dead band of the analog signal. Still, a general rule to keep in mind is that the dead band should be as small as possible while eliminating most of the noise in the analog signal. The above process summarizes the configuration of the linear motor using Lightning software.

A digital input signal is needed to switch between the primary and secondary operation mode, with its high and low states corresponding to different operating modes. To set up the I/O functionality of the motor drive, the user needs to click the "Conf./Tune->I/O Center" to open the I/O setting window. Fig.~\ref{fig:Lightning6} shows the overall look of the configured I/O ports. In the "Inputs" tab, the I1 port of the CN2 channel~(HIWIN D1-CN2-3) is configured as the hardware enable port. When the input of this channel is high, the linear motor will be hardware-enabled, and the user can achieve control of the linear motor. The I2 port of the drive~(HIWIN D1-CN2-4) is configured to decide the operating mode of the linear motor. When a high signal is detected, the drive will enter the secondary mode, which is velocity mode in our case. The I3 port~(HIWIN D1-CN2-5) is configured to control the homing function of the linear motor. The homing process will first drive the linear motor to the left and right limit switches. Then it will use the location information of the limit switch to find the center location between them and drive the linear motor to it. The homing function can only be activated when the motor drive is in stand-alone mode with high input received in the I3 port. Next, the I9 and I10 port~(HIWIN D1-CN2-10 and 11) are configured as the limit switch signal input channel. Table.~\ref{table:LimitSwitchPinout} shows the detailed connection between the limit switch and CN2 channel. 
\begin{figure}[t]
    \centering
    \includegraphics[width=\textwidth]{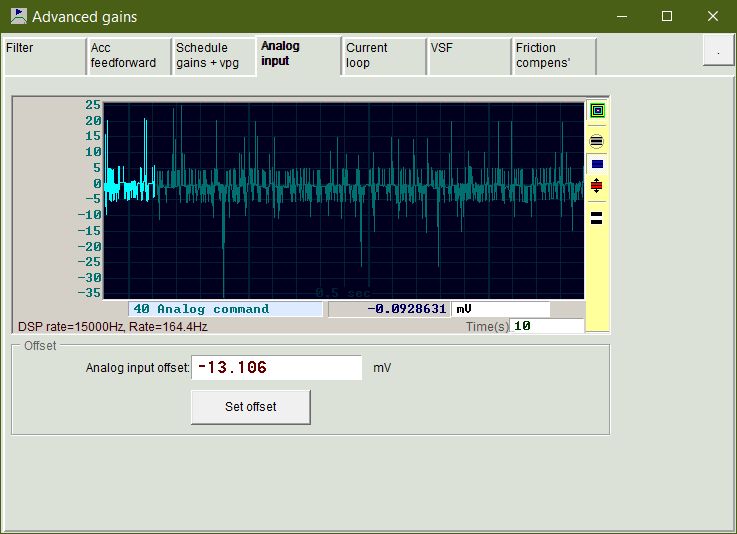}
    \caption{This figure illustrates the bias correction of the analog signal.}
    \label{fig:Lightning7}
\end{figure}
The above configuration helps the motor drive determine whether the linear motor reaches the limit switch. If so, the linear motor's movement beyond the limit switch is restricted. However, it will not deactivate the linear motor and stop its power supply. When an opposite-moving signal is supplied to the linear motor, it can still move in the opposite direction. This behavior is not desired, though. Once the linear motor hits the limit switches, a program should cut off all linear motor's power supply. The limit switch signal is supplied to the Speedgoat system to achieve this functionality. The output ports of the CN2 channel, HIWIN D1-CN2-13 and 14, are configured to be the limit switch output signal. Then, as Table.~\ref{table:IO-191-PinOut} shows, the output of the CN2 channel is connected to the Speedgoat system. Next, the limit switch signal is read in the Simulink model, and once it is high, meaning the linear motor hits the limit switch, the hardware enable is deactivated. This will cut off all the power supply to the linear motor. Moreover, it will also stop the Simulink model from executing, thus stopping the experiment. The above process completes the I/O configuration of the linear motor drive.

Due to the EMI noise in the electrical system, the analog signal that controls the linear motor will contain some level of DC bias. In velocity mode, this DC bias constantly moves the pendulum cart to one side of the linear rail and thus should be avoided. By selecting "Conf./Tune->Performance Center" to enter the performance center, the "Advanced Gain" button can be pressed, which will open up a window with multiple tabs. Selecting the "Analog Input" tab will result in Fig.~\ref{fig:Lightning7}, where the users can press the "Set offset" button to minimize the effect of the analog signal's bias.

\begin{figure}[t]
    \centering
    \includegraphics[width=\textwidth]{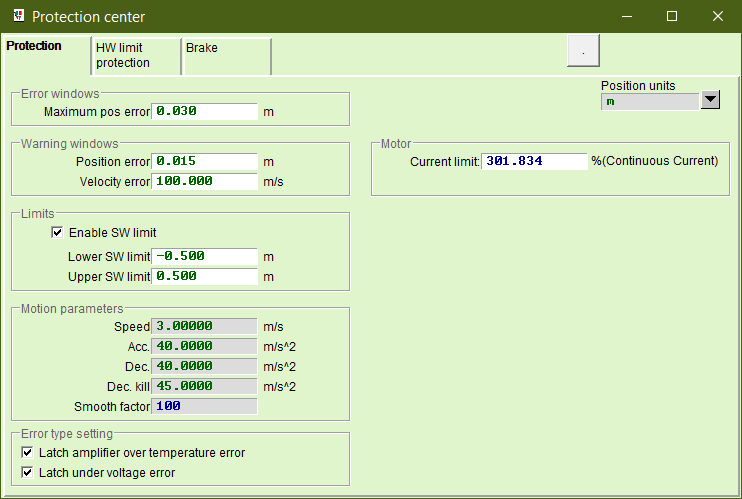}
    \caption{This figure illustrates the software safety configuration by using Lightning software.}
    \label{fig:Lightning8}
\end{figure}

The safety mechanism is the final configuration that needs to be made in the Lightning software. To do so, we need to select "Conf./Tune->Protection Center" which will result in Fig.~\ref{fig:Lightning8}, where one can set up the software position, velocity, and acceleration limit. Set up a narrow range for the limits and gradually increase it if the desired position, speed, or acceleration is out of range. This safety measure prevents unwanted high-speed movement of the linear motor during the development phase of the controller.

The configurations of Lightning software in this section allow quick setup of the linear motor. For more details on using the Lightning software and its other functionalities, please refer to sections 4 and 5 of HIWIN D1 drive's user manual~\cite{HIWIN_D1}.

\subsection{Software Set Up of the Simulink Model Details}
\label{Appendix:Simulink}

\begin{figure}[t]
    \centering
    \includegraphics[width=\textwidth]{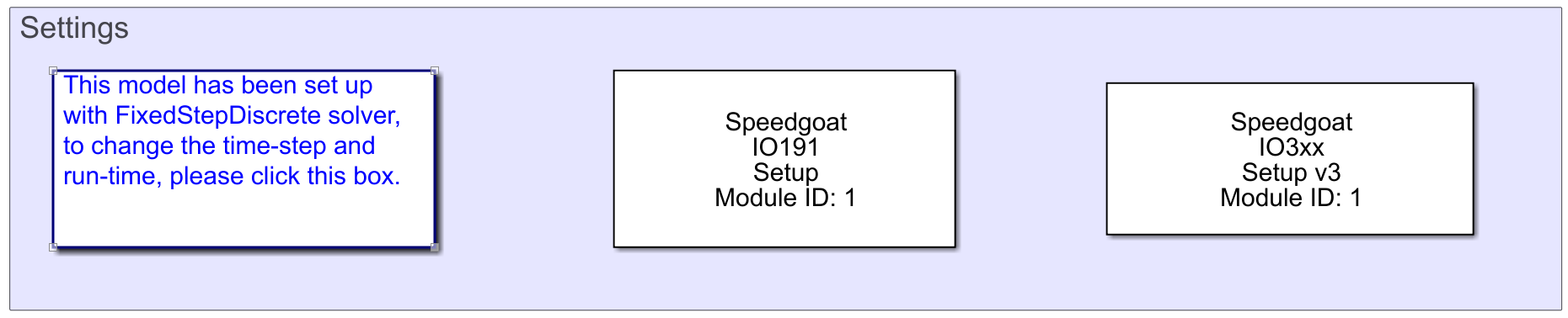}
    \caption{This figure illustrates the two block sets used in the Simulink model that configures the Speedgoat IO-191 and Speedgoat IO-392 modules.}
    \label{fig:SRT_Config1}
\end{figure}

The Simulink Real-Time is used to develop the controller used in the real-time pendulum experiments. The Simulink Real-Time and Speedgoat must be appropriately configured to read the sensors and output control signals. The first step of the configuration physically connects the Speeadgoat machine with the host computer using an Ethernet cable. This connection allows the user-programmed Simulink controller to be downloaded to the Speedgoat device and run in real-time. Moreover, this connection also enables the Speedgoat to send back the sensor signal at the run time and display it on the host computer using the "Data Inspector" function of the Simulink. Matlab and Speedgoat already have numerous tutorials on setting up the connection between the Speedgoat and the host computer. Thus we will skip them in this paper. The interested user should reference the \href{https://www.mathworks.com/help/slrealtime/getting-started-with-xpc-target-1.html?s_tid=CRUX_topnav}{documentation of the Speedgoat} for configuration details.

\begin{figure}[t]
    \centering
    \includegraphics[width=\textwidth]{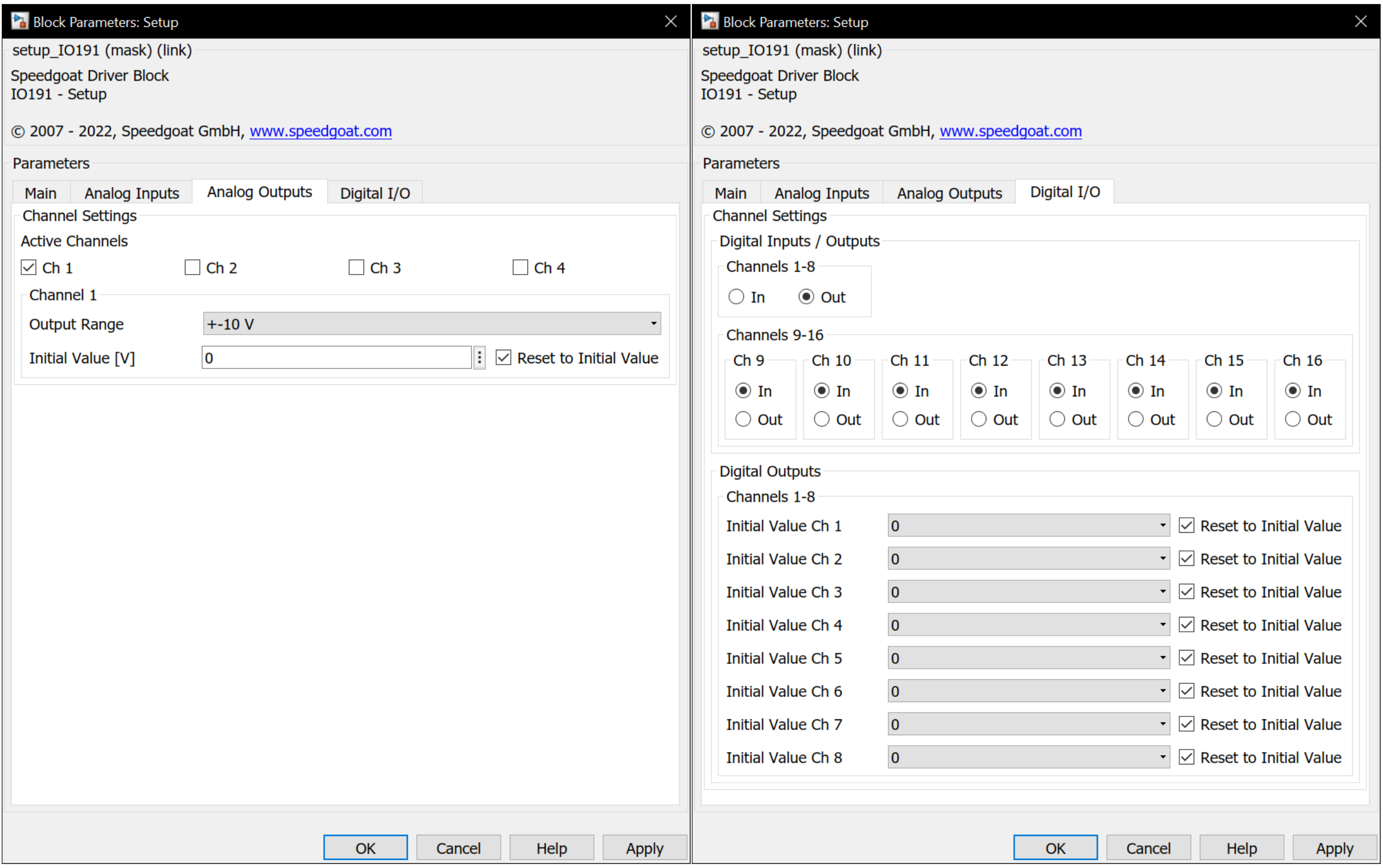}
    \caption{This figure illustrates the configuration of the IO-191 module's Simulink block. One of the analog signal outputs is selected to control the velocity of the pendulum cart in velocity mode. Meanwhile, eight ports of the digital I/O block are chosen as input ports, and the rest are set to output ports.}
    \label{fig:SRT_Config2}
\end{figure}

\begin{figure}[t]
    \centering
    \includegraphics[width=0.5\textwidth]{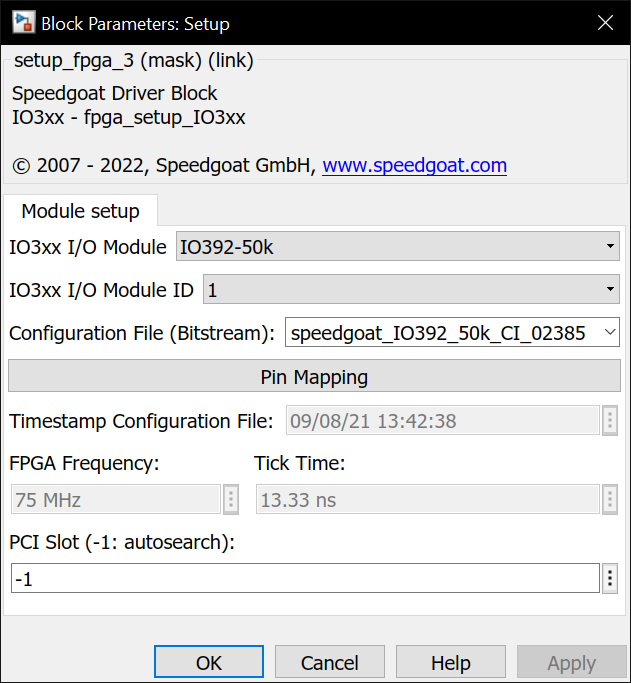}
    \caption{This figure illustrates the configuration of the IO-392 module's Simulink block. The bitstream file provided by Speedgoat is needed to set up the IO-392 block to read encoder data properly.}
    \label{fig:SRT_Config3}
\end{figure}

Two block sets must be set up and used in the Simulink file to configure the I/O module IO-191 and FPGA module IO-392 for I/O functionalities and encoder reading. They are the "Speedgoat IO-191 Setup" block and "Speedgoat IO-392 Setup" block shown in Fig.~\ref{fig:SRT_Config1}. Speedgoat provides those block sets when purchasing the Speedgoat machine through an active maintenance agreement. Due to the user agreement, the authors can not share those two block set installers. The interested reader should contact Speedgoat for purchase and installation guidance. To set up the I/O module, double click on the "Speedgoat IO-191 Setup" block and configure the "Analog Outputs" and "Digital I/Os" as Fig.~\ref{fig:SRT_Config2} shows. In the "Analog Outputs" tab, one of the channels is used as analog output with a range of $\pm 10V$. This analog output connects to the motor drive's analog input for the linear motor's speed control in velocity mode. In the digital I/O tab, the first eight channels are used as the digital output ports, while the rest are configured as digital input ports. Three of the digital outputs are used to control the "Hardware Enable," "Switch to Secondary Mode," and "Start Homing" functions of the motor drive, as Fig.~\ref{fig:Lightning6} shows. Two of the digital inputs are used to read the signals of limit switches. Table.~\ref{table:IO-191-PinOut} shows the detailed connection. The above process completes the IO-191 module configuration for real-time digital I/O functionalities. 

To set up the IO-392 block set, double click on the "Speedgoat IO3XX Setup v3" block and select the "IO-392-50k" for "IO3xx I/O Module" drop-down list. Moreover, the configuration file~(Bitstream) "speedgoat\_IO392\_50k\_CI\_02385" is used as Fig.~\ref{fig:SRT_Config3} shows. This bitstream file needs to be in the same folder as the Simulink file.
Moreover, this file can be thought of as the driver of the FPGA module. It is purchased separately and is provided by the Speedgoat according to the user's need. The interested reader should contact Speedgoat for a quote. 

\begin{figure}[t]
    \centering
    \includegraphics[width=\textwidth]{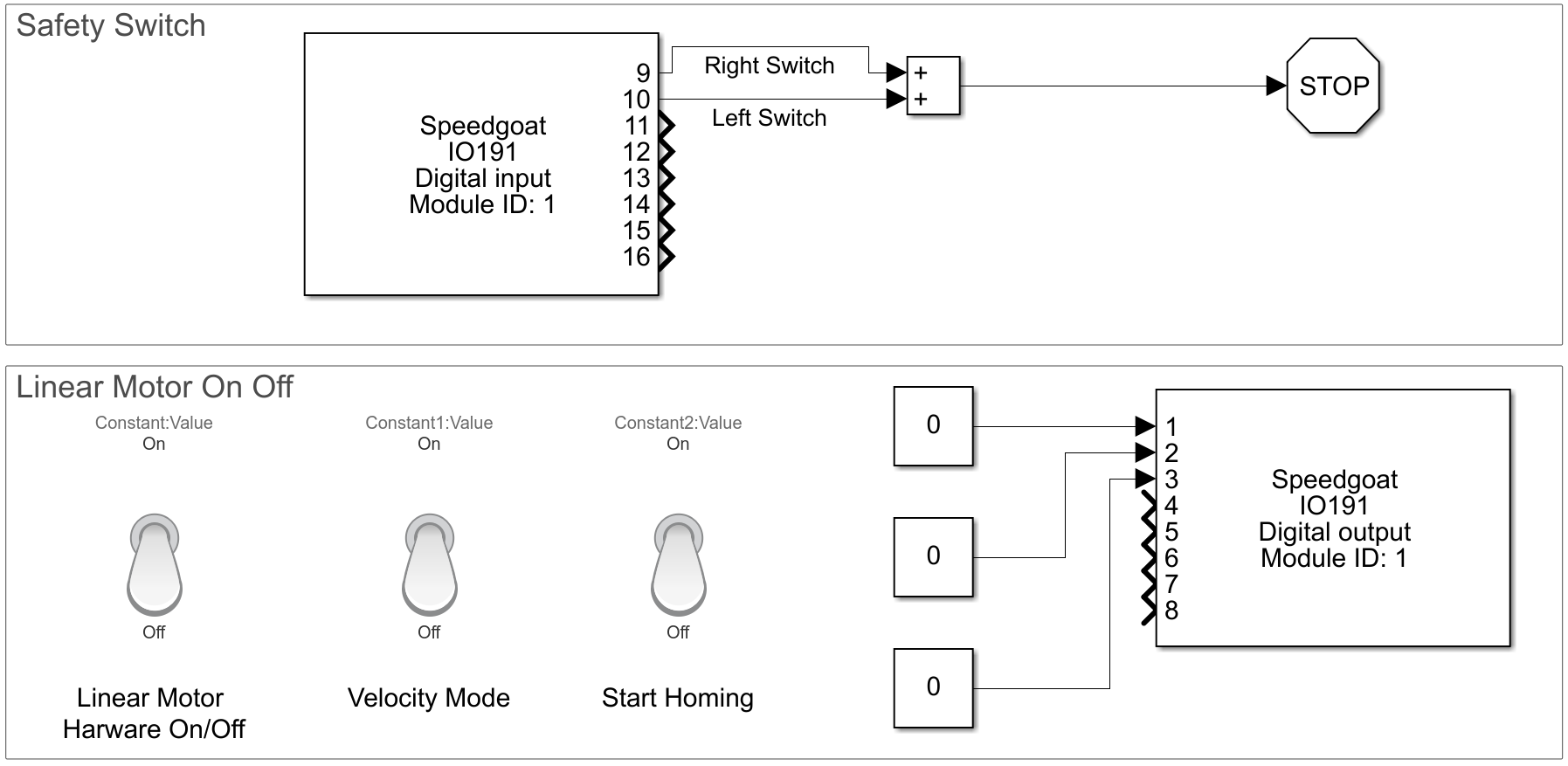}
    \caption{This figure shows enabling and disabling of the linear motor in Simulink. It also indicates how to switch the operating mode of the linear motor. By turning on the "Start Homing" toggle, the linear motor can start homing functionality in the stand-alone mode. The left and right limit switch signal is read and added up to stop the experiments if the limit switch is triggered. If one of the signals is turned to high, it will activate the "Stop Simulation" button of the Simulink, and the entire experiment is terminated.}
    \label{fig:SRT_Config4}
\end{figure}

Fig.~\ref{fig:SRT_Config4} shows the start and homing functionalities of the linear motor. A toggle switch is used in the simulation file to hardware start the linear motor. This toggle switch controls the "Digital I/O 01" output of the IO-191 module. Thus, when the toggle is switched on, meaning the production of pin 1b is high, the D1 drive will receive an "activate" signal on the third pin of the CN2, and thus hardware activates the linear motor as configured in Fig.~\ref{fig:Lightning6}. When hardware-enabled, the default operating mode the linear motor is in is the stand-alone mode. Users can switch on the "Velocity Mode" toggle to enter the velocity mode. This action will send out a high signal on the 2b pin of the Speedgoat machine, which is connected to the fourth pin of the D1 drive's CN2 channel, which will then set the motor drive to the velocity mode as Fig.~\ref{fig:Lightning6} configures. When turned on the first toggle and turned off the second toggle, the D1 drive will be in stand-alone mode. This action sets the "Digital I/O 03" to high so that the motor drive receives the "Start Homing" command and positions the linear motor to the middle point of the left and right limit switches. In Fig.~\ref{fig:SRT_Config4}, the "IO-191 Digital Input" block receives the signals of limit switches. In real-time execution, those signals are constantly checked. When the limit switch is not triggered, the reading is low and vice versa. Thus, by summing them up together and connecting the summation to the "Stop Simulation" button, the simulation can be terminated when any limit switches get activated. The termination of the run file will set the output of all the ports to the low state~(default), setting the linear motor to hardware disabled mode. This completes the setting up of the digital I/O functions in the Simulink file.

\begin{figure}[t]
    \centering
    \includegraphics[width=\textwidth]{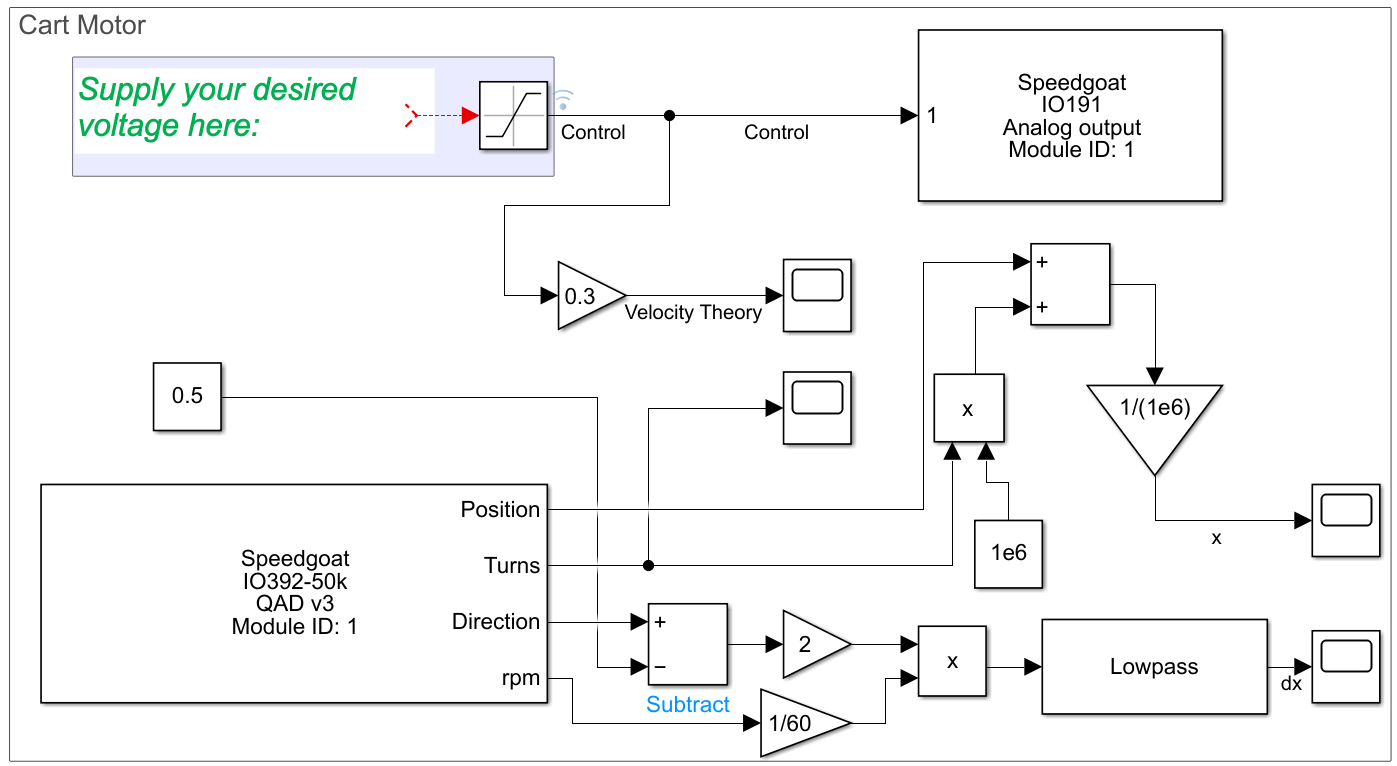}
    \caption{This figure illustrates the configuration of the pendulum cart in the Simulink model. To control the linear motor's velocity, the user should input desired analog voltage to the "Analog output" block. The output of the analog signal will be received by the D1 drive to control the linear motor to the user-defined speed. The "QAD V3" block is used to read the position and velocity of the linear motor. Moreover, some post-processing is needed to read the correct value of the linear motor's speed and velocity.}
    \label{fig:SRT_Config5}
\end{figure}

The desired analog voltage output of the IO-191 module needs to be supplied as Fig.~\ref{fig:SRT_Config5} shows to control the linear motor's motion. A saturation block limits the voltage input range in $\pm10V$ to avoid overflow issues. The desired voltage output is calculated using 
\begin{equation}
    \label{eq:DesiredVolatgeOutput}
    Volts=\frac{v}{K_a},
\end{equation}
where $Volts$ is the desired analog voltage output, $v$ is the desired velocity of the linear motor, and $K_a$ is the scaling factor between the two. In our set up, we used $K_a=0.3$ as Fig.~\ref{fig:Lightning5} shows. The analog output of the IO-191 module connects the analog input of the D1 drive so that the motor drive can receive the velocity command. The analog ground of the IO-191 and motor drive is connected as well. Table.~\ref{table:IO-191-PinOut} shows the detailed wiring, thus completing the controlling of the linear motor velocity. 
 
Fig.~\ref{fig:SRT_Config5} also shows the reading of the encoder signal. The "IO-392-50K QAD V3" block is used to read the encoder signal. This block is purchased along with the IO-392 module. To set up this block, one needs to double click on it and then set the correct channel vector number of the encoders one wants to read in the "Module setup" tab. As Table.~\ref{table:IO-392-PinOut} shows, the linear motor's encoder is wired to the fourth encoder channel. Thus, "[4]" should be entered in the "Channel vector" blank of the encoder block. Next, the reading mode of the encoder is configured in the "Quadrature decoder" tab by setting the "Operating mode" as "Quadrature decoder" and  "Sampling" as "$4\times$". The "Latch mode" is set to "Disabled" while "Interpolate inner position" is turned on. The correct "Steps per revolution vector" needs to be configured, which defines the counting range of the encoder counter. In the case of the linear motor's encoder, its resolution is $1\mu m$, which means there are $10^6$ counts per meter. Thus, this value is entered as $10^6$~(or $4\times10^6$). The default reading behavior of the encoder block wraps around to zero if the encoder reading value is larger than the value entered in the "Steps per revolution vector" blank. Some post-processing needs to be done to avoid this and achieve continuous reading that does not warp around. This is achieved by first selecting the "Show position," "Show turns output," "Show speed output," and "Show direction flag output" in the "Input and output port configuration" tab of the block. Fig.~\ref{fig:SRT_Config5} shows the resulted output ports of the encoder reader block. The "Turns" output will change accordingly whenever the encoder disk or linear motor performs a complete revolution. If the steps increase and a full revolution is made, then the "Turns" increase by one and vice versa. Thus, the total number of steps reflecting the absolute position of the linear motor or encoder disk can be calculate using
 \begin{equation}
     \label{eq:Encoder}
     Absolute\ Steps=Steps+Turns\times(Steps\ per\ Revolution),
 \end{equation}
where the "Steps" variable is the output of the "Position" port of the encoder reader block, "Turns" variable is the output of the "Turns" port. To transfer the $Absolute\ Steps$ into the "$Absolute Position$" of the linear motor, the "$Steps\ per\ Revolution$" is used to divide the $Absolute\ Steps$, which will result the absolute position value of the linear motor~(in $m$). This can be described as
  \begin{equation}
     \label{eq:EncoderValue}
     Absolute\ Position=\frac{Absolute\ Steps}{Steps\ per\ Revolution},
 \end{equation}
which completes the task of position reading of the linear motor. Sometimes, the velocity of the linear motor is also wanted, and it can be calculated using the "rpm" and "Direction" output port. The first thing to do is transfer the "rpm" into rounds per second. Dividing the output of the "rpm" port by $60$ achieves this task. However, the output of the "rpm" port is the absolute value of the velocity. Thus, to include directional information, the "Direction" port's output is transformed into $1$ or $-1$, where $1$ represents positive velocity while $-1$ represents the negative velocity. Then, it is multiplied with rounds per second to get the directional speed of the linear motor. The user can add a "Lowpass" filter block to denoise the encoder's velocity measurement, which completes the reading of the linear motor encoder.
 
 \begin{figure}[t]
    \centering
    \includegraphics[width=\textwidth]{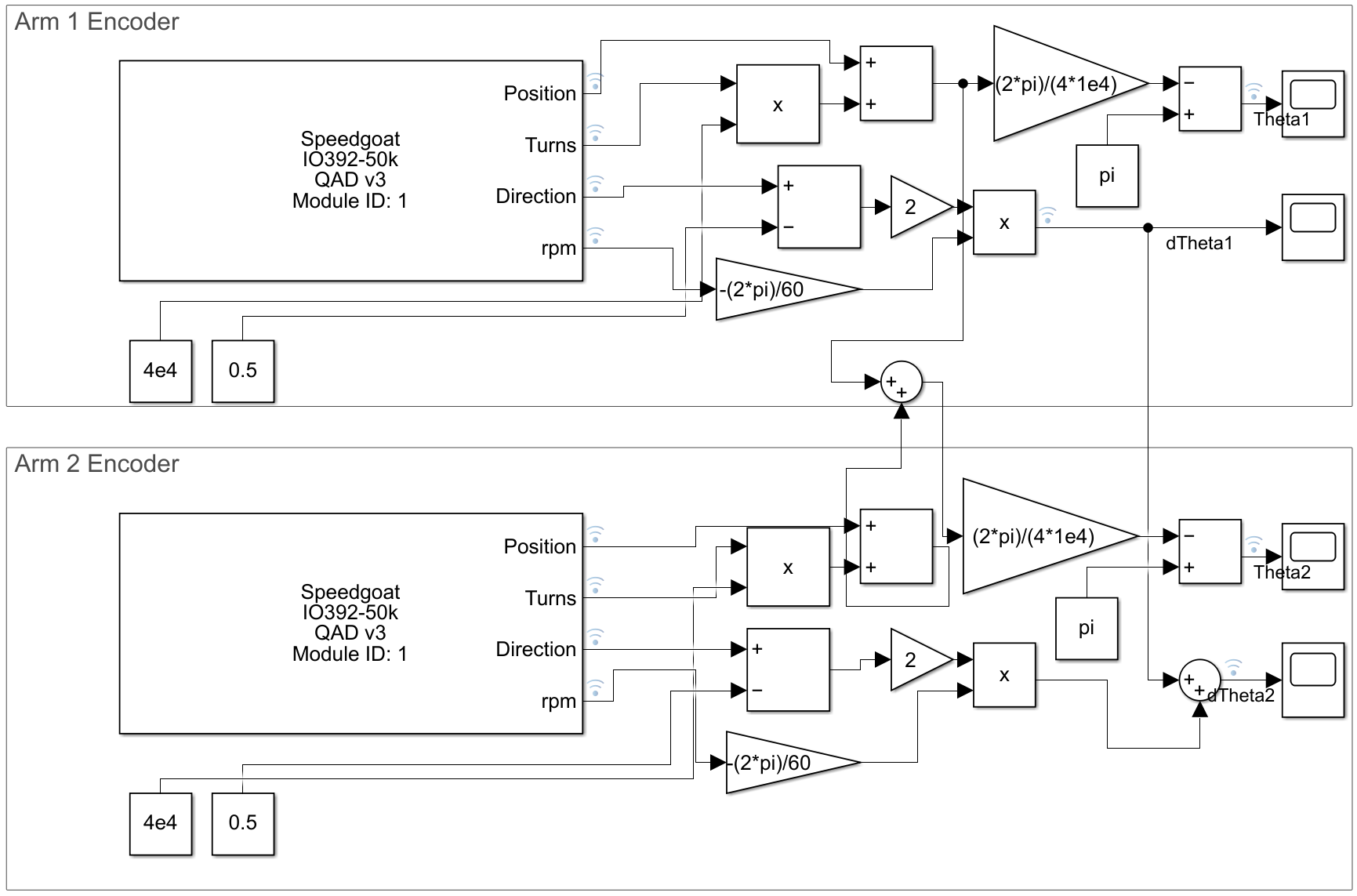}
    \caption{This figure illustrates how to read the angular position of the double pendulum shown in Fig.~\ref{fig:PendulumIllustration}. Conversion in Eq.~\eqref{eq:AngleConversion} and Eq.~\eqref{eq:AngularSpeedConversion} is needed to read the angular position of the second pendulum arm defined in Fig.~\ref{fig:PendulumIllustration}. A similar configuration of the encoder block is performed as we did in Fig.~\ref{fig:SRT_Config5} with a different value for the "Steps per revolution vector" blank.}
    \label{fig:SRT_Config6}
\end{figure}

Fig.~\ref{fig:SRT_Config6} shows the user can configure a similar setup to read the angular position of the pendulum arm, and we use double pendulum encoder reading as an example to illustrate this. First, as we did in the linear motor encoder reading configuration, the "Channel vector" value of two blocks is defined. As Table.~\ref{table:IO-392-PinOut} indicates, the first arm's encoder channel vector should be "[1]," the second arm's channel vector should be "[2]", and when using the triple pendulum arms, the third encoder's channel vector should be "[3]". Next, the "Operating mode" is selected as "Quadrature decoder," and  "Sampling" is selected as "$4\times$". Similarly, the "Latch mode" is set to "Disabled" while "Interpolate inner position" is turned on. The "Steps per revolution vector" is set to $4\times10^4$. To read the absolute rotational angle of the pendulum arm in radius, the following equation is used
   \begin{equation}
     \label{eq:EncoderValuePenArm}
     Absolute\ Rotational\ Angle=\frac{Absolute\ Steps}{Steps\ per\ Revolution}\times 2\pi.
 \end{equation}
Moreover, the user can calculate the angular speed of the pendulum arm in radius by multiplying the output of the "rpm" port with $1/60$ to transfer it to the rounds per second. Next, it can be multiplied by $\pm2\pi$ to transfer the velocity into $rad/s$. The choice of plus or minus sign determines whether the pendulum arm's clock-wise rotation increases or decreases the encoder reading. In our case, we define the clock-wise rotation of the pendulum arm increases the value of the encoder reading. Finally, it is vital to note that the reading of the second pendulum arm does not correspond to the $\theta_2$ value shown in Fig.~\ref{fig:PendulumIllustration}. Define the encoder reading of the first, second, and third arm as $\alpha$, $\beta$ and $\gamma$. Then the relationship between the encoder reading and pendulum arm's angle $\theta_1$, $\theta_2$, and $\theta_3$ can be written as
\begin{subequations}
    \label{eq:AngleConversion}
    \begin{align}
        \theta_1 & = \alpha, \\
        \theta_2 & = \alpha+\beta, \\
        \theta_3 & = \alpha+\beta+\gamma.
    \end{align}
\end{subequations}
Thus, the angular speed of the pendulum arm can be written as
\begin{subequations}
    \label{eq:AngularSpeedConversion}
    \begin{align}
        \dot{\theta}_1 & = \dot{\alpha}, \\
        \dot{\theta}_2 & = \dot{\alpha}+\dot{\beta}, \\
        \dot{\theta}_3 & = \dot{\alpha}+\dot{\beta}+\dot{\gamma}.
    \end{align}
\end{subequations}
We provide a template file that achieves all the encoder reading and linear motor control tasks\footnote{Code availalbe at \href{https://github.com/dynamicslab/MultiArm-Pendulum}{https://github.com/dynamicslab/MultiArm-Pendulum}}. The reader can directly modify this file to include their controller without performing the encoder reading configuration. This conversion is also shown in Fig.~\ref{fig:SRT_Config6}, which completes the setup of the Simulink configuration.


\begin{figure}
    \centering
    \includegraphics[width=0.8\textwidth]{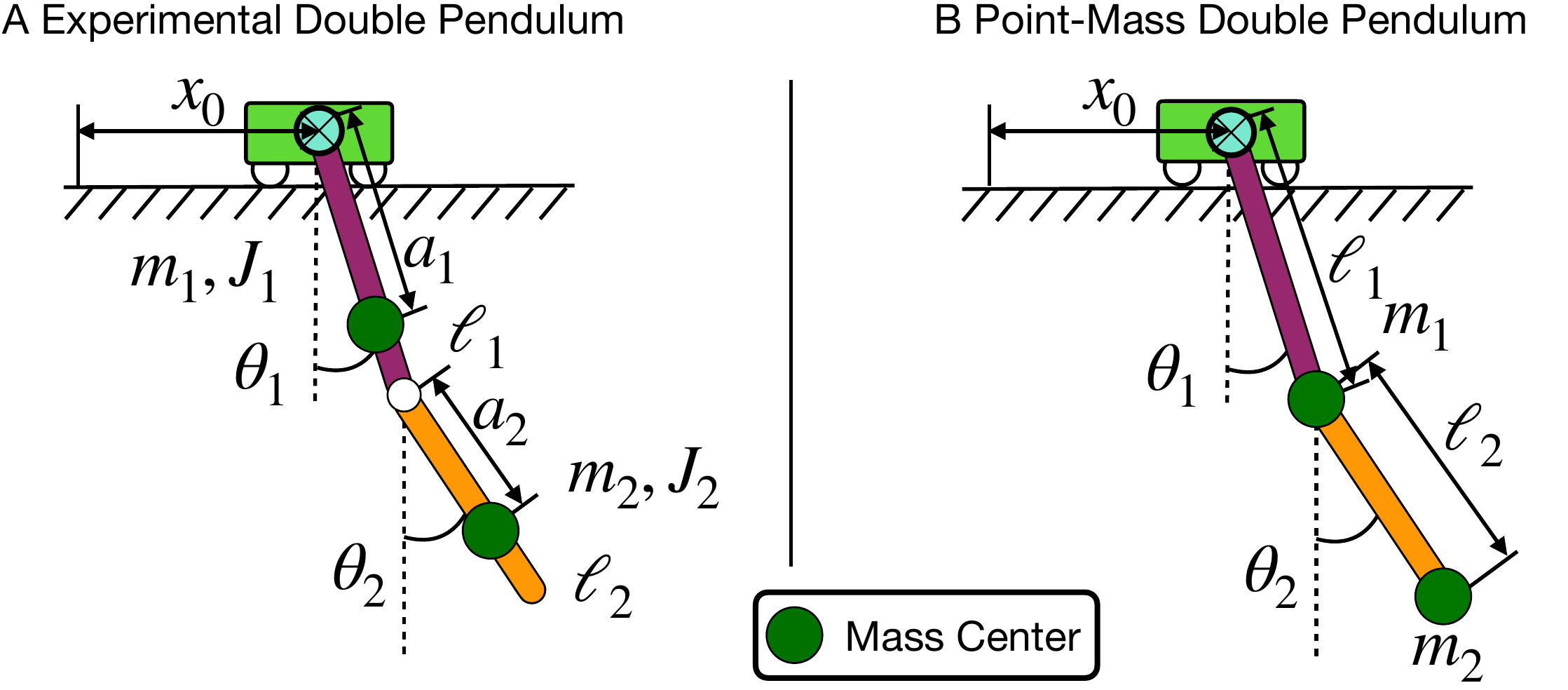}
    \caption{This figure illustrates the difference between the point-mass model of the double pendulum (left) and the more realistic experimental double pendulum (right) studied herein. The point-mass model ignores the mass of each arm of the pendulum, while the experimental model considers both the inertial ($J_i$) and mass $(m_i)$ of the pendulum arms.}
    \label{Fig:DP_Illustrate}
\end{figure}

\section{Step by Step Operation Details}
\label{Appendix:OperationDetail}
\subsection{Pre-Experiment Preparations}
Before starting the experiments, the following operations are needed:
\begin{enumerate}
    \item Put on safety glasses. 
    \item Determine the overall sampling frequency of the setup, control input range, cart velocity range, and cart position range. These parameters can help determine whether the purposed experiments is physically realizable. If the desired cart velocity and acceleration range is higher than the limit introduced in Fig.~\ref{fig:Lightning8}, change the limits in Fig.~\ref{fig:Lightning8} accordingly.
    \item Develop the controller Simulink file, and make sure it compiles correctly. Rewire the connection of the IO-191 module if the custom I/O functionalities are needed. 
    \item Double check the wiring between the Speedgoat terminal block and pendulum arm encoder and linear motor drive.
    \item Check the connection between the slip-ring and slip-ring brush. Make sure the slip-ring and slip-ring brush are contacting each other.
    \item Connect the host computer and Speedgoat machine using an Ethernet cable. Turn on the Speedgoat machine and setup the Simulink Real-Time to establish communication between the host computer and Speedgoat machine. This step and those prior to it complete the connection between the pendulum arm and Speedgoat machine.
    \item Check all the wires that connect the linear motor's electrical parts mentioned in Fig.~\ref{fig:MotorEletricalPartOverview}. Make sure all cables are connected properly according to the wiring diagram shown in Fig.~\ref{fig:MotorWiring}. 
    \item Release the emergency stop button and connect the circuit breaker.
    \item Plug in the power cord of the emergency stop button module. This power cord is used to supply power to the linear motor, motor drive, and all other components in the electrical parts of the linear motor.
    \item Press the green button of the emergency stop button module to supply AC single phase power to the electrical box of the system. Wait a moment to observe if any shortage has occurred. If so, press the emergency stop button to cut off the power supply. If not, proceed to the next step.
    \item Press the green momentary switch to self-lock the magnetic contactor. This will provide the power supply to the line reactor and DC power supplied module. Wait a moment to observe if any shortage has occurred. If so, press the emergency stop button to cut off the power supply. If not, proceed to the next step.
    \item Set the toggle switch of the electrical box to on. This provides power to the motor drive, and in turn the motor drive will provide power to the linear motor. Wait a moment to observe if any shortage has occurred. If so, press the emergency stop button to cut off the power supply. If not, proceed to the next step.
    \item Remove any and all obstacles from the motor rail.
    \item Execute the Lightning software and connect the D1 motor drive to the host computer. Make sure all the parameters shown in the Lightning software are correct. 
    \item Open the Simulink model with developed controller. Compile it and download it to the Speedgoat machine.
    \item Execute the compiled controller, start the desired experiments.
\end{enumerate}

\subsection{Peri-Experiment Operations and Precautions}

One should adhere to the following operations and precautions below when performing experiments:
\begin{enumerate}
    \item Never get close to the moving pendulum cart just in case any mistakes in the designed controller file make the linear motor perform undesired motion and injure the operator.
    \item The operator should stand on the front or back side of the experimental setup. Never stand on the side of the linear motor as one may be struck by a pendulum arm.
    \item The operator should closely monitor the behavior of the experimental setup. If for any reasons there is an electical shortage, smoke, heat, or unwanted noise, immediately press the emergency stop button to cut off the power supply to the linear motor to avoid further damage.
    \item The operator should closely monitor the surroundings of the experimental setup. If any person, animal, or unwanted object enters the experimental area the experiment should be stopped immediately to avoid injury and/or prevent the damage of the experimental setup.
    \item The operator should continually check the data being collected from the experiment. If an unwanted measurement drift is apparent in the plotted data, immediate stop the experiment by pressing the stop simulation button of the Simulink file.
\end{enumerate}

\subsection{Post-Experiment Operations}

After an experiment, the following operation should be performed:
\begin{enumerate}
    \item Check all wiring related to the electrical component of the pendulum setup. Reconnect any wires that may have become disconnected during the experiment.
    \item Check for damage to the system hardware.
    \item Go back to the pre-experiment operations to perform another experiment or follow the proceeding steps to turn off the system.
    \item Close the Lightning software to cut off the communication between the D1 drive and host computer.
    \item Switch the toggle switch to off to cut off the power to the motor drive.
    \item Press the red momentary switch to cut off the power to the DC power supply and line reactor.
    \item Press the off button of the emergency stop button module. This will cut off all power supply to the linear motor. Then unplug the power cord.
    \item Save the experimental files and data to a secured location. After saving close Matlab and Simulink.
    \item Turn off the Speedgoat machine. This will cut off the power supply to the pendulum arm encoder.
    \item Turn off the host computer.
\end{enumerate}
With the above post-experiment operations, all of the power supplied to the experimental system is now cut off.

\section{Safety Mechanism and Notes Details}
\label{Appendix:SafetyNotes}

The following electrical safety mechanisms should further be implemented:
\begin{enumerate}
    \item Use the circuit breaker protect the motor from over current failure.
    \item Install the limit switch on the linear motor to prevent the out of range motion of the linear motor.
    \item Use the regenerative resistors to absorb extra kinetic energy during the breaking of linear motor.
    \item Use the noise filter to reduce the ground current protecting the appliances that share the same ground.
    \item The noise filter, shielded cable, edge filter, and EMI cores are used to reduce the side effect of EMI noise. This in turn provides a cleaner signal and prevents the possible damage caused by faulty measurements signal.
    \item Obtain better control of the electrical system using the emergency stop button, momentary switches, while allowing for rapid disconnection from the power supply, thus improving safety in the case of any electrical shortage.
    \item Use an electrical box to prevent potential mechanical damage to the electrical components.
    \item Properly ground electrical components to avoid a shock to the operator. 
    \item Use the linear motor's temperature sensor to prevent overheating and over current failure.
\end{enumerate}

The Lighting and Simulink software should be modified as follows to guarantee user safety:
\begin{enumerate}
    \item Use the Lightning software to internally restrict the range of linear motor's position, velocity, and acceleration. 
    \item Use the Lightning software to stop the linear motor when the limit switch is triggered.
    \item Program the Simulink model to limit the analog output, which restricts the velocity range of the linear motor.
    \item Use the Simulink model to stop the experiment when the limit switch is triggered.
\end{enumerate}

In addition to the above safety procedures, one should also use the following points to ensure the safety of the operator and surrounding personnel.
\begin{enumerate}
    \item Always check the wire connections for damage or disconnections before starting the experimentto avoid an electrical shortage and fire.
    \item Clear the linear motor rail of obstacles before supplying power to the motor. 
    \item Do not touch the linear motor rail by hand. This could potentially rust the motor rail. 
    \item Apply general purpose grease every month to protect the motor rails from rusting.
    \item Never touch the linear motor while it is in operation.
    \item Never perform any wiring of electrical components with the power supply on.
    \item Keep a safe distance from the system while it is operating. We suggest the operator and all surrounding personnel should be at least one meter from the experimental setup while performing experiments.
    \item Do not stand to the side of the pendulum arms as this increases the risk of being hit by a pendulum arm.
    \item Do not expose the electrical components to moisture or liquid as they are not waterproof. Furthermore, do not operator the equipment if hands are wet.
    \item Wear safety glasses for all construction and operation steps for the experiment.
\end{enumerate}

\begin{table}[t]
\centering
\caption{Comparison of the estimated parameters and CAD model estimated parameters of the single, double, and triple pendulum. The length $l_1$, $l_2$ and $l_3$ are not estimated for the single, double, and triple pendulum respectively, since they are not present in the corresponding EOM. It is interesting to note that the estimated values of the local gravity constant $g$ are different for all three pendulums. This is caused by the fact the $g$ is an optimization parameter during the parameter estimation and is allowed to vary to best fit the data. Units: mass $kg$, length $m$, inertia $kgm^2$, and local gravity constant $m/s^2$.}
\label{table:EstimatedParameters}
\vspace{-.05in}
\resizebox{\textwidth}{!}{%
\begin{tabular}{|c|cc|cc|cc|}
\hline
\begin{tabular}[c]{@{}c@{}}Pendulum\\ Type\end{tabular} &
  \multicolumn{2}{c|}{Single} &
  \multicolumn{2}{c|}{Double} &
  \multicolumn{2}{c|}{Triple} \\ \hline
\backslashbox{Vars}{Source} &
  \multicolumn{1}{c|}{CAD} &
  \begin{tabular}[c]{@{}c@{}}Estimated \\ Parameters\end{tabular} &
  \multicolumn{1}{c|}{CAD} &
  \begin{tabular}[c]{@{}c@{}}Estimated \\ Parameters\end{tabular} &
  \multicolumn{1}{c|}{CAD} &
  \begin{tabular}[c]{@{}c@{}}Estimated \\ Parameters\end{tabular} \\ \hline
$m_1$ &
  \multicolumn{1}{c|}{$0.1183$} &
  $0.1476$ &
  \multicolumn{1}{c|}{$0.1199$} &
  $0.0938$ &
  \multicolumn{1}{c|}{$0.11$} & 0.2582
   \\ \hline
$m_2$ &
  \multicolumn{1}{c|}{\diagbox[width=\dimexpr \textwidth/8+2\tabcolsep\relax, height=0.47cm]{}{}} &
  \diagbox[width=\dimexpr \textwidth/8+4\tabcolsep\relax, height=0.47cm]{}{} &
  \multicolumn{1}{c|}{$0.1183$} &
  $0.1376$ &
  \multicolumn{1}{c|}{$0.12$} & 0.2794
   \\ \hline
$m_3$ &
  \multicolumn{1}{c|}{\diagbox[width=\dimexpr \textwidth/8+2\tabcolsep\relax, height=0.47cm]{}{}} &
  \diagbox[width=\dimexpr \textwidth/8+4\tabcolsep\relax, height=0.47cm]{}{} &
  \multicolumn{1}{c|}{\diagbox[width=\dimexpr \textwidth/8+2\tabcolsep\relax, height=0.47cm]{}{}} &
  \diagbox[width=\dimexpr \textwidth/8+2\tabcolsep\relax, height=0.47cm]{}{} &
  \multicolumn{1}{c|}{$0.1$} & 0.1186
   \\ \hline
$l_1$ &
  \multicolumn{1}{c|}{$0.17272$} &
  \diagbox[width=\dimexpr \textwidth/8+4\tabcolsep\relax, height=0.47cm]{}{} &
  \multicolumn{1}{c|}{$0.17272$} &
  $0.1727$ &
  \multicolumn{1}{c|}{$0.1727$} & 0.1728
   \\ \hline
$l_2$ &
  \multicolumn{1}{c|}{\diagbox[width=\dimexpr \textwidth/8+2\tabcolsep\relax, height=0.47cm]{}{}} &
  \diagbox[width=\dimexpr \textwidth/8+4\tabcolsep\relax, height=0.47cm]{}{} &
  \multicolumn{1}{c|}{$0.2286$} &
  \diagbox[width=\dimexpr \textwidth/8+2\tabcolsep\relax, height=0.47cm]{}{} &
  \multicolumn{1}{c|}{$0.2286$} & 0.2287
   \\ \hline
$l_3$ &
  \multicolumn{1}{c|}{\diagbox[width=\dimexpr \textwidth/8+2\tabcolsep\relax, height=0.47cm]{}{}} &
  \diagbox[width=\dimexpr \textwidth/8+4\tabcolsep\relax, height=0.47cm]{}{} &
  \multicolumn{1}{c|}{\diagbox[width=\dimexpr \textwidth/8+2\tabcolsep\relax, height=0.47cm]{}{}} &
  \diagbox[width=\dimexpr \textwidth/8+2\tabcolsep\relax, height=0.47cm]{}{} &
  \multicolumn{1}{c|}{$0.2413$} &
  \diagbox[width=\dimexpr \textwidth/8+2\tabcolsep\relax, height=0.47cm]{}{} \\ \hline
$a_1$ &
  \multicolumn{1}{c|}{$0.08638$} &
  $0.1478$ &
  \multicolumn{1}{c|}{$0.08755$} &
  $0.1086$ &
  \multicolumn{1}{c|}{$0.08755$} & 0.16
   \\ \hline
$a_2$ &
  \multicolumn{1}{c|}{\diagbox[width=\dimexpr \textwidth/8+2\tabcolsep\relax, height=0.47cm]{}{}} &
  \diagbox[width=\dimexpr \textwidth/8+4\tabcolsep\relax, height=0.47cm]{}{} &
  \multicolumn{1}{c|}{$0.1256$} &
  $0.1168$ &
  \multicolumn{1}{c|}{$0.127$} & 0.2029
   \\ \hline
$a_3$ &
  \multicolumn{1}{c|}{\diagbox[width=\dimexpr \textwidth/8+2\tabcolsep\relax, height=0.47cm]{}{}} &
  \diagbox[width=\dimexpr \textwidth/8+4\tabcolsep\relax, height=0.47cm]{}{} &
  \multicolumn{1}{c|}{\diagbox[width=\dimexpr \textwidth/8+2\tabcolsep\relax, height=0.47cm]{}{}} &
  \diagbox[width=\dimexpr \textwidth/8+2\tabcolsep\relax, height=0.47cm]{}{} &
  \multicolumn{1}{c|}{$0.12$} & 0.1837
   \\ \hline
$J_1$ &
  \multicolumn{1}{c|}{$0.001578$} &
  $1.0912\times10^{-4}$ &
  \multicolumn{1}{c|}{$0.001627$} &
  $10^{-4}$ &
  \multicolumn{1}{c|}{$0.001578$} & $1\times10^{-4}$ 
   \\ \hline
$J_2$ &
  \multicolumn{1}{c|}{\diagbox[width=\dimexpr \textwidth/8+2\tabcolsep\relax, height=0.47cm]{}{}} &
  \diagbox[width=\dimexpr \textwidth/8+4\tabcolsep\relax, height=0.47cm]{}{} &
  \multicolumn{1}{c|}{$0.002908$} &
  $10^{-4}$ &
  \multicolumn{1}{c|}{$0.0029$} & $3.1416\times10^{-4}$
   \\ \hline
$J_3$ &
  \multicolumn{1}{c|}{\diagbox[width=\dimexpr \textwidth/8+2\tabcolsep\relax, height=0.47cm]{}{}} &
  \diagbox[width=\dimexpr \textwidth/8+4\tabcolsep\relax, height=0.47cm]{}{} &
  \multicolumn{1}{c|}{\diagbox[width=\dimexpr \textwidth/8+2\tabcolsep\relax, height=0.47cm]{}{}} &
  \diagbox[width=\dimexpr \textwidth/8+2\tabcolsep\relax, height=0.47cm]{}{} &
  \multicolumn{1}{c|}{$0.002157$} & $1.6874\times10^{-4}$
   \\ \hline
$\varepsilon_1$ &
  \multicolumn{1}{c|}{\diagbox[width=\dimexpr \textwidth/8+2\tabcolsep\relax, height=0.47cm]{}{}} &
  $2.2394\times10^{-4}$ &
  \multicolumn{1}{c|}{\diagbox[width=\dimexpr \textwidth/8+2\tabcolsep\relax, height=0.47cm]{}{}} &
  $10^{-5}$ &
  \multicolumn{1}{c|}{\diagbox[width=\dimexpr \textwidth/8+2\tabcolsep\relax, height=0.47cm]{}{}} & 0.0015
   \\ \hline
$\varepsilon_2$ &
  \multicolumn{1}{c|}{\diagbox[width=\dimexpr \textwidth/8+2\tabcolsep\relax, height=0.47cm]{}{}} &
  \diagbox[width=\dimexpr \textwidth/8+4\tabcolsep\relax, height=0.47cm]{}{} &
  \multicolumn{1}{c|}{\diagbox[width=\dimexpr \textwidth/8+2\tabcolsep\relax, height=0.47cm]{}{}} &
  $10^{-5}$ &
  \multicolumn{1}{c|}{\diagbox[width=\dimexpr \textwidth/8+2\tabcolsep\relax, height=0.47cm]{}{}} & $2.5958\times10^{-4}$
   \\ \hline
$\varepsilon_3$ &
  \multicolumn{1}{c|}{\diagbox[width=\dimexpr \textwidth/8+2\tabcolsep\relax, height=0.47cm]{}{}} &
  \diagbox[width=\dimexpr \textwidth/8+4\tabcolsep\relax, height=0.47cm]{}{} &
  \multicolumn{1}{c|}{\diagbox[width=\dimexpr \textwidth/8+2\tabcolsep\relax, height=0.47cm]{}{}} &
  \diagbox[width=\dimexpr \textwidth/8+2\tabcolsep\relax, height=0.47cm]{}{} &
  \multicolumn{1}{c|}{\diagbox[width=\dimexpr \textwidth/8+2\tabcolsep\relax, height=0.47cm]{}{}} & $1\times10^{-5}$
   \\ \hline
$g$ &
  \multicolumn{1}{c|}{$9.8083$} &
  $9.81001$ &
  \multicolumn{1}{c|}{$9.8083$} &
  $9.808$ &
  \multicolumn{1}{c|}{$9.8083$} & 9.8083
   \\ \hline
\end{tabular}%
}
\end{table}

\section{Parameter Estimation Details}
\label{Appendix:ParameterEstimation}
This Appendix. gives the details of parameters estimation process shown in Table.~\ref{table:EstimatedParameters}. In this Appendix., we will first use the EOM of the double pendulum on the cart as an example to explain why it is desired to use the acceleration of the pendulum cart as our control input. Then, we will show more details on the parameter estimation process we used to identify the parameters in Table.~\ref{table:EstimatedParameters}. Since the model derivation task of single, double and triple pendulum are standard text book example, we refer the reader to check~\cite{graichen2007swing,gluck2013swing} for more details on deriving the EOM of pendulum arm on the cart.

\subsection{Equations of Motion} 
\label{sec:EOM}
This section shows why it is desired to use the acceleration of the pendulum cart as control input and we illustrate this using double pendulum as an example. Let $x_0$ be the position of pendulum cart and $v_{x_0}$ its velocity, and the rotational angle of pendulum arm as $\theta_1$ and $\theta_2$ respectively. Following the modeling process shown in~\cite{graichen2007swing}, the second derivative of the double pendulum's rotational angle can be written as $\ddot\theta_1$ and $\ddot\theta_2$, giving
\begin{equation}\label{eq:edp_ddtheta}
    \begin{split}
        \ddot{\theta}_1&=\frac{A_{10}\sin(\theta_1)+A_{11}\sin(\theta_1-\theta_2)+A_{12}\sin(\theta_1-2\theta_2)+A_{22}\sin(2\theta_1-2\theta_2)+\varepsilon_1F_{11}+\varepsilon_2F_{12}+C_1\dot{v}_{x_0}}{D(\theta_1 - \theta_2)}\\
            \ddot{\theta}_2&=\frac{B_{01}\sin(\theta_2)+B_{11}\sin(\theta_1-\theta_2)+B_{21}\sin(2\theta_1-\theta_2)+B_{22}\sin(2\theta_1-2\theta_2)+\varepsilon_1F_{21}+\varepsilon_2F_{22}+C_2\dot{v}_{x_0}}{D(\theta_1 - \theta_2)}
    \end{split}
\end{equation}
where 
\begin{equation}
    \begin{split}
        A_{10} &= - l_1ga_2^2m_2^2 - 2a_1gm_1a_2^2m_2 - 2J_2l_1gm_2 - 2J_2a_1gm_1 \\
        A_{11} &= - 2l_1a_2^3\dot{\theta}_2^2m_2^2 - 2J_2l_1a_2\dot{\theta}_2^2m_2 \\
        A_{12} &= -l_1a_2^2gm_2^2 \\
        A_{22} &= -l_1^2a_2^2\dot{\theta}_1^2m_2^2 \\
        B_{01} &= -a_2m_2(gm_2l_1^2 - gm_1l_1a_1 + 2gm_1a_1^2 + 2J_1g)\\
        B_{11} &= a_2m_2(2m_2l_1^3\dot{\theta}_1^2 + 2m_1l_1a_1^2\dot{\theta}_1^2 + 2J_1l_1\dot{\theta}_1^2)\\
        B_{21} &= a_2m_2(gm_2l_1^2 + a_1gm_1l_1)\\
        B_{22} &= 2m_2(gm_2l_1^2 + a_1gm_1l_1)\\
        F_{11} &= -2\dot{\theta}_1m_2a_2^2 - 2J_2\dot{\theta}_1 \\ 
        F_{12} &= 2J_2\dot{\theta}_2 - 2J_2\dot{\theta}_1 - 2a_2^2\dot{\theta}_1m_2 + 2a_2^2\dot{\theta}_2m_2\\
            &\quad - 2l_1a_2\dot{\theta}_1m_2\cos(\theta_1 - \theta_2) + 2l_1a_2\dot{\theta}_2m_2\cos(\theta_1 - \theta_2)\\
        F_{21} &= 2l_1a_2\dot{\theta}_1m_2\cos(\theta_1 - \theta_2) \\
        F_{22} &= 2J_1\dot{\theta}_1 - 2J_1\dot{\theta}_2 + 2l_1^2\dot{\theta}_1m_2 - 2l_1^2\dot{\theta}_2m_2 + 2a_1^2\dot{\theta}_1m_1 - 2a_1^2\dot{\theta}_2m_1 \\
            &\quad + 2l_1a_2\dot{\theta}_1m_2\cos(\theta_1 - \theta_2) - 2l_1a_2\dot{\theta}_2m_2\cos(\theta_1 - \theta_2) \\
        C_1 &= l_1a_2^2m_2^2\cos(\theta_1 - 2\theta_2) - 2J_2l_1m_2\cos(\theta_1) - l_1a_2^2m_2^2\cos(\theta_1) \\
        &\quad - 2J_2a_1m_1\cos(\theta_1) - 2a_1a_2^2m_1m_2\cos(\theta_1) \\
        C_2 &= 2l_1^2a_2m_2^2\cos(\theta_1 - \theta_2) \cos(\theta_1) - 2J_1a_2m_2\cos(\theta_2) - 2l_1^2a_2m_2^2\cos(\theta_2) \\
            &\quad - 2a_1^2a_2m_1m_2\cos(\theta_2) + 2l_1a_1a_2m_1m_2\cos(\theta_1 - \theta_2) \cos(\theta_1) \\
        D(\theta_1 - \theta_2) &=-l_1^2a_2^2m_2^2cos(\theta_1-\theta_2)^2 + l_1^2a_2^2m_2^2 +  J_2l_1^2m_2 + m_1a_1^2a_2^2m_2 \\&\quad + J_2m_1a_1^2 + J_1a_2^2m_2 + J_1J_2
    \end{split}.
\end{equation}
Here we use the convention that $A_{ij},B_{ij}$ are the coefficients of the energy preserved Hamiltonian system for the $\ddot\theta_1$ and $\ddot\theta_2$ equations, respectively, while the subscripts denote the coefficients on the $\theta_1$ and $\theta_2$ terms inside the sine function they are multiplied against. The $F_{ij}$ are the $(\theta_1,\theta_2,\dot\theta_1,\dot\theta_2)$-dependent friction terms in $\ddot\theta_i$ equation with friction coefficient $\varepsilon_j$. Finally, the $C_i$ terms correspond to the influence of the cart velocity, $\dot v_{x_0}$, for the $\ddot \theta_i$ equation and $D(\theta_1 - \theta_2)$ represents the denominator function. By writing the equation of the of the experimental double pendulum in the form of Eq.~\eqref{eq:edp_ddtheta}, one can see the acceleration of the pendulum cart $\dot{v}_{x_0}$ can be nicely used as control input of the system. By \textbf{assuming} perfect velocity control can be provided by the motor drive, the mass of the pendulum cart $M$ can be avoided in the equation of motion of the pendulum cart. 

In experiments with the physical double pendulum on a cart the velocity of the pendulum cart is controlled by the servo drive by applying an analog voltage input as we mentioned in Sec.~\ref{sec:Software}. This voltage input is calculated by first taking the integral of the acceleration of the pendulum cart $\dot{v}_{x_0}$. The integration will result the desired velocity of the pendulum cart $v_{x_0}$. Then the desired velocity is scaled to the corresponding voltage using the scaling factor shown in Fig.~\ref{fig:Lightning5}. Thus, in the experiments, the acceleration of the double pendulum is chosen as the control input of the system. To account for this control input, we write out the complete first order ODE of the pendulum on the cart system using acceleration of the cart as control input, denoted $u$. The final result is the ODE
\begin{equation}\label{eq:edp_ode}
    \begin{split}
        \dot{\theta}_1&=\omega_1,\\
        \dot{\theta}_2&=\omega_2,\\
        \dot{\omega}_1&=\frac{A_{10}\sin(\theta_1)+A_{11}\sin(\theta_1-\theta_2)+A_{12}\sin(\theta_1-2\theta_2)+A_{22}\sin(2\theta_1-2\theta_2)+\varepsilon_1F_{11}+\varepsilon_2F_{12}+C_1u}{D(\theta_1 - \theta_2)},\\
        \dot{\omega}_2&=\frac{B_{01}\sin(\theta_2)+B_{11}\sin(\theta_1-\theta_2)+B_{21}\sin(2\theta_1-\theta_2)+B_{22}\sin(2\theta_1-2\theta_2)+\varepsilon_1F_{21}+\varepsilon_2F_{22}+C_2u}{D(\theta_1 - \theta_2)},\\
        \dot{x}_0&=v_{x_0},\\
        \dot{v}_{x_0}&=u.\\
    \end{split}
\end{equation}
System~\ref{eq:edp_ode} therefore represents the full equations of motion for the double pendulum on a cart that we investigate for the parameter estimation task. The equations of motions shown in Eq.~\eqref{eq:edp_ode} made various assumptions. For example, it did not consider the the air drag of the pendulum arm, it did not consider the effect of rolling bearing balls during the rotational movement as shown in~\cite{myers2020low}, it did not consider the time-lag between the control signal and movement of the linear motor, it did not consider the oscillation of the pendulum, it did not consider the fact that linear motor might not be moving in a horizontal line, etc. However, it does a pretty good job on capturing most of the system dynamics and is the most widely used equation of motion for the experimental double pendulum on the cart system. If the user consider more factors, a more accurate and complicated model of the double pendulum on the cart can be derived. The same procedures shown above can also be used to derive the equation of motion of the single and triple pendulum on the cart. A good reference on deriving the equation of motion of the triple pendulum can be seen in~\cite{gluck2013swing}. 

\subsection{Parameter Estimation}\label{sec:Estimation}
Having now derived the equations of motion for the experimental double pendulum on the cart, we now seek to identify the values of the numerous parameters to arise in it. Our goal is to align the parameters of the model \eqref{eq:edp_ode} with those of our constructed double pendulum on a cart. Of course, the parameters of the physical model are uncertain and we expect a difference between the actual and theoretical parameter values due to imperfect manufacturing, installation tolerance, etc. In this subsection, we will present the method we have used for parameter estimation of the double pendulum. The same approach is also used to estimate the parameters of the double and triple pendulum.

To estimate the parameters of the double pendulum we fix the position of the cart. The result will be that $u = 0$ in \eqref{eq:edp_ode}, thus reducing the ODE to one in the four variables ($\theta_1,\theta_2,\dot{\theta}_1,\dot{\theta}_2$), given by
\begin{equation}\label{eq:edp_pares_ode}
    \begin{split}
        \dot{\theta}_1&=\omega_1,\\
        \dot{\theta}_2&=\omega_2,\\
        \dot{\omega}_1&=\frac{A_{10}\sin(\theta_1)+A_{11}\sin(\theta_1-\theta_2)+A_{12}\sin(\theta_1-2\theta_2)+A_{22}\sin(2\theta_1-2\theta_2)+\varepsilon_1F_{11}+\varepsilon_2F_{12}}{D(\theta_1 - \theta_2)},\\
        \dot{\omega}_2&=\frac{B_{01}\sin(\theta_2)+B_{11}\sin(\theta_1-\theta_2)+B_{21}\sin(2\theta_1-\theta_2)+B_{22}\sin(2\theta_1-2\theta_2)+\varepsilon_1F_{21}+\varepsilon_2F_{22}}{D(\theta_1 - \theta_2)}
    \end{split}
\end{equation}
Next, the experimental data of the pendulum arms is collected by swinging them from an arbitrary initial condition. The relative angles of pendulum arms with respect to its joint is measured using the encoders, with $10000$ CPR accuracy in quadrature mode, which results in $40000$ PPR (pulse per revolution). 
Suppose that we have collected $n \geq 1$ measurements of the values of $(\theta_1,\theta_2,\dot\theta_1,\dot\theta_2)$, each over $L \geq 1$ time values. We arrange each measure into the data matrices $\bY_i \in \mathbb{R}^{4 \times L}$ for each $i = 1,\dots,n$ and denote $\hat\bY_i(p) \in\mathbf{R}^{4 \times L}$ the data generated by simulating \eqref{eq:edp_pares_ode} with the same initial conditions for each $i = 1,\dots,n$ and a given choice of the parameter values 
\begin{equation}
    p = [ m_1, \ m_2, \ a_1, \ a_2, \ \ell_1, \ J_1, \ J_2, \ \varepsilon_1, \ \varepsilon_2, \ g].
\end{equation}
Then, the goal of our parameter estimation is to solve the optimization problem
\begin{equation}\label{eq:pares_opt_eq}
    \min_p \sum_{i=1}^{n} {\|{\bY}_i-\hat{\bY}_i(p)\|}_2.
\end{equation}
Of course, a (global) minimum of the above error identifies a choice of parameters for which the trajectories of the simulated double pendulum using \eqref{eq:edp_pares_ode} most closely aligns with that of the physical model.


To perform our parameter estimation we have used $n = 26$ different measurements, all sampled with sampling rate of $1kHz$. Each $\bY_i$ consists of $L = 2667$ consecutive data points. We solve the optimization problem \eqref{eq:pares_opt_eq} using MATLAB's optimization toolbox. Specifically, we employ the \textit{particleswarm} algorithm to achieve the global minimum. To better assist the convergence to the global optimum, we initialize with upper and lower bound constraints on the parameter values. These bounding values and their units are given component-wise by
\begin{equation}\label{eq:param_es_constraints}
    \begin{split}
        p_{l} &=[ 0.09\mathrm{kg}, \ 0.09\mathrm{kg}, \ 0.09\mathrm{m}, \ 0.10\mathrm{m}, \ 0.172719\mathrm{m}, \ 10^{-4}\mathrm{kgm}^2, \ 10^{-4}\mathrm{kgm}^2, \ 10^{-5}, \ 10^{-5}, \ 9.808\mathrm{m}/\mathrm{s}^2 ]\\
        p_u &= [0.14\mathrm{kg}, \ 0.14\mathrm{kg}, \ 0.16\mathrm{m}, \ 0.22\mathrm{m}, \ 0.172721\mathrm{m}, \ 0.002\mathrm{kgm}^2, \ 0.002\mathrm{kgm}^2, \ 10^{-3}, \ 10^{-3}, \ 9.809\mathrm{m}/\mathrm{s}^2].
    \end{split}
\end{equation}
As is apparent from \eqref{eq:param_es_constraints}, we begin the optimization process by being more certain of some parameter values than the other from the aid of simple measurement devices such as electric scales. In contrast, there is not a good way to directly measure the the inertial and friction coefficients of the pendulum arms, and so we have provided a larger variance in our upper and lower bounds. However, the CAD models of the pendulum arms can be used to have a rough guess of the inertial parameters. The result of the parameter estimation process using the minimization scheme \eqref{eq:pares_opt_eq} is 
\begin{equation}\label{eq:param_value}
    \begin{split}
    p_* &= [0.0938\mathrm{kg},\ 0.1376\mathrm{kg},\ 0.1086\mathrm{m},\ 0.1168\mathrm{m},\ 0.1727\mathrm{m},\\ & \qquad \qquad 10^{-4}\mathrm{kgm}^2,\ 10^{-4}\mathrm{kgm^2},\ 10^{-5},\ 10^{-5},\ 9.808\mathrm{m}/\mathrm{s}^2].
    \end{split}
\end{equation}
It should be noted that the parameters we identified in Eq.~\eqref{eq:param_value} is \textbf{NOT} the real physical parameters of the experimental double pendulum. Rather, it should be think of as a set of optimization variables that optimize the minimization scheme \eqref{eq:pares_opt_eq} while at the same time is numerically \textbf{close to} the actual experimental parameters. This is due to the fact that model we used does not capture all the dynamics of the pendulum arm and the data we collected is not perfectly clean\footnote{The estimated angular velocity of the pendulum arm deviates from the true velocity. This makes the forward simulation of Eq.~\ref{eq:edp_pares_ode} accumulates error due to the chaotic behavior of the double pendulum. This deviation of estimated velocity of the pendulum arm makes the estimated parameters deviates from the true physical value as well.}. Thus, there's rooms for the optimization algorithms to fine tune the optimization variables to compensate for unconsidered factors during the modeling. This is why the global optimization algorithm is used to find out a set of parameters that best compensate those unmodeled factors. Another important thing to note is, the identified model parameters provides a good simulated forward prediction performance if the dynamics of the double pendulum under prediction is similar to those used in parameters estimation. In other words, using data set that covers different parts of the double pendulum's phase space will result various optimal solution for Eq.~\eqref{eq:param_value}. This happens quite frequently when the control input is applied to the pendulum on the cart system, where often times an extra step of parameter tuning is needed to compensate the unmodeled behavior of the pendulum as shown in~\cite{graichen2007swing,gluck2013swing,jahn2021design}. Another interesting thing we noticed is that the parameters estimated using a higher energy level of double pendulum's motion requires a longer time to converge to the optimal solution. Moreover, it might not provide a good forward prediction performance for the low energy motions. We suspect the increase of the training time is due to the fact that the higher energy level motion is more sensitive to the change of parameters. As for the worse prediction performance on the low energy data, we believe it is caused by too much tuning been made in the high energy level data set to compensate for the unmodeled factors, since the unmodeled factors plays more role in the high energy data set. This is why we included $n=26$ different measurements, including low, medium, and high energy level motion of the pendulum to get a balance performance on prediction performance. The parameter estimation task of the single and triple pendulum is similar to the one for double pendulum and is omitted here. Instead, the estimated parameter is directly given in Table.~\ref{table:EstimatedParameters}.

\section{Bill of Materials}
\label{Appendix:BillofMaterials}

\subsection{Bill of Materials for Building Pendulum Arm}
All the materials needed to build the pendulum arm shown in Fig.~\ref{fig:PendulumArmOverview} can be seen in Table.~\ref{table:PendulumArmBill}. The price of some of the parts are not shown and the readers need to request a quote from the company to get its price. Other parts such as screws are sold in pack, for those parts the price for one pack is listed and is indicated by "*" symbol. Besides the materials mentioned in Table.~\ref{table:PendulumArmBill}, we also recommend ordering 1008-1010 carbon steel ring shim set~(1/4" ID, McMaster 3088A928) to accommodate for the manufacturing error if needed.

\small
\begin{longtable}[c]{|c||c|c|c|c|c|}
\caption{Materials needed to build the pendulum arm shown in Fig.~\ref{fig:PendulumArmOverview}.}
\label{table:PendulumArmBill}\\
\hline
\textbf{Designator} &
  \textbf{Component} &
  \textbf{Number} &
  \textbf{\begin{tabular}[c]{@{}c@{}}Total\\ cost-currency\end{tabular}} &
  \textbf{\begin{tabular}[c]{@{}c@{}}Source \\ of materials\end{tabular}} &
  \textbf{Material type} \\ \hline\hline
\endfirsthead
\multicolumn{6}{c}%
{{\bfseries Table \thetable\ continued from previous page}} \\
\hline
\textbf{Designator} &
  \textbf{Component} &
  \textbf{Number} &
  \textbf{\begin{tabular}[c]{@{}c@{}}Total\\ cost-currency\end{tabular}} &
  \textbf{\begin{tabular}[c]{@{}c@{}}Source \\ of materials\end{tabular}} &
  \textbf{Material type} \\ \hline\hline
\endhead
Arm1Shaft~($1$) &
  \begin{tabular}[c]{@{}c@{}}McMaster\\ 1346K17\end{tabular} &
  $1$ &
  $9.61$ &
  \href{https://www.mcmaster.com/1346K17/}{McMaster } &
  1566 Carbon Steel \\ \hline
\begin{tabular}[c]{@{}c@{}}SlipRing\\ 8 Channel~($2$)\end{tabular} &
  \begin{tabular}[c]{@{}c@{}}Moog\\ AC2690-8\end{tabular} &
  $1$ &
  NA &
  \href{https://www.moog.com/products/slip-rings/commercial-industrial-slip-rings/separates/ac2690.html}{Moog} &
  Gold and Others \\ \hline
Arm1~($3$) &
  \begin{tabular}[c]{@{}c@{}}McMaster\\ 8975K224\end{tabular} &
  $1$ &
  $27.4$ &
  \href{https://www.mcmaster.com/8975K224/}{McMaster} &
  Aluminum \\ \hline
91251A077~($4$) &
  \begin{tabular}[c]{@{}c@{}}McMaster\\ 91251A077\end{tabular} &
  $16$ &
  $8.74^*$ &
  \href{https://www.mcmaster.com/91251A077/}{McMaster} &
  \begin{tabular}[c]{@{}c@{}}Black-Oxide\\ Alloy Steel\end{tabular} \\ \hline
WireClip~($5$) &
  3D Printed &
  $4$ &
  NA &
  NA &
  PLA \\ \hline
EncoderShim~($6$) &
  3D Printed &
  $2$ &
  NA &
  NA &
  PLA \\ \hline
Encoder~($7$) &
  \begin{tabular}[c]{@{}c@{}}US Digital\\ EM2-2-10000-I\end{tabular} &
  $2$ &
  $87.02$ &
  \href{https://www.usdigital.com/products/encoders/incremental/modules/em2/?q=EM2-2-10000-I}{US Digital} &
  Others \\ \hline
91251A081~($8$) &
  \begin{tabular}[c]{@{}c@{}}McMaster\\ 91251A081\end{tabular} &
  $4$ &
  $9.01^*$ &
  \href{https://www.mcmaster.com/91251A081/}{McMaster} &
  \begin{tabular}[c]{@{}c@{}}Black-Oxide\\ Alloy Steel\end{tabular} \\ \hline
BrushShim~($9$) &
  3D Printed &
  $1$ &
  NA &
  NA &
  PLA \\ \hline
BrushBlock~($10$) &
  Moog AC259-5 &
  $1$ &
  NA &
  \href{https://www.moog.com/products/slip-rings/commercial-industrial-slip-rings/separates/ac2690.html}{Moog} &
  Gold and Others \\ \hline
3088A449~($11$) &
  \begin{tabular}[c]{@{}c@{}}McMaster\\ 3088A449\end{tabular} &
  $2$ &
  $7.13^*$ &
  \href{https://www.mcmaster.com/3088A449/}{McMaster} &
  Steel \\ \hline
\begin{tabular}[c]{@{}c@{}}R188~\footnote{R188 Hybrid Ceramic Bearings (1/4 X 1/2 X 1/8).\\ Bearing Material: Chrome Steel 52100. Rolling Element: Silicon Nitride. Cage: PTFE. Enclosure: Open. Rating: ABEC 7. Lube: Dry.}~($12$)\end{tabular} &
  \begin{tabular}[c]{@{}c@{}}R188 Hybrid\\ Ceramic Bearing\end{tabular} &
  $4$ &
  $105.60$ &
  \href{https://www.ortechceramics.com/products/ceramic-bearings/hybrid-ceramic-bearings/hybrid-deep-groove-ball-bearings-inch/r188-hybrid-ceramic-bearings-14-x-12-x-18/}{Ortech} &
  \begin{tabular}[c]{@{}c@{}}Ceramic, Steel,\\ and Others\end{tabular} \\ \hline
Arm2Shaft~($13$) &
  \begin{tabular}[c]{@{}c@{}}McMaster\\ 1346K11\end{tabular} &
  $1$ &
  $8.62$ &
  \href{https://www.mcmaster.com/1346K11/}{McMaster} &
  1566 Carbon Steel \\ \hline
\begin{tabular}[c]{@{}c@{}}SlipRing\\ 5 Channel~($14$)\end{tabular} &
  \begin{tabular}[c]{@{}c@{}}Moog\\ AC2690-5\end{tabular} &
  $1$ &
  NA &
  \href{https://www.moog.com/products/slip-rings/commercial-industrial-slip-rings/separates/ac2690.html}{Moog} &
  Gold and Others \\ \hline
Arm2~($15$) &
  \begin{tabular}[c]{@{}c@{}}McMaster\\ 8975K224\end{tabular} &
  $1$ &
  $27.4$ &
  \href{https://www.mcmaster.com/8975K224/}{McMaster} &
  Aluminum \\ \hline
97022A213~($16$) &
  \begin{tabular}[c]{@{}c@{}}McMaster\\ 97022A213\end{tabular} &
  $2$ &
  $5.34^*$ &
  \href{https://www.mcmaster.com/97022A213/}{McMaster} &
  316 Stainless Steel \\ \hline
EncoderDisk~($17$) &
  \begin{tabular}[c]{@{}c@{}}US Digital\\ HUBDISK-2\\ -10000-250-IE\end{tabular} &
  $2$ &
  $74.16$ &
  \href{https://www.usdigital.com/products/encoders/incremental/components/hubdisk-2/?q=HUBDISK-2-10000-250-IE}{US Digital} &
  Others \\ \hline
98126A447~($18$) &
  \begin{tabular}[c]{@{}c@{}}McMaster \\ 98126A447\end{tabular} &
  $2$ &
  $6.82^*$ &
  \href{https://www.mcmaster.com/98126A447/}{McMaster} &
  18-8 Stainless Steel \\ \hline
97633A130~($19$) &
  \begin{tabular}[c]{@{}c@{}}McMaster\\ 97633A130\end{tabular} &
  $2$ &
  $9.77^*$ &
  \href{https://www.mcmaster.com/97633A130/}{McMaster} &
  \begin{tabular}[c]{@{}c@{}}1060-1090\\ Spring Steel\end{tabular} \\ \hline
BearingPlate~($20$) &
  \begin{tabular}[c]{@{}c@{}}McMaster\\ 89015K255\end{tabular} &
  $2$ &
  $31.7$ &
  \href{https://www.mcmaster.com/89015K255 /}{McMaster} &
  Aluminum \\ \hline
Arm1Cover~($21$) &
  3D Printed &
  $1$ &
  NA &
  NA &
  PLA \\ \hline
Arm3Shaft~($23$) &
  \begin{tabular}[c]{@{}c@{}}McMaster\\ 1346K11\end{tabular} &
  $1$ &
  $8.62$ &
  \href{https://www.mcmaster.com/1346K11/}{McMaster} &
  1566 Carbon Steel \\ \hline
Arm3~($23$) &
  \begin{tabular}[c]{@{}c@{}}McMaster\\ 8975K224\end{tabular} &
  $1$ &
  $27.4$ &
  \href{https://www.mcmaster.com/8975K224/}{McMaster} &
  Aluminum \\ \hline
Arm2Cover~($24$) &
  3D Printed &
  $1$ &
  NA &
  NA &
  PLA \\ \hline
\end{longtable}
\normalsize

\subsection{Bill of Materials for Building Pendulum Cart}
All the materials needed to build the above mentioned pendulum cart can be seen in Table.~\ref{table:CartBill}. Some of the components do have a public price information, thus we refer the reader to contact the corresponding manufacture and ask for a quote. As for some bolts and nuts, they are sold in packs. Thus, the price of one pack is listed and we indicate this using a "$^*$" symbol. Besides the materials mentioned in Table.~\ref{table:CartBill}, we also recommend ordering 1074-1095 Spring Steel Ring Shim~(McMaster 98055A117: 0.1mm thick, 10mm ID. McMaster 98055A121: 0.5mm Thick, 10mm ID) to serve as replacement of 3D printed shim~($35$) if the printer precision is not high.

\small
\begin{longtable}[c]{|c||c|c|c|c|c|}
\caption{Materials needed to build the pendulum cart in Fig.~\ref{fig:PendulumCartOverview}.}
\label{table:CartBill}\\
\hline
\textbf{Designator} &
  \textbf{Component} &
  \textbf{Number} &
  \textbf{\begin{tabular}[c]{@{}c@{}}Total\\ cost-currency\end{tabular}} &
  \textbf{\begin{tabular}[c]{@{}c@{}}Source \\ of materials\end{tabular}} &
  \textbf{Material type} \\ \hline\hline
\endfirsthead
\multicolumn{6}{c}%
{{\bfseries Table \thetable\ continued from previous page}} \\
\hline
\textbf{Designator} &
  \textbf{Component} &
  \textbf{Number} &
  \textbf{\begin{tabular}[c]{@{}c@{}}Total\\ cost-currency\end{tabular}} &
  \textbf{\begin{tabular}[c]{@{}c@{}}Source \\ of materials\end{tabular}} &
  \textbf{Material type} \\ \hline\hline
\endhead
91251A077~($4$) &
  \begin{tabular}[c]{@{}c@{}}McMaster\\ 91251A077\end{tabular} &
  $6$ &
  $8.74$ &
  \href{https://www.mcmaster.com/91251A077/}{McMaster} &
  \begin{tabular}[c]{@{}c@{}}Black-Oxide\\ Alloy Steel\end{tabular} \\ \hline
Encoder~($7$) &
  \begin{tabular}[c]{@{}c@{}}US Digital\\ EM2-2-10000-I\end{tabular} &
  $1$ &
  $43.51$ &
  \href{https://www.usdigital.com/products/encoders/incremental/modules/em2/?q=EM2-2-10000-I}{US Digital} &
  Others \\ \hline
DriverClip~($25$) &
  3D Printed &
  $1$ &
  NA &
  NA &
  PLA \\ \hline
CableDriver~($26$) &
  \begin{tabular}[c]{@{}c@{}}US Digital\\ PC4-H10\end{tabular} &
  $3$ &
  $51.09$ &
  \href{https://www.usdigital.com/products/accessories/interfaces/cable-drivers/pc4/?q=PC4-H10}{US Digital} &
  Others \\ \hline
96144A111~($27$) &
  \begin{tabular}[c]{@{}c@{}}McMaster\\ 96144A111\end{tabular} &
  $6$ &
  $21.24$ &
  \href{https://www.mcmaster.com/96144A111/}{McMaster} &
  \begin{tabular}[c]{@{}c@{}}Black-Oxide\\ Alloy Steel\end{tabular} \\ \hline
3329A310~($28$) &
  \begin{tabular}[c]{@{}c@{}}McMaster\\ 3329A310\end{tabular} &
  $1$ &
  $28.1$ &
  \href{https://www.mcmaster.com/3329A310/}{McMaster} &
  \begin{tabular}[c]{@{}c@{}}Chrome-Plated \\ Brass and Others\end{tabular} \\ \hline
CartPlate~($29$) &
  \begin{tabular}[c]{@{}c@{}}McMaster\\ 9246K486\end{tabular} &
  $1$ &
  $26.69$ &
  \href{https://www.mcmaster.com/9246K486/}{McMaster} &
  Aluminum \\ \hline
LinearMotor~($30$) &
  \begin{tabular}[c]{@{}c@{}}HIWIN LMX1K\\ -SA12-1-2000\\ -PGS1-V103+HS\end{tabular} &
  $1$ &
  NA &
  \href{https://motioncontrolsystems.hiwin.us/request/all-categories}{HIWIN} &
  \begin{tabular}[c]{@{}c@{}}Aluminum and\\ Others\end{tabular} \\ \hline
61900~\footnote{61900 Hybrid Ceramic Bearings (10 X 22 X 6). Bearing Material: Chorme Steel 52100. Ball Material: Silicon Nitride. Cage/Retainer: PTFE. Enclosure: Open. Rating(ABEC): 7. Bearing Internal Clearance: Tight. Radial Play C2. Lube: Dry.}~($31$) &
  \begin{tabular}[c]{@{}c@{}}61900 Hybrid \\ Ceramic Bearing\end{tabular} &
  $2$ &
  $81.9$ &
  \href{https://www.ortechceramics.com/products/ceramic-bearings/hybrid-ceramic-bearings/hybrid-deep-groove-ball-bearings/61900-hybrid-ceramic-bearings-10x22x6/}{Ortech} &
  \begin{tabular}[c]{@{}c@{}}Ceramic, Steel,\\ and Others\end{tabular} \\ \hline
Spacer~($32$) &
  3D Printed &
  $1$ &
  NA &
  NA &
  PLA \\ \hline
9140T273~($33$) &
  \begin{tabular}[c]{@{}c@{}}McMaster\\ 9140T273\end{tabular} &
  $1$ &
  $39.23$ &
  \href{https://www.mcmaster.com/9140T273/}{McMaster} &
  Aluminum \\ \hline
98055A122~($34$) &
  \begin{tabular}[c]{@{}c@{}}McMaster\\ 98055A122\end{tabular} &
  $1$ &
  $10.69^*$ &
  \href{https://www.mcmaster.com/98055A122/}{McMaster} &
  \begin{tabular}[c]{@{}c@{}}1074-1095\\ Spring Steel\end{tabular} \\ \hline
Shim~($35$) &
  3D Printed &
  $1$ &
  NA &
  NA &
  PLA \\ \hline
EncoderDisk2~($36$) &
  \begin{tabular}[c]{@{}c@{}}US Digital \\ HUBDISK-2\\ -10000-394-IE\end{tabular} &
  $1$ &
  $37.06$ &
  \href{https://www.usdigital.com/products/encoders/incremental/components/hubdisk-2/?q=HUBDISK-2-10000-394-IE}{US Digital} &
  Others \\ \hline
90326A105~($37$) &
  \begin{tabular}[c]{@{}c@{}}McMaster\\ 90326A105\end{tabular} &
  $1$ &
  $6.35$ &
  \href{https://www.mcmaster.com/90326A105/}{McMaster} &
  Steel \\ \hline
BlockBase~($38$) &
  3D Printed &
  $1$ &
  NA &
  NA &
  NA \\ \hline
9125A471~($39$) &
  \begin{tabular}[c]{@{}c@{}}McMaster\\ 91251A471\end{tabular} &
  $2$ &
  $8.35^*$ &
  \href{https://www.mcmaster.com/91251A471/}{McMaster} &
  \begin{tabular}[c]{@{}c@{}}Black-Oxide\\ Alloy Steel\end{tabular} \\ \hline
BrushBlock-8~($40$) &
  \begin{tabular}[c]{@{}c@{}}Moog\\ AC259-8\end{tabular} &
  $1$ &
  NA &
  \href{https://www.moog.com/products/slip-rings/commercial-industrial-slip-rings/separates/ac2690.html}{Moog} &
  \begin{tabular}[c]{@{}c@{}}Gold and \\ Others\end{tabular} \\ \hline
\begin{tabular}[c]{@{}c@{}}LimitSwitch\\ Plate1~($41$)\end{tabular} &
  \begin{tabular}[c]{@{}c@{}}McMaster\\ 89015K113\end{tabular} &
  $1$ &
  $4.35^*$ &
  \href{https://www.mcmaster.com/89015K113/}{McMaster} &
  Aluminum \\ \hline
90751A110~($42$) &
  \begin{tabular}[c]{@{}c@{}}McMaster\\ 90751A110\end{tabular} &
  $4$ &
  $8.86^*$ &
  \href{https://www.mcmaster.com/90751A110/}{McMaster} &
  \begin{tabular}[c]{@{}c@{}}18-8\\ Stainless Steel\end{tabular} \\ \hline
\begin{tabular}[c]{@{}c@{}}LimitSwitch\\ Plate2~($43$)\end{tabular} &
  \begin{tabular}[c]{@{}c@{}}McMaster\\ 89015K113\end{tabular} &
  $1$ &
  $4.35^*$ &
  \href{https://www.mcmaster.com/89015K113/}{McMaster} &
  Aluminum \\ \hline
6000N312~($44$) &
  \begin{tabular}[c]{@{}c@{}}McMaster\\ 6000N312\end{tabular} &
  $2$ &
  $3.96$ &
  \href{https://www.mcmaster.com/6000N312/}{McMaster} &
  \begin{tabular}[c]{@{}c@{}}Stainless\\ Steel\end{tabular} \\ \hline
\begin{tabular}[c]{@{}c@{}}LimitSwitch\\ Base~($45$)\end{tabular} &
  3D Printed &
  $1$ &
  NA &
  NA &
  PLA \\ \hline
90592A085~($46$) &
  \begin{tabular}[c]{@{}c@{}}McMaster\\ 90592A085\end{tabular} &
  $4$ &
  $1.44^*$ &
  \href{https://www.mcmaster.com/90592A085/}{McMaster} &
  Steel \\ \hline
91290A318~($47$) &
  \begin{tabular}[c]{@{}c@{}}McMaster\\ 91290A318\end{tabular} &
  $2$ &
  $14.37^*$ &
  \href{https://www.mcmaster.com/91290A318/}{McMaster} &
  Steel \\ \hline
91290A121~($48$) &
  \begin{tabular}[c]{@{}c@{}}McMaster\\ 91290A121\end{tabular} &
  $2$ &
  $14.33^*$ &
  \href{https://www.mcmaster.com/91290A121/}{McMaster} &
  Alloy Steel \\ \hline
LimitSwitch~($49$) &
  \begin{tabular}[c]{@{}c@{}}Omron\\ EE-SX671\\ (PNP)\end{tabular} &
  $2$ &
  $64.06$ &
  \href{https://www.newark.com/omron-industrial-automation/ee-sx671/optical-sensor-transmissive-slotted/dp/52F3828}{Omron} &
  Ohters \\ \hline
\end{longtable}
\normalsize



\subsection{Bill of Materials for System Frame}
The materials needed to build the system frame can be summarized in Table.~\ref{table:SystemFrame}. It is worth noting that component $56, 57, 62$ and $63$ already comes with fastener, so there's no need to purchase additional fasteners. Moreover, the component $50, 55, 58$ and $61$ can be directly purchased with the desired length by ordering McMaster 5537T907. This can ease the manufacturing process but will increase the overall price.

\small
\begin{longtable}[c]{|c||c|c|c|c|c|}
\caption{Materials needed to build the pendulum cart in Fig.~\ref{fig:FrameOverview}.}
\label{table:SystemFrame}\\
\hline
\textbf{Designator} &
  \textbf{Component} &
  \textbf{Number} &
  \textbf{\begin{tabular}[c]{@{}c@{}}Total\\ cost-currency\end{tabular}} &
  \textbf{\begin{tabular}[c]{@{}c@{}}Source \\ of materials\end{tabular}} &
  \textbf{Material type} \\ \hline\hline
\endfirsthead
\multicolumn{6}{c}%
{{\bfseries Table \thetable\ continued from previous page}} \\
\hline
\textbf{Designator} &
  \textbf{Component} &
  \textbf{Number} &
  \textbf{\begin{tabular}[c]{@{}c@{}}Total\\ cost-currency\end{tabular}} &
  \textbf{\begin{tabular}[c]{@{}c@{}}Source \\ of materials\end{tabular}} &
  \textbf{Material type} \\ \hline\hline
\endhead
Extrusion1-4~($50$) &
  \begin{tabular}[c]{@{}c@{}}McMaster\\ 5537T103-4\end{tabular} &
  $8$ &
  $261.52$ &
  \href{https://www.mcmaster.com/5537T103/}{McMaster} &
  \begin{tabular}[c]{@{}c@{}}6560 \\ Aluminum\end{tabular} \\ \hline
X-Connector~($51$) &
  \begin{tabular}[c]{@{}c@{}}McMaster\\ 89015K255\end{tabular} &
  $4$ &
  $63.4$ &
  \href{https://www.mcmaster.com/89015K255/}{McMaster} &
  \begin{tabular}[c]{@{}c@{}}6061 \\ Aluminum\end{tabular} \\ \hline
AlignTool1~($52$) &
  3D Printed &
  $8$ &
  NA &
  NA &
  PLA \\ \hline
AlignTool2~($53$) &
  3D Printed &
  $8$ &
  NA &
  NA &
  PLA \\ \hline
Drop-In Nut~($54$) &
  \begin{tabular}[c]{@{}c@{}}McMaster\\ 5537T271\end{tabular} &
  $16$ &
  $34.56$ &
  \href{https://www.mcmaster.com/5537T271/}{McMaster} &
  \begin{tabular}[c]{@{}c@{}}Zinc-Plated \\ Steel\end{tabular} \\ \hline
Extrusion2-4~($55$) &
  \begin{tabular}[c]{@{}c@{}}McMaster\\ 5537T103-4\end{tabular} &
  $4$ &
  $130.76$ &
  \href{https://www.mcmaster.com/5537T103/}{McMaster} &
  \begin{tabular}[c]{@{}c@{}}6560 \\ Aluminum\end{tabular} \\ \hline
47065T998~($56$) &
  \begin{tabular}[c]{@{}c@{}}McMaster\\ 47065T998\end{tabular} &
  $10$ &
  $166.1$ &
  \href{https://www.mcmaster.com/47065T998/}{McMaster} &
  \begin{tabular}[c]{@{}c@{}}Anodized \\ Aluminum\end{tabular} \\ \hline
47065T984~($57$) &
  \begin{tabular}[c]{@{}c@{}}McMaster\\ 47065T948\end{tabular} &
  $2$ &
  $32.48$ &
  \href{https://www.mcmaster.com/47065T948/}{McMaster} &
  \begin{tabular}[c]{@{}c@{}}Anodized \\ Aluminum\end{tabular} \\ \hline
Extrusion3~($58$) &
  \begin{tabular}[c]{@{}c@{}}McMaster\\ 5537T103-2\end{tabular} &
  $2$ &
  $35.84$ &
  \href{https://www.mcmaster.com/5537T103/}{McMaster} &
  \begin{tabular}[c]{@{}c@{}}6560 \\ Aluminum\end{tabular} \\ \hline
Extrusion4~($59$) &
  \begin{tabular}[c]{@{}c@{}}Leftover material \\ when producing~($58$)\end{tabular} &
  $2$ &
  NA &
  \href{https://www.mcmaster.com/5537T271/}{McMaster} &
  \begin{tabular}[c]{@{}c@{}}6560 \\ Aluminum\end{tabular} \\ \hline
5537T323~($60$) &
  \begin{tabular}[c]{@{}c@{}}McMaster\\ 5537T323\end{tabular} &
  $20$ &
  $101$ &
  \href{https://www.mcmaster.com/5537T323/}{McMaster} &
  Zinc \\ \hline
Extrusion5~($61$) &
  \begin{tabular}[c]{@{}c@{}}McMaster\\ 5537T103-2\end{tabular} &
  $2$ &
  $35.84$ &
  \href{https://www.mcmaster.com/5537T103/}{McMaster} &
  \begin{tabular}[c]{@{}c@{}}6560 \\ Aluminum\end{tabular} \\ \hline
Wheel~($62$) &
  \begin{tabular}[c]{@{}c@{}}McMaster\\ 4706T83\end{tabular} &
  $4$ &
  $83.52$ &
  \href{https://www.mcmaster.com/47065T83/}{McMaster} &
  \begin{tabular}[c]{@{}c@{}}Zinc-Plated Steel\\ and Rubber\end{tabular} \\ \hline
5537T37~($63$) &
  \begin{tabular}[c]{@{}c@{}}McMaster\\ 5537T37\end{tabular} &
  $4$ &
  $105.76$ &
  \href{https://www.mcmaster.com/5537T37/}{McMaster} &
  \begin{tabular}[c]{@{}c@{}}Anodized \\ Aluminum\end{tabular} \\ \hline
Extrusion6~($64$) &
  \begin{tabular}[c]{@{}c@{}}McMaster\\ 5537T103-6\end{tabular} &
  $2$ &
  $92.54$ &
  \href{https://www.mcmaster.com/5537T103/}{McMaster} &
  \begin{tabular}[c]{@{}c@{}}6560 \\ Aluminum\end{tabular} \\ \hline
\end{longtable}
\normalsize


\subsection{Bill of Materials for Real-Time System}
This section summarizes the materials needed to build the real-time system shown in Fig.\ref{fig:RealTimeSysOverview}. All materials needed is shown in Table.~\ref{table:RealTimeSys}. Since monitor~($69$) does not affect the entire system's performance, we will not include the detailed information on which monitor we used. As long as the monitor the readers choose to use has a display port, it can connect with the target system. The host computer we used is quite old, so we decided not to put its price. The readers can use any computer they like as a host computer if it supports running Matlab and has an Ethernet port. As for the price of Speedgoat product, please send a quote to Speedgoat for more details.

\small
\begin{longtable}[c]{|c||c|c|c|c|c|}
\caption{Materials needed to build the real-time system.}
\label{table:RealTimeSys}\\
\hline
\textbf{Designator} &
  \textbf{Component} &
  \textbf{Number} &
  \textbf{\begin{tabular}[c]{@{}c@{}}Total\\ cost-currency\end{tabular}} &
  \textbf{\begin{tabular}[c]{@{}c@{}}Source \\ of materials\end{tabular}} &
  \textbf{Material type} \\ \hline\hline
\endfirsthead
\multicolumn{6}{c}%
{{\bfseries Table \thetable\ continued from previous page}} \\
\hline
\textbf{Designator} &
  \textbf{Component} &
  \textbf{Number} &
  \textbf{\begin{tabular}[c]{@{}c@{}}Total\\ cost-currency\end{tabular}} &
  \textbf{\begin{tabular}[c]{@{}c@{}}Source \\ of materials\end{tabular}} &
  \textbf{Material type} \\ \hline\hline
\endhead
\begin{tabular}[c]{@{}c@{}}Terminal\\ Block~($65$)\end{tabular} &
  \begin{tabular}[c]{@{}c@{}}Speedgoat \\ Terminal \\ Board \\ 2$\times$17 Pin\end{tabular} &
  $2$ &
  NA &
  \href{https://www.speedgoat.com/company/contact-us}{Speedgoat} &
  Others \\ \hline
Cable~($66$) &
  \begin{tabular}[c]{@{}c@{}}M12M to \\ M12F 17 Pin \\ 3 feet Cable\end{tabular} &
  $4$ &
  NA &
  \href{https://www.speedgoat.com/company/contact-us}{Speedgoat} &
  Others \\ \hline
\begin{tabular}[c]{@{}c@{}}IO-191\\ ($67-1$)\end{tabular} &
  \begin{tabular}[c]{@{}c@{}}IO-191\\ -Baseline\end{tabular} &
  $1$ &
  NA &
  \href{https://www.speedgoat.com/company/contact-us}{Speedgoat} &
  Others \\ \hline
\begin{tabular}[c]{@{}c@{}}IO-392\\ ($67-2$)\end{tabular} &
  \begin{tabular}[c]{@{}c@{}}IO-392\\ -Baseline\end{tabular} &
  $1$ &
  NA &
  \href{https://www.speedgoat.com/company/contact-us}{Speedgoat} &
  Others \\ \hline
\begin{tabular}[c]{@{}c@{}}TargetMachine\\ ($67-3$)\end{tabular} &
  \begin{tabular}[c]{@{}c@{}}Baseline \\ Education\\ Real-Time \\ Target Machine\end{tabular} &
  $1$ &
  NA &
  \href{https://www.speedgoat.com/company/contact-us}{Speedgoat} &
  Others \\ \hline
HostComputer~($68$) &
  NA &
  $1$ &
  NA &
  NA &
  Others \\ \hline
Monitor~($69$) &
  NA &
  $2$ &
  NA &
  NA &
  Others \\ \hline
5537T698~($70$) &
  \begin{tabular}[c]{@{}c@{}}McMaster\\ 5537T698\end{tabular} &
  $2$ &
  $2.14$ &
  \href{https://www.mcmaster.com/5537T698/}{McMaster} &
  \begin{tabular}[c]{@{}c@{}}Silver \\ Zinc-Plated\\ Steel\end{tabular} \\ \hline
8961K102~($71$) &
  \begin{tabular}[c]{@{}c@{}}McMaster\\ 8961K102\end{tabular} &
  $1$ &
  $3.9$ &
  \href{https://www.mcmaster.com/8961K102/}{McMaster} &
  \begin{tabular}[c]{@{}c@{}}Zinc-Plated\\ Steel\end{tabular} \\ \hline
\begin{tabular}[c]{@{}c@{}}IO-392\\ DAQ4RS422~($72$)\end{tabular} &
  \begin{tabular}[c]{@{}c@{}}IO-392 \\ Configuration \\ File\end{tabular} &
  $1$ &
  NA &
  \href{https://www.speedgoat.com/company/contact-us}{Speedgoat} &
  NA \\ \hline
\end{longtable}
\normalsize

\subsection{Bill of Materials for the Electronic Parts}

This section summarize necessary components needed in the electrical system of the multi-link pendulum on the cart system. All the components needed in the electrical parts are shown in Table.~\ref{table:ElectricalParts}. Some of the components needed such as cables, DIN rail, DIN rail terminal blocks , wires, etc, are not shown in Fig.~\ref{fig:MotorEletricalPartOverview} and is included in Table.~\ref{table:ElectricalParts}. 

\small
\begin{longtable}[c]{|c||c|c|c|c|c|}
\caption{Materials needed to build the electrical part of the system.}
\label{table:ElectricalParts}\\
\hline
\textbf{Designator} &
  \textbf{Component} &
  \textbf{Number} &
  \textbf{\begin{tabular}[c]{@{}c@{}}Total\\ cost-currency\end{tabular}} &
  \textbf{\begin{tabular}[c]{@{}c@{}}Source \\ of materials\end{tabular}} &
  \textbf{Material type} \\ \hline\hline
\endfirsthead
\multicolumn{6}{c}%
{{\bfseries Table \thetable\ continued from previous page}} \\
\hline
\textbf{Designator} &
  \textbf{Component} &
  \textbf{Number} &
  \textbf{\begin{tabular}[c]{@{}c@{}}Total\\ cost-currency\end{tabular}} &
  \textbf{\begin{tabular}[c]{@{}c@{}}Source \\ of materials\end{tabular}} &
  \textbf{Material type} \\ \hline\hline
\endhead
\begin{tabular}[c]{@{}c@{}}EmergencyStop\\ Module~($75$)\end{tabular} &
  \begin{tabular}[c]{@{}c@{}}Powertec \\ 71354\end{tabular} &
  $1$ &
  $19.99$ &
  \href{https://powertecproducts.com/71354-magnetic-switch-120-v/}{Powertec} &
  Others \\ \hline
\begin{tabular}[c]{@{}c@{}}Circuit\\ Breaker~($76$)\end{tabular} &
  \begin{tabular}[c]{@{}c@{}}Eaton \\ FAZ-B3\\ -2-NA\end{tabular} &
  $1$ &
  $48$ &
  \begin{tabular}[c]{@{}c@{}}\href{https://www.automationdirect.com/adc/shopping/catalog/circuit_protection_-z-_fuses_-z-_disconnects/circuit_breakers_-a-_circuit_protectors/miniature_circuit_breakers_(mcb)/faz-b3-2-na}{Automation} \\ \href{https://www.automationdirect.com/adc/shopping/catalog/circuit_protection_-z-_fuses_-z-_disconnects/circuit_breakers_-a-_circuit_protectors/miniature_circuit_breakers_(mcb)/faz-b3-2-na}{Direct}\end{tabular} &
  Others \\ \hline
\begin{tabular}[c]{@{}c@{}}Magnetic\\ Contactor~($77$)\end{tabular} &
  \begin{tabular}[c]{@{}c@{}}WEG \\ CWC016\\ -00-22V18\end{tabular} &
  $1$ &
  $23.5$ &
  \begin{tabular}[c]{@{}c@{}}\href{https://www.automationdirect.com/adc/shopping/catalog/motor_controls/iec_magnetic_contactors/4-pole_miniature_contactors_(7_to_16_amp)/cwc016-00-22v18}{Automation}\\\href{https://www.automationdirect.com/adc/shopping/catalog/motor_controls/iec_magnetic_contactors/4-pole_miniature_contactors_(7_to_16_amp)/cwc016-00-22v18} {Direct}\end{tabular} &
  Others \\ \hline
\begin{tabular}[c]{@{}c@{}}Noise\\ Filter~($78$)\end{tabular} &
  \begin{tabular}[c]{@{}c@{}}Schaffner \\ FN2090\\ -10-06\end{tabular} &
  $1$ &
  $31.5$ &
  \begin{tabular}[c]{@{}c@{}}\href{https://www.digikey.com/en/products/detail/schaffner-emc-inc/FN2090-10-06/1928911}{Automation}\\ \href{https://www.digikey.com/en/products/detail/schaffner-emc-inc/FN2090-10-06/1928911}{Direct}\end{tabular} &
  Others \\ \hline
\begin{tabular}[c]{@{}c@{}}Line\\ Reactor~($78$)\end{tabular} &
  \begin{tabular}[c]{@{}c@{}}LR-11P0\\ -1PH\end{tabular} &
  $1$ &
  $139$ &
  \begin{tabular}[c]{@{}c@{}}\href{https://www.automationdirect.com/adc/shopping/catalog/retired_products/drives_-a-_soft_starters/lr-11p0-1ph}{Automation}\\ \href{https://www.automationdirect.com/adc/shopping/catalog/retired_products/drives_-a-_soft_starters/lr-11p0-1ph}{Direct}\end{tabular} &
  Others \\ \hline
\begin{tabular}[c]{@{}c@{}}Motor\\ Drive~($80$)\end{tabular} &
  \begin{tabular}[c]{@{}c@{}}HIWIN \\ D1 Drive\end{tabular} &
  $1$ &
  NA &
  \href{https://motioncontrolsystems.hiwin.us/request/all-categories}{HIWIN} &
  Others \\ \hline
\begin{tabular}[c]{@{}c@{}}Edge\\ Filter~($81$)\end{tabular} &
  \begin{tabular}[c]{@{}c@{}}HIWIN \\ MF-40-S\end{tabular} &
  $1$ &
  NA &
  \href{https://motioncontrolsystems.hiwin.us/request/all-categories}{HIWIN} &
  Others \\ \hline
\begin{tabular}[c]{@{}c@{}}Regenerative\\ Resistor~($82$)\end{tabular} &
  \begin{tabular}[c]{@{}c@{}}Yohii \\ 500W \\ 30 $\Omega$ \\ Resistor\end{tabular} &
  $2$ &
  $2.14$ &
  Amazon &
  Others \\ \hline
\begin{tabular}[c]{@{}c@{}}DC \\ Supply~($83$)\end{tabular} &
  \begin{tabular}[c]{@{}c@{}}MEAN WELL \\ MDR-60-24\end{tabular} &
  $1$ &
  $24.84$ &
  \begin{tabular}[c]{@{}c@{}}\href{https://www.meanwell-web.com/en-gb/ac-dc-industrial-din-rail-power-supply-output-mdr--60--24}{MEAN}\\ \href{https://www.meanwell-web.com/en-gb/ac-dc-industrial-din-rail-power-supply-output-mdr--60--24}{WELL}\end{tabular} &
  Others \\ \hline
Enclosure~($84$) &
  \begin{tabular}[c]{@{}c@{}}Hammond \\ CSKO242410\end{tabular} &
  $1$ &
  $171$ &
  \begin{tabular}[c]{@{}c@{}}\href{https://www.automationdirect.com/adc/shopping/catalog/enclosures_-z-_subpanels_-z-_thermal_management_-z-_lighting/enclosures/junction_boxes/csko242410}{Automation}\\\href{https://www.automationdirect.com/adc/shopping/catalog/enclosures_-z-_subpanels_-z-_thermal_management_-z-_lighting/enclosures/junction_boxes/csko242410}{Direct}\end{tabular} &
  \begin{tabular}[c]{@{}c@{}}Carbon \\ Steel\end{tabular} \\ \hline
\begin{tabular}[c]{@{}c@{}}Toggle\\ Switch~($85$)\end{tabular} &
  \begin{tabular}[c]{@{}c@{}}20A \\ 12VDC / \\ 120VAC \\ Toggle \\ Switch\end{tabular} &
  $1$ &
  $7.99$ &
  Amazon &
  Others \\ \hline
\begin{tabular}[c]{@{}c@{}}Momentary\\ Switch~($86$)\end{tabular} &
  \begin{tabular}[c]{@{}c@{}}LC LICTOP \\ AC 660V \\ Push Button \\ Switch\end{tabular} &
  $1$ &
  $6.99$ &
  Amazon &
  Others \\ \hline
\begin{tabular}[c]{@{}c@{}}EMI \\ Cores~($87$)\end{tabular} &
  \begin{tabular}[c]{@{}c@{}}HIWIN \\ D1-EMC1\end{tabular} &
  $2$ &
  NA &
  \href{https://motioncontrolsystems.hiwin.us/request/all-categories}{HIWIN} &
  Others \\ \hline
\begin{tabular}[c]{@{}c@{}}Enclosure\\ Panel~($88$)\end{tabular} &
  \begin{tabular}[c]{@{}c@{}}Hammond \\ CSFC2424\end{tabular} &
  $1$ &
  $21$ &
  \begin{tabular}[c]{@{}c@{}} \href{https://www.automationdirect.com/adc/shopping/catalog/enclosures_-z-_subpanels_-z-_thermal_management_-z-_lighting/enclosure_parts_-a-_accessories/flush_covers/csfc2424}{Automation}\\\ \href{https://www.automationdirect.com/adc/shopping/catalog/enclosures_-z-_subpanels_-z-_thermal_management_-z-_lighting/enclosure_parts_-a-_accessories/flush_covers/csfc2424}{Direct}\end{tabular}
  &
  \begin{tabular}[c]{@{}c@{}}Carbon \\ Steel\end{tabular} \\ \hline
\begin{tabular}[c]{@{}c@{}}DIN \\ Rail~($89$)\end{tabular} &
  DN-R35S1-2 &
  $1$ &
  $12.5$ &
  \begin{tabular}[c]{@{}c@{}} \href{https://www.automationdirect.com/adc/shopping/catalog/enclosures_-z-_subpanels_-z-_thermal_management_-z-_lighting/enclosure_parts_-a-_accessories/din_rails/dn-r35s1-2}{Automation}\\\ \href{https://www.automationdirect.com/adc/shopping/catalog/enclosures_-z-_subpanels_-z-_thermal_management_-z-_lighting/enclosure_parts_-a-_accessories/din_rails/dn-r35s1-2}{Direct}\end{tabular}
  &
  \begin{tabular}[c]{@{}c@{}}Plated \\ Steel\end{tabular} \\ \hline
\begin{tabular}[c]{@{}c@{}}14AWG \\ Cable~($90$)\end{tabular} &
  \begin{tabular}[c]{@{}c@{}}Amazon 14 \\ AWG Cable\end{tabular} &
  $1$ &
  NA &
  Amazon &
  Copper \\ \hline
\begin{tabular}[c]{@{}c@{}}12AWG \\ Cable~($91$)\end{tabular} &
  \begin{tabular}[c]{@{}c@{}}Amazon 12 \\ AWG Cable\end{tabular} &
  $1$ &
  NA &
  Amazon &
  Copper \\ \hline
\begin{tabular}[c]{@{}c@{}}10AWG \\ Cable~($92$)\end{tabular} &
  \begin{tabular}[c]{@{}c@{}}Amazon 10 \\ AWGCable\end{tabular} &
  $1$ &
  NA &
  Amazon &
  Copper \\ \hline
\begin{tabular}[c]{@{}c@{}}DIN Rail \\ Terminal \\ Block~($93$)\end{tabular} &
  \begin{tabular}[c]{@{}c@{}}Amazon \\ DIN Rail\\ Block\end{tabular} &
  $10$ &
  NA &
  Amazon &
  Others \\ \hline
\begin{tabular}[c]{@{}c@{}}DIN Rail \\ Ground\\ Terminal \\ Block~($94$)\end{tabular} &
  \begin{tabular}[c]{@{}c@{}}Amazon \\ DIN Rail \\ Terminal \\ Block\end{tabular} &
  $10$ &
  NA &
  Amazon &
  Others \\ \hline
\begin{tabular}[c]{@{}c@{}}DC\\ Connector\\ ($95$)\end{tabular} &
  \begin{tabular}[c]{@{}c@{}}Wago \\ 721-103/\\ 026-000\end{tabular} &
  $1$ &
  $4.03$ &
  \begin{tabular}[c]{@{}c@{}}\href{https://www.newark.com/wago/721-103-026-000/terminal-block-pluggable-3pos/dp/29K2205\\ }{Newark}\end{tabular} &
  Others \\ \hline
\begin{tabular}[c]{@{}c@{}}AC\\ Connector\\ ($96$)\end{tabular} &
  \begin{tabular}[c]{@{}c@{}}Wago \\ 721-204/\\ 026-000\end{tabular} &
  $1$ &
  $6.91$ &
  \href{https://www.digikey.com/en/products/detail/wago-corporation/721-204-026-000/15545040}{Digikey} &
  Others \\ \hline
\begin{tabular}[c]{@{}c@{}}MotorPower\\ Connector\\ ($97$)\end{tabular} &
  \begin{tabular}[c]{@{}c@{}}Wago \\ 721-104/\\ 026-000\end{tabular} &
  $1$ &
  $5.26$ &
  \href{https://www.digikey.com/en/products/detail/wago-corporation/721-204-026-000/15545040}{Digikey} &
  Others \\ \hline
\begin{tabular}[c]{@{}c@{}}EdgeFilter\\ Input\\ Connector\\ ($98$)\end{tabular} &
  \begin{tabular}[c]{@{}c@{}}Phoenix \\ Contact \\ 1718517\end{tabular} &
  $1$ &
  $11.3$ &
  \begin{tabular}[c]{@{}c@{}}Mouser \\ Electronics \\ Phoenix \\ Contact \\ 1718517\end{tabular} &
  Others \\ \hline
\begin{tabular}[c]{@{}c@{}}EdgeFilter\\ Output\\ Connector~($99$)\end{tabular} &
  \begin{tabular}[c]{@{}c@{}}Phoenix \\ Contact \\ 1718504\end{tabular} &
  $1$ &
  $11.3$ &
  \begin{tabular}[c]{@{}c@{}}Mouser \\ Electronics \\ Phoenix \\ Contact \\ 1718504\end{tabular} &
  Others \\ \hline
\begin{tabular}[c]{@{}c@{}}Regen\\ Connector~($100$)\end{tabular} &
  \begin{tabular}[c]{@{}c@{}}Wago \\ 723-603\end{tabular} &
  $1$ &
  $3.85$ &
  \href{https://www.galco.com/buy/Wago/723-603}{Galco} &
  Others \\ \hline
\begin{tabular}[c]{@{}c@{}}Communication\\ Cable~($101$)\end{tabular} &
  \begin{tabular}[c]{@{}c@{}}D1-DNT07A \\ USB232 \\ to RJ11 \\ Adapter Cable\end{tabular} &
  $1$ &
  $22$ &
  Amazon &
  Others \\ \hline
\begin{tabular}[c]{@{}c@{}}HookUp\\ Wire~($102$)\end{tabular} &
  \begin{tabular}[c]{@{}c@{}}24 AWG \\ Hook Up Wire\end{tabular} &
  $1$ &
  NA &
  Amazon &
  Copper \\ \hline
\begin{tabular}[c]{@{}c@{}}Shielding\\ Wire~($103$)\end{tabular} &
  \begin{tabular}[c]{@{}c@{}}Electriduct 1/2" \\ Tinned Copper \\ Metal Braid \\ Sleeving Flexible \\ EMI RFI \\ Shielding \\ Wire Mesh \\ (0.32" Diameter) \\ - 10 Feet\end{tabular} &
  $1$ &
  $21.88$ &
  Amazon &
  \begin{tabular}[c]{@{}c@{}}Tinned \\ Copper\\ Metal\end{tabular} \\ \hline
\begin{tabular}[c]{@{}c@{}}Cable\\ Sleeve\\ ($104$)\end{tabular} &
  \begin{tabular}[c]{@{}c@{}}Keco 100ft \\ – 1/2 inch \\ PET \\ Expandable\\  Braided \\ Cable Sleeve\end{tabular} &
  $1$ &
  $16.99$ &
  Amazon &
  Others \\ \hline
\begin{tabular}[c]{@{}c@{}}CN2\\ Connector\\ ($105$)\end{tabular} &
  \begin{tabular}[c]{@{}c@{}}3M \\ 10126-3000\end{tabular} &
  $1$ &
  $11.62$ &
  \href{https://www.digikey.com/en/products/detail/3m/10126-3000PE/773864}{Digikey} &
  Others \\ \hline
\begin{tabular}[c]{@{}c@{}}CN2\\ Connector\\ Cover~($106$)\end{tabular} &
  \begin{tabular}[c]{@{}c@{}}3M \\ 10326-52F0-008\end{tabular} &
  $1$ &
  $10.62$ &
  \href{https://www.digikey.com/en/products/detail/3m/10326-52F0-008/703410}{Digikey} &
  Others \\ \hline
\begin{tabular}[c]{@{}c@{}}Limit\\ Switch\\ Cable~($107$)\end{tabular} &
  \begin{tabular}[c]{@{}c@{}}HIWIN \\ D-Sub \\ 9-Pin Female \\ Limit Switch \\ Extension \\ Cable\end{tabular} &
  $1$ &
  NA &
  \href{https://motioncontrolsystems.hiwin.us/request/all-categories}{HIWIN} &
  Others \\ \hline
\end{longtable}
\normalsize

\end{document}